\documentclass[a4paper, 11pt]{article} 
\pdfoutput=1
\usepackage[utf8]{inputenc}
\usepackage[T1]{fontenc}
\usepackage[english]{babel}
\usepackage{lmodern, geometry,
	jheppub, amsmath, graphicx, xcolor, subfigure,
	feynmp, slashed, adjustbox, rotating, pdflscape}
\usepackage{subfigure}
\usepackage{hyperref}
\usepackage[pscoord]{eso-pic}
\usepackage{soul}

\graphicspath{{fig/}}

\makeatletter\g@addto@macro\bfseries{\boldmath}\makeatother

\allowdisplaybreaks

\def\TeV{\ifmmode {\mathrm{Te\kern -0.1em V}}\else
                   \textrm{Te\kern -0.1em V}\fi}%
\def\GeV{\ifmmode {\mathrm{Ge\kern -0.1em V}}\else
                   \textrm{Ge\kern -0.1em V}\fi}%
\def\MeV{\ifmmode {\mathrm{Me\kern -0.1em V}}\else
                   \textrm{Me\kern -0.1em V}\fi}%
\def\keV{\ifmmode {\mathrm{ke\kern -0.1em V}}\else
                   \textrm{ke\kern -0.1em V}\fi}%
\def\eV{\ifmmode  {\mathrm{e\kern -0.1em V}}\else
                   \textrm{e\kern -0.1em V}\fi}%
\let\tev=\TeV
\let\gev=\GeV

\def\iab{\mbox{ab$^{-1}$}}
\def\ifb{\mbox{fb$^{-1}$}}
\def\ipb{\mbox{pb$^{-1}$}}

\newcommand{\FDF}[1][]{\varphi^\dagger #1\!\overleftrightarrow{D}\!_\mu\varphi}
\newcommand{\FDFI}[1][]{\varphi^\dagger #1\!\overleftrightarrow{D}^I\!\!\!_\mu\:\varphi}

\newcommand{\ckm}{\ensuremath{V_\text{\tiny CKM}}}
\newcommand{\pmns}{\ensuremath{V_\text{\tiny PMNS}}}

\DeclareMathAlphabet{\mathsfit}{\encodingdefault}{\sfdefault}{m}{sl}
\newcommand{\ges}[1]{\mathsfit{#1}}

\definecolor{Darkgreen}{RGB}{30,120,30}
\definecolor{mypink}{RGB}{219, 48, 122}

\title{The electro-weak couplings of the top and bottom quarks -- global fit and future prospects}

\author[a]{Gauthier Durieux}
\author[b]{Adrian Irles}
\author[c]{V\'ictor Miralles}
\author[c]{Ana Pe\~nuelas}
\author[c]{Mart{\'i}n Perell{\'o}}
\author[b]{Roman P\"oschl}
\author[c]{Marcel~Vos}

\affiliation[a]{Department of Physics, Technion, Haifa, Israel}
\affiliation[b]{LAL, CNRS/IN2P3 et Université de Paris-Sud XI, Orsay, France}
\affiliation[c]{IFIC (UV/CSIC) Valencia, Spain}

\emailAdd{durieux@campus.technion.ac.il}
\emailAdd{irles@lal.in2p3.fr}
\emailAdd{victor.miralles@ific.uv.es}
\emailAdd{ana.penuelas@ific.uv.es}
\emailAdd{martin.perello@ific.uv.es}
\emailAdd{poeschl@lal.in2p3.fr}
\emailAdd{marcel.vos@ific.uv.es}

\abstract{
We evaluate the implications of LHC and LEP/SLC measurements for the electro-weak couplings of the top and bottom quarks.
We derive global bounds on the Wilson coefficients of ten two-fermion operators in an effective field theory description.
The combination of hadron collider data with $Z$-pole measurements is found to yield tight limits on the operator coefficients that modify the left-handed couplings of the bottom and top quark to the $Z$ boson.
We also present projections for the high-luminosity phase of the LHC and for future electron-positron colliders. 
The bounds on the operator coefficients are expected to improve substantially during the remaining LHC programme, by factors of $1$ to $5$ if systematic uncertainties are scaled as statistical ones.
The operation of an $e^+e^-$ collider at a center-of-mass energy above the top-quark pair production threshold is expected to further improve the bounds by one to two orders of magnitude.
The combination of measurements in $pp$ and $e^+e^-$ collisions allows for a percent-level determination of the top-quark Yukawa coupling, that is robust in a global fit.
}

\preprint{\begin{tabular}{@{}r@{}}IFIC/19-33\\FTUV/19-0724\end{tabular}}

\begin{document}

\maketitle

\section{Introduction}
\label{sec:intro}
With the discovery of the Higgs boson~\cite{Aad:2012tfa,Chatrchyan:2012xdj}
at the LHC, the particle content of the Standard Model (SM) is experimentally confirmed. 
Measurements are performed in a very broad range of production processes to characterize the interactions among all currently known particles.
Precision measurements may be affected by the presence of new particles or interactions and thus provide an indirect probe of new physics.
While the SM seems to stand all tests so far, experiments keep searching for subtle deviations from its predictions.

In this paper, we study the electro-weak (EW) couplings of the third-generation quarks which have particular relevance in many extensions of the SM.
In particular, the EW couplings of the top and bottom quarks have an exquisite sensitivity to a broad class of composite Higgs/extra dimension scenarios~\cite{Richard:2014upa, Durieux:2018ekg}. 

As the top quark escaped scrutiny at the previous generation of electron-positron colliders, the LHC measurements analyzed in this paper provide the first constraints on its EW couplings. We include measurements by ATLAS and CMS at a 
center-of-mass energy of 13~\tev{} of the associated $t\bar{t}X$ production rate
(with $X=\gamma, W, Z, H$), the single top-quark production cross section in
the $t$-channel, $Wt$ associated production and $tZq$ production as well as the $W$ helicity fractions in top-quark decay. 

As the left-handed top and bottom quarks are part of the same doublet, their couplings are related~\cite{Englert:2017dev, Durieux:2018tev}. 
We include measurements in bottom-quark pair production
at LEP and SLC in the fit. The precise measurements at the $Z$-pole of the ratio $R_b$ and the $b$-quark asymmetry parameter, $\mathcal{A}_b$, which is extracted from measurements of the left-right and forward-backward asymmetries in bottom-quark pair production, provide strong constraints.

An effective field theory (EFT) is employed to parameterize the effects of new physics arising at scales higher than that of the considered measurements.
The ten CP-conserving operator coefficients modifying $t\bar{t}Z$, $b\bar{b}Z$, $t\bar{b}W$ and $t\bar{t}H$ interactions are simultaneously constrained in a global analysis of LHC and LEP/SLC data.
Our results apply to new-physics models in which deviations to the measurements considered are dominated by these ten parameters.
Other contributions are sometimes already well constrained.
Those of four-fermion operators are notably not tightly bound yet, but inclusion of the effect of all dimension-six operators in a fully global analysis is beyond the scope of this work.

The fitting code, publicly available~\cite{hepfitgit}, is implemented in the {\scshape{\tt HEPfit}}~\cite{deBlas:2016ojx} package which uses a Markov-Chain Monte-Carlo implementation based on the Bayesian Analysis Toolkit~\cite{Caldwell:2008fw}.

In the next decade the LHC program is expected to sharpen the limits 
considerably. We define several scenarios for the expected measurement precision
after completing the LHC program, including the high-luminosity phase (HL-LHC~\cite{ApollinariG.:2017ojx}). We also assess the potential of a 
future $e^+e^-$ collider (either linear colliders, such as the 
International Linear Collider ILC~\cite{Bambade:2019fyw}, the Compact Linear Collider
CLIC~\cite{Charles:2018vfv}, or circular colliders such as FCCee~\cite{Abada:2019zxq}
or CEPC~\cite{CEPCStudyGroup:2018rmc}).
A Higgs factory operated at center-of-mass energy of 250~\gev{} will improve the constraints on the bottom-quark 
operator coefficients~\cite{Bilokin:2017lco} significantly.
Operation above the top-quark pair production threshold is part of the initial stage
of the CLIC project~\cite{Abramowicz:2018rjq} and later stages of the ILC and FCCee.  
In this paper the potential of operation at a center-of-mass energy of 500~\gev{}
is studied, where very tight constraints on the top-quark operators are
expected~\cite{Amjad:2015mma,Durieux:2018tev}. 

This study represents the most complete characterization of the EW interactions of the bottom and top quarks to date.
Our fit yields more stringent constraints than previous work~\cite{Buckley:2015nca,deBeurs:2018pvs,Hartland:2019bjb}.
We moreover present the first comparison of the HL-LHC and ILC~\cite{Bambade:2019fyw} potential for precision measurements that constrain the top and bottom-quark EW couplings.

This paper is organized as follows.
The effective-field-theory and fitting frameworks are presented in \autoref{sec:eft}.
The measurements at the LHC and LEP/SLC that are included in the fit are introduced in \autoref{sec:top_production}. 
The results from the fit to existing data are presented in \autoref{sec:results}.
Projections for the potential of the HL-LHC and the ILC are presented in \autoref{sec:prospects}.
The results and prospects for the extraction of the top-quark Yukawa coupling are discussed in \autoref{sec:yukawa}.
The findings are summarized in \autoref{sec:conclusions}.

\section{Effective field theory and fit setup}
\label{sec:eft}

This section presents the framework in which we develop our fit to the data.

\subsection{Effective field theory}

We adopt an EFT approach to parameterize systematically the effects of physics beyond-the-SM (BSM) at a high scale.
The Wilson coefficients of each higher-dimensional operator can be related to the parameters of concrete BSM realizations with a matching procedure (i.e.\ bounds on Wilson coefficients are mapped onto the coupling and mass of new heavy states).
The EFT description preserves the gauge symmetries of the SM and is a proper quantum field theory.
As such, the EFT predictions can be improved systematically in a perturbative order-by-order expansion.

The EFT expands the SM Lagrangian in terms of a new physics
scale $\Lambda$:
\begin{equation}
\mathcal{L}_\text{eff} = \mathcal{L}_\text{SM} +   \left( \frac{1}{\Lambda^2} \sum_i C_i O_i + \text{h.c.} \right)  + \mathcal{O}\left(\Lambda^{-4} \right) .
\end{equation}
Operators of odd dimension violate baryon or lepton numbers and are ignored.
The interferences of SM amplitudes with those involving an insertion of dimension-six operators gives rise to the leading $\Lambda^{-2}$ terms.
We also include terms of order $\Lambda^{-4}$ arising from the squares of amplitudes where dimension-six operators are inserted once, or from the interference of amplitudes featuring two dimension-six operator insertions with SM ones.
The contributions of dimension-eight operators are not included, even though they first arise at the same $\Lambda^{-4}$ order.

The convergence of the EFT expansion hinges on the smallness of $C_i/\Lambda^2$.
For {\em typical} choices of the coefficient $C_i \sim 1$ the new physics scale $\Lambda$ has to exceed several \tev{} for the effective operator paradigm to hold. 
Following the recommendation of the LHC TOP Working Group~\cite{AguilarSaavedra:2018nen}, fits with and without $\Lambda^{-4}$ contributions are compared to assess the convergence of the expansion.
A strong impact of the $\Lambda^{-4}$ terms on the fit results is an indication that one must carefully check the validity of the EFT expansion when recasting the bounds on concrete SM extensions~\cite{Contino:2016jqw}.
We therefore discuss their impact explicitly in \hyperref[sec:results]{sections~\ref{sec:results}} and~\ref{sec:prospects}.

\subsection{Operator basis}
\label{sec:op_basis}

The number of operators involved in the most general EFT description is daunting even at the first order of the expansion.
We therefore isolate a smaller subset that provides an adequate basis for a study of BSM effects in the top and bottom-quark EW couplings.
This analysis is relevant for scenarios where the dominant BSM effects in the measurements considered appear in these operators.
We focus on the set of operators with leading contributions to the available measurements, restricting the study to dimension-six operators.
We also limit the fit to two-fermion operators, as a fully general treatment including the four-fermion operators is impossible with the current data set.\footnote{The reason for this omission is purely practical: the current data offer insufficient constraints for a global fit including these operator coefficients. We discuss the possibility of extending the fit to CP-conserving four-fermion operators in \autoref{subsec:fourfermion}.}
Finally, we ignore the imaginary part of the operator coefficients.
These lead to CP-violating interactions of the top quark and are efficiently constrained using dedicated analyses at colliders~\cite{Birman:2016jhg,Bernreuther:2017cyi} and low-energy probes~\cite{Cirigliano:2016njn}. 
\begin{equation}
\begin{array}{@{}rlcc@{}}
	O_{\varphi Q}^1
		&\equiv \frac{y_t^2}{2}
		&\bar{\ges q}\gamma^\mu \ges q
		&\FDF[i]
	,\\
	O_{\varphi Q}^3
		&\equiv \frac{y_t^2}{2}
		&\bar{\ges q}\tau^I\gamma^\mu \ges q
		&\FDFI[i]
	,\\
	O_{\varphi u}
		&\equiv \frac{y_t^2}{2}
		&\bar{\ges u}\gamma^\mu \ges u
		&\FDF[i]
		,\\
	O_{\varphi d}
		&\equiv \frac{y_t^2}{2}
		&\bar{\ges d}\gamma^\mu \ges d
		&\FDF[i]
        ,\\
    O_{\varphi ud} 
        &\equiv \frac{y_t^2}{2}
        &\bar{\ges u} \gamma^\mu \ges d  
        &\varphi^T \epsilon i D_\mu \varphi
        ,
\end{array}
\quad
\begin{array}{@{}rlcc@{}}
	O_{uW}
		&\equiv y_t g_W
		&\bar{\ges q}\tau^I\sigma^{\mu\nu} \ges u
		&\epsilon\varphi^* W_{\mu\nu}^I
	,\\
        O_{dW}
		&\equiv y_t g_W
		&\bar{\ges q}\tau^I\sigma^{\mu\nu} \ges d
    	&	\varphi W_{\mu\nu}^I
	,\\
	O_{uB}
		&\equiv y_t g_Y
		&\bar{\ges q}\sigma^{\mu\nu} \ges u
	 & \varphi B_{\mu\nu}
		,\\
	O_{dB}
		&\equiv y_t g_Y
		&\bar{\ges q}\sigma^{\mu\nu} \ges d
		&\epsilon\varphi^* B_{\mu\nu}
	,
\end{array}
\quad
\begin{array}{@{}rlcc@{}}
	O_{u\varphi}
		&\equiv  
		&\bar{\ges q} \ges u
		&\epsilon\varphi^* \; \varphi^\dagger\varphi
	,\\
	O_{d\varphi}
		&\equiv  
		&\bar{\ges q} \ges d
		&\epsilon\varphi^* \; \varphi^\dagger\varphi
	,
\end{array}
\label{eq:op_2q}
\end{equation}

In the Warsaw basis~\cite{Grzadkowski:2010es} (see also
Refs.~\cite{AguilarSaavedra:2008zc,Zhang:2010dr}), the two-fermion
operators that affect top and bottom-quark interactions with vector,
tensor, or scalar Lorentz structures are listed in \autoref{eq:op_2q}, where we have defined $ \ges q \equiv (u_L,\ckm d_L)^T,	\ges u \equiv u_R$, and $\ges d \equiv d_R$. The matrix \ckm~ is the Cabibbo-Kobayashi-Maskawa~\cite{Cabibbo:1963yz, Kobayashi:1973fv} matrix, while $\epsilon\equiv(^{\;\;\:0}_{-1}{}^1_0)$ acts on $SU(2)_L$ indices.

The operators $O_{\varphi Q}^1$ and $O_{\varphi Q}^3$ modify the left-handed couplings of the $Z$ boson to down-type and up-type quarks.
At leading order, the effect on the left-handed coupling of the top quark is proportional to the difference of the Wilson coefficients, $\delta g_L^t = - (C_{\varphi Q}^1 -
C_{\varphi Q}^3) {m_{t}^2}/{\Lambda^2}$, that on the left-handed
coupling of the bottom quark depends on the sum: $\delta g_L^b = - (C_{\varphi Q}^1 +
C_{\varphi Q}^3) {m_{t}^2}/{\Lambda^2}$. The simultaneous fit of the coefficients
$C_{\varphi Q}^1$ and $C_{\varphi Q}^3$ is the main rationale to combine
the bottom and top-quark operators in the fit.

Two further operators $O_{\varphi u}$
and $O_{\varphi d}$ modify the right-handed couplings of the bottom and top quark to the $Z$ boson, 
respectively, $\delta g_R^t = - C_{\varphi u}\: {m_{t}^2}/{\Lambda^2}$ and $\delta g_R^b = - C_{\varphi d}\: {m_{t}^2}/{\Lambda^2}$.

The operators labeled $O_{uW}$, $O_{dW}$, $O_{uB}$ and $O_{dB}$ in \autoref{eq:op_2q} are EW dipole operators.
The $O_{uW}$ and $O_{uB}$ give rise to tensor couplings of the photon and $Z$ boson to the up-type quarks.
Non-zero values of the Wilson coefficients $C_{uW}$ and $C_{uB}$ induce an anomalous dipole moment of the top quark.
Similarly, the operators $O_{dW}$ and $O_{dB}$ give rise to tensor couplings of down-type quark to the photon and $Z$ boson and induce an anomalous dipole moment in the bottom quark. 

The $O^3_{\varphi Q}$ and $O_{uW}$ operators also modify the charged-current interactions of the top quark with a $W$ boson and left-handed $b$-quark.
The $O_{\varphi ud}$ and $O_{dW}$ operators, give rise to interactions between the top quark, the right-handed $b$-quark, and the $W$ boson.

Finally, the last two operators, $O_{u\varphi}$ and $O_{d\varphi}$, lead
to a shift in the Yukawa couplings of up-type and down-type quarks. The operator $O_{u\varphi}$ affects several observables included in the analysis. We discuss their potential to constrain $C_{u\varphi}$ in \autoref{sec:yukawa}. A truly global treatment of this operator must take advantage of the
measurements of the Higgs boson production and decay rates. 
Such a combined fit of the top-quark, EW and Higgs EFTs 
is beyond the scope of the current work and is left for a future publication. The observables included in the analysis are not sensitive to $O_{d\varphi}$, so this operator is ignored in the following.  

We do not consider the chromo-magnetic dipole operators 
$O_{uG} \equiv y_t g_s
\bar{\ges q}\sigma^{\mu\nu} \ges u \epsilon\varphi^* G_{\mu\nu}$ and 
$O_{dG} \equiv y_t g_s
\bar{\ges q}\sigma^{\mu\nu} \ges d \epsilon\varphi^* G_{\mu\nu}$, or the four-fermion operators of the $q\bar{q}t\bar{t}$ type.
The former and a certain number of combinations of the later are better constrained by measurements of the $pp \rightarrow t\bar{t}/b\bar{b}$ processes not considered here.
Top and bottom-quark pair production may however not be able to tightly constrain all the numerous $q\bar qt\bar t$ operators simultaneously.
Their contributions to associated $pp \rightarrow t\bar{t}X$ production processes considered here could then be sizeable.
Reciprocally, the measurement of associated production processes could play an important role in probing all combinations of $q\bar{q}t\bar{t}$ operator coefficients.
Again, since we do not include such operators in our analysis, our results will apply to BSM scenarios in which they induce subleading contributions.

In the context of the fit to top and bottom-quark data we use the notation $O_{tW}$, $O_{tB}$ and $O_{bW}$, $O_{bB}$ for the dipole operators.
We will use the notation $O_{\varphi t}$, $O_{\varphi b}$, and $O_{\varphi t b}$ when referring to the operators that modify the right-handed couplings of the top and bottom quark and the notation $O_{t\varphi}$ for the operator that modifies the top-quark Yukawa coupling.

The Wilson coefficients are normalized to the TeV scale.

The top-quark EFT conventions adopted here are different from the standard established by the LHC TOP Working Group in Ref.~\cite{AguilarSaavedra:2018nen}.
In Appendix \ref{sec:lhc_top_wg} we provide the conversion to these standards.

\subsection{Fit setup}
\label{sec:fit}

The dependence of the observables included in the fit on the Wilson coefficients is calculated at leading order with the Monte Carlo generator \texttt{MG5\_aMC@NLO} \cite{Alwall:2014hca}. The \texttt{TEFT\_EW} UFO model  \cite{Bylund:2016phk} is used for most of the operators.
Exceptions are $C_{t\varphi}$ for which the \texttt{dim6top} UFO model \cite{AguilarSaavedra:2018nen} is used, and $C_{bW}$ and $C_{bB}$ for which we use the \texttt{SMEFTsim} UFO model \cite{Brivio:2017btx}.
The following values of the input parameters are used in the calculation:
\begin{equation}
\begin{array}{@{}rlcc@{}}
	
	\alpha		&= 1/127.9 \, ,	\\
	G_F		&= 1.16637\times10^{-5} \text{ GeV}^{-2} \, ,
	\end{array}
\quad
\begin{array}{@{}rlcc@{}}
	m_Z	&= 91.1876 \text{ GeV}\, ,	\\
	m_H	&= 125 \text{ GeV}	\, , 
	\end{array}
\quad
\begin{array}{@{}rlcc@{}}
	m_b	&= 0 \text{ GeV} \, ,	\\
	m_t	&= 172.5 \text{ GeV}\, .
\end{array}
\nonumber
\end{equation}

The dependence of observables on the Wilson coefficients admits the following expansion:
\begin{equation}
o = o_{SM} + \frac{1}{\Lambda^2} \sum_i C_i o_i + \frac{1}{\Lambda^4} \sum_j \sum_k C_j C_k o_{jk}
+ \mathcal{O}(\Lambda^{-4})
.
\label{eq:param_eq}
\end{equation}
The leading EFT term proportional to $\Lambda^{-2}$ reflects the interference of SM amplitudes with those featuring one dimension-six operator insertion.
The terms proportional to $\Lambda^{-4}$ stem from the square of the amplitudes involving one insertion of dimension-six operators, or from amplitudes involving two such insertions in interference with SM ones.
Terms of order $\Lambda^{-4}$ due to dimension-eight operators are ignored.
The parameterized relations between observables and Wilson coefficients are given in Appendix~\ref{sec:parameterization}.

For several combinations of operators and observables the term proportional to $\Lambda^{-2}$ in \autoref{eq:param_eq} is suppressed.
The $\Lambda^{-4}$ terms then plays an important role and the EFT expansion is not valid in full generality.

A well-known example is the  dependence of the associated production processes $pp \rightarrow t\bar{t}X$ on the top-quark dipole operators.
The $\sigma^{\mu \nu}q_{\nu}$ structure involves the momentum of the $Z$ boson or photon, which leads to a suppression because the radiated $Z$ boson or photon tends to be soft~\cite{Bylund:2016phk}.
In this case, other processes can be found, where the $\Lambda^{-2}$ term dominates the sensitivity: the inclusion of charged-current interactions and $e^+e^- \rightarrow t\, \bar{t}$ production restores the validity of the fit for $C_{tW}$ and $C_{tB}$.

Several operators affecting the bottom-quark EW couplings lead to amplitudes whose interferences with SM ones are suppressed by the small bottom-quark mass.
The $O_{bW}$ and $O_{\varphi tb}$ operators induce a $t\bar{b}W$ interaction involving a right-handed bottom quark.
The $O_{bB}$ operator also generate a chirality flipping $b\bar bZ$ dipolar interactions.
The interferences of the amplitudes they generate with SM ones thus vanish in the $m_b \rightarrow 0$ approximation adopted in this paper.
For these operators, a strong dependence of the fit results on the $\Lambda^{-4}$ terms remains even after the ILC programme.

\subsection{Implementation of the fit}

The fit to data is performed using the open source {\scshape{\tt HEPfit}} package~\cite{deBlas:2019okz, hepfitsite}.
{\scshape{\tt HEPfit}} is a general tool designed to combine direct and indirect constraints, in EFTs or particular SM extensions.
Its flexibility allows to easily implement any BSM model or observable.
{\scshape{\tt HEPfit}} is available under the GNU General Public License. 
The developers' version can be downloaded at \cite{hepfitgit}.

The fit is performed as a Bayesian statistical analysis of the model. 
{\scshape{\tt HEPfit}} includes a Markov-Chain Monte-Carlo implementation provided by the Bayesian Analysis Toolkit \cite{Caldwell:2008fw} to explore
the parameter space. Similar fits 
 using the {\scshape{\tt HEPfit}} package have been performed for different models~\cite{Ana:2019, Victor:2019} and for effective field theories~\cite{deBlas:2018tjm, Ciuchini:2019usw}.

The results in this paper were verified with an independent fitting code based on the {\em Minuit} minimization package in ROOT~\cite{James:2004xla}.
The results for individual limits agree to 1\%.
For the comparison of the global limits we perform an ad-hoc fit in which we reduce the number of parameters and observables.
In this case the results agree to 10\%.
In general we find {\scshape{\tt HEPfit}} is more robust when dealing with several local minima, so all final results are obtained using {\scshape{\tt HEPfit}}.

The fit is based on the Bayesian approach of statistics and the interpretation differs slightly from the frequentist interpretation. The fit results are given as intervals on the operator coefficients with a given posterior probability, typically 68\%.

\section{Measurements}
\label{sec:top_production}

The measurements that form the input to the fit are presented in this section.

\subsection{Top-quark neutral-current interactions}

\begin{itemize}
\item $ p p \rightarrow t \bar{t} h $ production.
The production of a Higgs boson in association with a top-quark pair was observed by ATLAS and CMS in 2018~\cite{Aaboud:2018urx,Sirunyan:2018hoz}.
The production rate is sensitive to the coefficient $C_{t\varphi}$ of the operator that shifts the value of the top-quark Yukawa coupling.

\item $ p p \rightarrow t \bar{t} Z/W $ production.
The associated production of top quarks with a $Z$ boson gives access to all operators that modify the 
coupling of the top quark with neutral EW gauge bosons and is therefore a key channel in a combined fit~\cite{Bylund:2016phk}. 
The ATLAS and CMS measurements of the inclusive cross section using 36~\ifb{} of data 
at 13~\tev{} have reached a precision of approximately 15-20\%~\cite{Aaboud:2019njj,Sirunyan:2017uzs}. 
The results on $pp \rightarrow t\bar{t}W$ production are also included
in the fit. A recent preliminary result~\cite{CMS:2019nos}, with an integrated luminosity of
78~\ifb{} and a relative uncertainty of less than 10\%, is not included.

\item $ p p \rightarrow t \bar{t} \gamma $ production. The rate of the $pp \rightarrow t\bar{t}\gamma$ process depends on the $C_{tW}$ and $C_{tB}$ coefficients of EW dipole operators. ATLAS has published a measurement 
of the $pp \rightarrow t\bar{t}\gamma$ fiducial cross-section~\cite{Aaboud:2018hip} at
$\sqrt{s}=$ 13~\tev. 

\item Single top-quark production in association with a $Z$ boson has been observed by ATLAS and CMS. 
For the $pp \rightarrow tZq$ process the first 
cross-section measurements have reached a precision of approximately 
15-35\% \cite{Sirunyan:2018zgs, Aaboud:2017ylb}. 

\item $ pp \rightarrow \gamma^*/Z^* \rightarrow t\bar{t}$ production.
The neutral-current pair production process $q\bar{q} \rightarrow
Z/\gamma \rightarrow t \bar{t}$ is overwhelmed by the QCD process and
has not been isolated. This contribution to the inclusive $pp \rightarrow t\bar{t}$ process
leads to a dependence of the rate on the EW operators considered, but in practice
this contribution can be ignored.
\end{itemize}

\subsection{Top-quark charged-current interactions}
\label{subsec:charged}

\begin{itemize}
\item Top-quark decay, $ t \rightarrow W b$.
The charged-current $t\bar{b}W$ vertex is accessible at hadron colliders in top-quark decay. 
The $t\rightarrow Wb$ decay has a branching ratio of nearly 100\%. The helicity fractions
of the $W$ boson produced in top-quark decay can be predicted to excellent 
precision~\cite{Czarnecki:2010gb}. The measurements by ATLAS and 
CMS~\cite{Aaboud:2016hsq,Khachatryan:2016fky,Chatrchyan:2013jna,Aad:2012ky} at $\sqrt{s}=$ 7 and 
8~\tev{} reach a precision of several
percent. The combination of precise predictions and measurements converts these measurements in 
true hadron collider precision measurements and in sensitive probes to new physics
affecting the $t\bar{b}W$ vertex~\cite{Birman:2016jhg}. We include the 
8~\tev{} measurements of $F_{L}$ and $F_{0}$, that yield a tight limit on $C_{tW}$.

\item Single-top-quark production.
A second handle on the $t\bar{b}W$ vertex is found in charged-current single top-quark production.
The $t$-channel process has a sizeable cross section, which has been measured
to better than 10\% precision~\cite{Sirunyan:2018rlu,Aaboud:2016ymp} at $\sqrt{s}=$ 13~\tev.
ATLAS and CMS have also published precise measurements of the rate for the $Wt$ associated production channel~\cite{Sirunyan:2018lcp,Aaboud:2016lpj}. 

\item Top-quark decay in single top-quark production. A measurement of the $W$-boson helicity in a sample of polarized top quarks yields further limits on anomalous top-quark couplings~\cite{Aaboud:2017yqf,Aaboud:2017aqp,Khachatryan:2014vma}. These are however not considered here, as they are primarily competitive for the (CP-violating) imaginary parts of the operator coefficients that we do not include in our study.
\end{itemize}

\subsection{Measurements in bottom-quark production}
\label{subsec:bottom}

\begin{itemize}
    \item $e^+e^- \rightarrow b\,\bar{b}$ production.
The LEP and SLC measurements of bottom-quark pair production provide a powerful, complementary handle on the operator coefficients $C_{\varphi Q}^1$ and $C_{\varphi Q}^3$.
Combining measurements of bottom-quark production at LEP/SLC with measurements in top-quark production yield solid constraints on both operator coefficients in a global fit~\cite{Durieux:2018tev}.
We consider the measurements of $R_b$ and $A_{FBLR}^{bb}$ at the $Z$ pole \cite{ALEPH:2005ab}. 

\item $pp \rightarrow b\bar{b} Z$ production. The associated production processes $pp \rightarrow b \bar{b} Z$ and $pp \rightarrow b \bar{b} \gamma$ 
at the Tevatron and LHC probe the $b\bar{b}Z$ and $b\bar{b}\gamma$ vertices. The ATLAS and CMS experiments
have measured the cross section for the associated production of a $Z$ boson and at least one 
$b$-quark~\cite{Aad:2014dvb, Chatrchyan:2012vr} in early LHC runs. 
The constraints derived from these measurements
are not currently competitive with the LEP and SLC measurements. We therefore ignore them in 
the following.
\end{itemize}

\subsection{Indirect constraints}
\label{subsec:flavour}

For reference, we collect here several observables that can be used to derive indirect constraints to top-quark couplings.
A more complete discussion can be found in appendix~A of Ref.~\cite{AguilarSaavedra:2018nen}.

\begin{itemize}
\item Data from $B$-factories can provide stringent bounds. For instance, the rare meson decays $B_s \rightarrow \mu^+ \mu^-$ and $K \rightarrow \pi \nu \bar{\nu}$ give access to the $t\bar{t}Z$ vertex~\cite{Brod:2014hsa} and yield constraints on the coefficients of the $O_{\varphi Q}^3$, $O_{\varphi Q}^1$ and $O_{\varphi t}$ operators. The $b\rightarrow s\gamma$ decays give access to the $t\bar{t}\gamma$ vertex~\cite{Bissmann:2019gfc} and bounds on $C_{tW}$ and $C_{tB}$ are derived from the $\bar{B} \rightarrow X_s \gamma$ decay rate measured by the BaBar, Belle and CLEO experiments. $B$-meson decays are used in Ref.~\cite{Grzadkowski:2008mf,Drobnak:2011aa} to access the $Wtb$ vertex and FCNC interactions in Ref.~\cite{Fox:2007in}. Dimension-six operators involving top quark are also studied in the Standard Model effective field theory (SMEFT) matching onto Weak Effective Theory (WET) for $\Delta F=1$~\cite{Aebischer:2015fzz}, $\Delta F=0$ \cite{Feruglio:2018fxo}, $\Delta F=2$ \cite{Endo:2018gdn}.

Beside top-quark operators, bottom-quark ones are obviously also indirectly constrained by $B$-factory measurements.
    
\item Electro-weak precision measurements also provide indirect sensitivity to top-quark operators through loop effects~\cite{Greiner:2011tt, Zhang:2012cd}.
The impact of top-quark operators in Higgs production and decay at both hadron and lepton colliders was examined in Ref.~\cite{Vryonidou:2018eyv, Boselli:2018zxr}.
Including also the dependence of diboson production at future lepton colliders, a combined analysis of the electroweak, Higgs, and top-quark effective field theories was performed in Ref.~\cite{Durieux:2018ggn}.
Further studies are also in preparation~\cite{Sunghoon}.

\item Measurements of electric dipole moments offer a complementary constraint on the top quark electro-weak couplings. In particular the CP-violating operators, that are not included in this work, receive stringent individual limits~\cite{Cirigliano:2016nyn}. The inclusion of these bounds in a fully global analysis remains to be done.
\end{itemize}


Indirect bounds on a single coefficient, or small systems of a few coefficients, are often competitive in comparison with the individual, direct bounds from direct LHC data that we present in \autoref{sec:results}. With the inclusion of prospects for top-quark pair production at future electron-positron colliders of \autoref{sec:prospects}, the bounds become less relevant. A global fit including these results requires consideration of an extended set of operator coefficients and is beyond the scope of the current work. We therefore do not consider indirect measurements in the fit. 

\subsection{Summary of measurements}

\begin{table*}[ht]
\centering
\begin{tabular}{|l|c|c|c|c|c|}
\hline
Process & observable & $\sqrt{s}$  & $\int \cal{L}$  & SM  & Ref.\\ \hline
$p p \rightarrow t \bar{t} H$ & cross section & 13 TeV & 36~\ifb  & 
- &  \cite{Aaboud:2018urx} \\ 
$p p \rightarrow t \bar{t} Z/W$ &  cross section & 13 TeV &  36~\ifb  & 
\cite{Bylund:2016phk} & \cite{Aaboud:2019njj} \\ 
$p p \rightarrow t \bar{t} \gamma$ & fid. x-sec. & 13 TeV & 36~\ifb &
\cite{Bylund:2016phk} & \cite{Aaboud:2018hip}  \\ 
single-top (t-ch) & cross section & 13~\tev{} & 36~\ifb{} &
- & \cite{Sirunyan:2018rlu} \\ 
single-top (Wt) & cross section & 13~\TeV & 36~\ifb{}  & 
 - &  \cite{Sirunyan:2018lcp}   \\ 
single-top (tZq) & cross section & 13 TeV & 36~\ifb{}  & 
 \cite{Sirunyan:2017nbr} & \cite{CMS:2019bke} \\ 
$t \rightarrow W^{+} b $ & $F_0$, $F_L$  & 8~\tev{} & 20~\ifb{} & 
\cite{Czarnecki:2010gb}  & \cite{Aaboud:2016hsq} \\
$e^{-} e^{+} \rightarrow b \bar{b} $ & $R_{b}$ ,  $A_{FBLR}^{bb}$ & $\sim$ 91~\GeV & 202.1~\ipb  &
 - & \cite{ALEPH:2005ab}  \\ \hline
\end{tabular}
\caption{Measurements included in the EFT fit of the top and bottom-quark EW sector.
For each measurement, the process, the measured observable, the center-of-mass energy and 
the integrated luminosity are listed. The last column lists the references for the measurement that is included in the fit.}
\label{tab:measurements}
\end{table*}

The selected measurements that are included in our fit are 
summarized in \autoref{tab:measurements}.
For all LHC observables, ATLAS and CMS measurements are available at  $\sqrt{s}= 13~\tev$ for an  integrated luminosity of 36~\ifb{}.
As the measurements have not yet been combined, and a proper correlation of uncertainties requires harmonization of the definitions of the
systematics, we include only one measurement for each observable. We
select the most precise measurement among the 13~\tev{} measurements.
The measurements of the same quantities at 8~\tev{} are not included except for the measurement of the $W$-boson 
helicity fractions in top-quark decay, that is only available at 8~\tev.

The LEP and SLC measurements of $R_b$ and $A_{FBLR}^{bb}$ at the $Z$ pole have been combined in the EW fit of Ref.~\cite{ALEPH:2005ab}.
The fit correlates the measurements of several quantities 
and reports a complete covariance matrix.

Even with a single measurement for each observable included in
the fit, the systematic uncertainties are expected to lead
to correlations among the measurements. Also the theory predictions are
correlated, through the parton density functions and the similarity of the matrix elements of the several associated production processes. We have cross-checked the effect of correlated systematics on the fit results explicitly.
The full covariance matrix of the LEP/SLC electro-weak fit is taken into account.
These correlations have a negligible effect on our results.
Also the introduction of an ad-hoc correlation of 50\% between the results for associated top-quark production has a minor effect on the fit.
We therefore expect that a full treatment of all correlations, once the combinations of ATLAS and CMS measurements are made available, will lead to only a slight improvement of the limits.

\subsection{Sensitivity to operator coefficients}

The set of measurements in \autoref{tab:measurements} provides sensitivity to all operators listed in \autoref{sec:eft}. 
Associated production of top quarks with a $Z$ boson at
the LHC alone is sensitive to all five top-quark operators in our basis.
Associated production with a photon gives access to the dipole operators
$C_{tB}$ and $C_{tW}$.  
Charged-current processes, such as $t\bar{t}W$ production, EW single 
top-quark production and top-quark decay are sensitive to $C_{tW}$,
$C_{\varphi Q}^3$, $C_{bW}$ and $C_{\varphi tb}$. Results on $ e^+ e^- \rightarrow b\bar{b}$
production are sensitive to $C_{\varphi Q}^1$ and $C_{\varphi Q}^3$ and the 
pure bottom-quark operators included in the fit.

To explore the relative sensitivity of the existing measurements, the results of single-parameter fits are shown in \autoref{fig:individual_limits}.
For each of the operators, the first column displays the {\em individual} limit
on the Wilson coefficients of the complete data set presented in 
\autoref{tab:measurements}. The second column shows the result
of the most constraining measurement. The third column displays the 
second-best constraint. 

For most operators, there is a strong hierarchy in the sensitivity of the measurements. For a majority 
of operator coefficients a single measurement drives the individual sensitivity.
Typically, the limit of the most sensitive measurement is a factor 2-5 better 
than that of the second-best measurement for most operators. For  
$C_{\varphi Q}^1$ and $C_{\varphi Q}^3$ the precise LEP/SLC measurement of $R_b$ yields
a constraint 30 times better than that of the associated production processes at the LHC.
For $C_{tB}$, the associated $t\bar{t}\gamma$ and $t\bar{t}Z$ production modes provide similar sensitivity and the combined results is significantly stronger than the limit derived from a single observable.
Also in the case of $C_{\varphi tb}$, the different single top-quark measurements provide similar sensitivity.

\begin{figure*}
    \centering
    \includegraphics[scale=0.75]{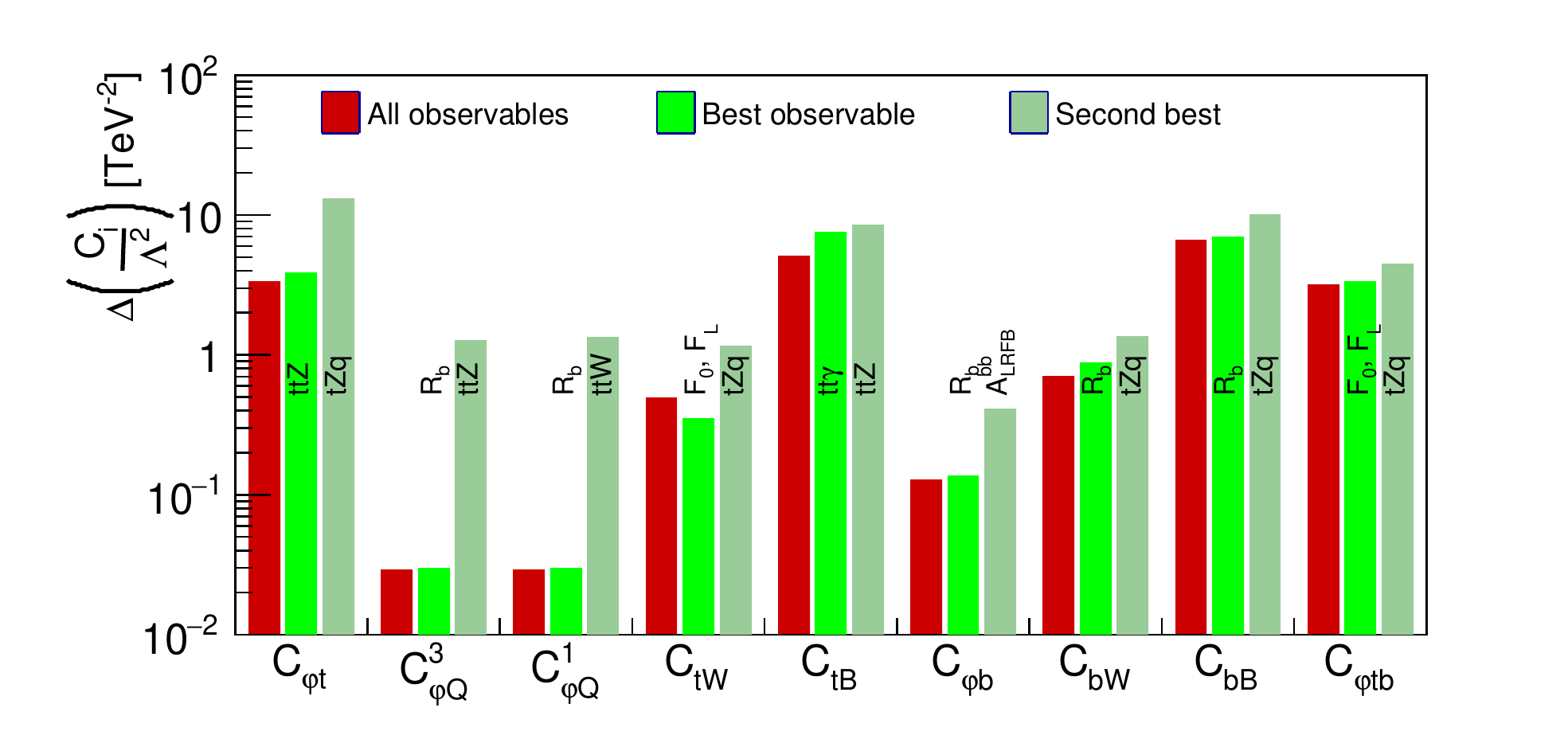}
    \caption{Results of single-parameter individual fits to the Wilson
    coefficients of the dimension-six operators introduced in \autoref{sec:eft}. For each operator the 1$\sigma$ uncertainty is shown. The three bars respectively correspond to the result of the combined fit using all data (red), to the constraint obtained from the most sensitive single measurement (light green), and to that of the second-best measurement (greyish green)}.
    \label{fig:individual_limits}
\end{figure*}

A few observables are sensitive to a large number of operators: the measurement 
of $R_{b}$ at LEP yields the best individual limits on five different operators.
The $t\bar{t}Z$ and $tZq$ cross sections are also sensitive to several operators:
they score among the two most sensitive measurements for six operators. The most
specific observables are the helicity fractions of $W$ boson in top-quark decay.
They provide a stringent limit on $C_{tW}$ and are not strongly affected by the other operators.

\section{Present constraints}
\label{sec:results}
The result of a fit to currently available data are presented in this section. 

\subsection{Fit to LHC and LEP/SLC data}

The main result of this paper is a ten-parameter fit to the LHC and LEP/SLC measurements of \autoref{tab:measurements}. The 68\% probability bounds on ten Wilson coefficients are presented in \autoref{fig:results}.
Global or marginalized limits are obtained when all coefficients are varied simultaneously.
These are shown as blue continuous lines.
The individual limits from a single-parameter fit are presented as red dashed lines.
The global limits are also given in \autoref{tab:quadratic}.

\begin{figure}[ht!]
\centering%
\subfigure[]{\includegraphics[width=0.75\linewidth, trim=50 5 0 10, clip]{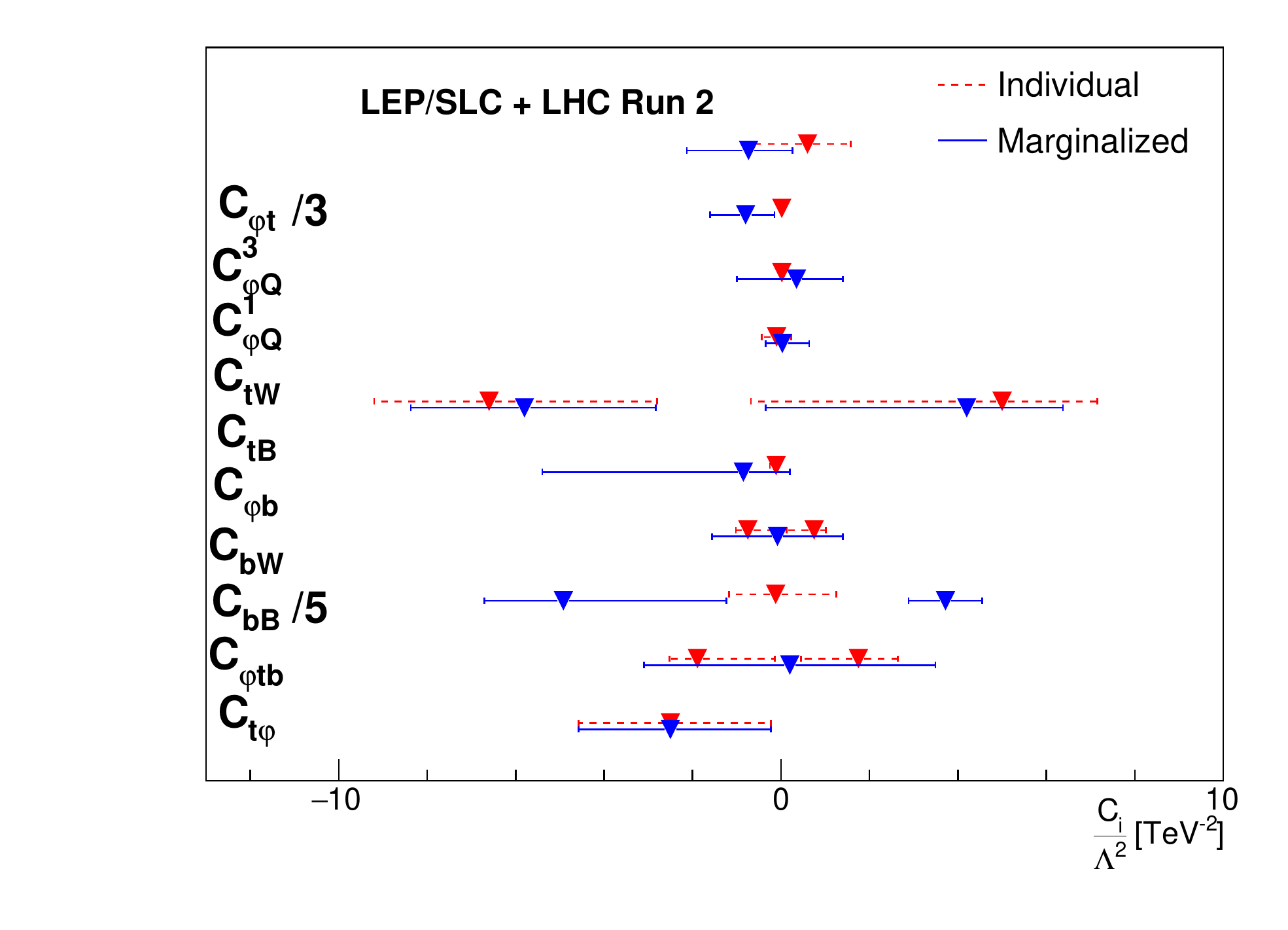}}
\\
\subfigure[]{\includegraphics[width=0.75\linewidth]{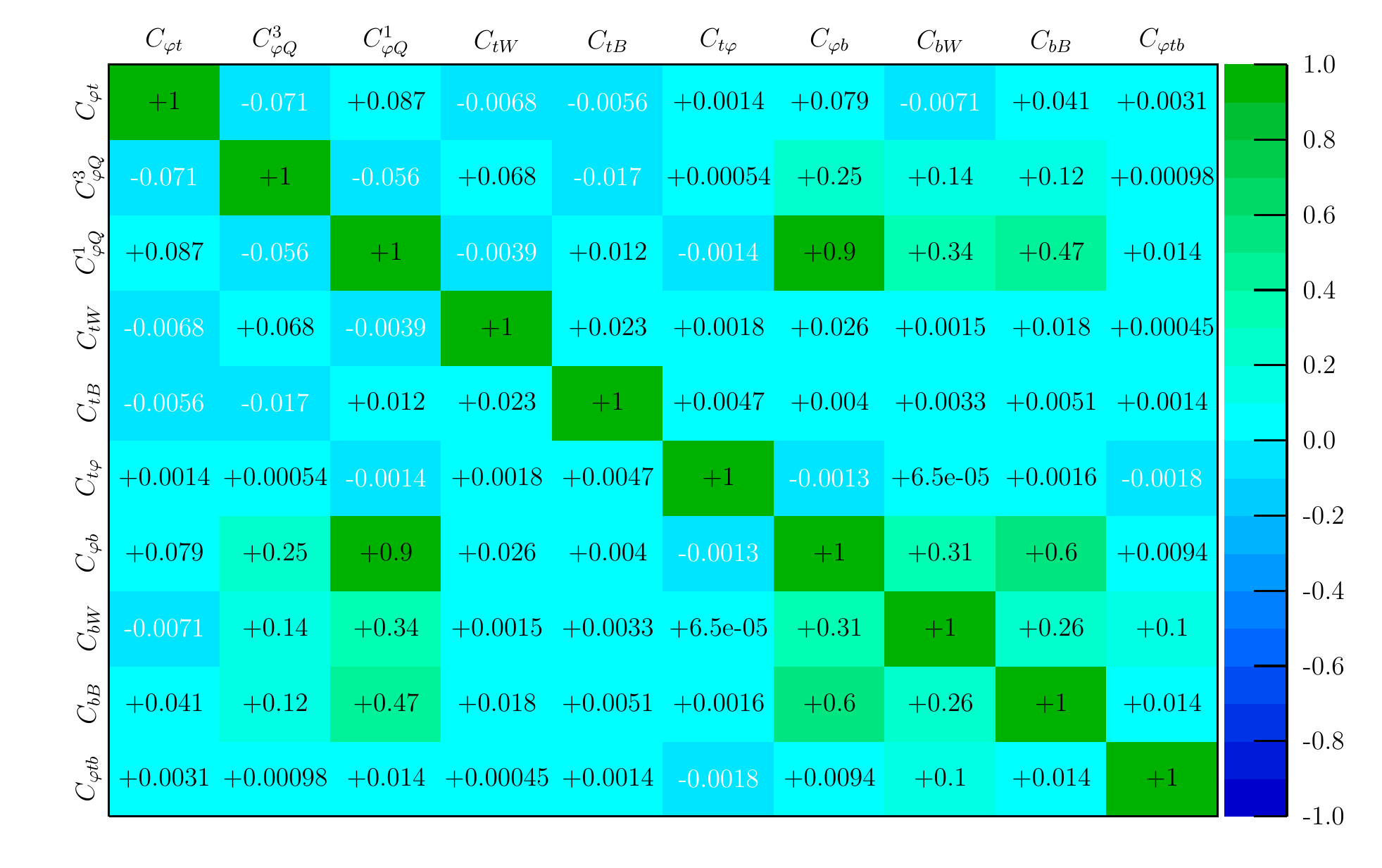}}
\caption{The 68\% probability intervals (upper panel) and correlation matrix (lower panel) for the Wilson coefficients of the ten effective operators that modify the EW couplings of top and bottom quarks derived from a fit of the data included in \autoref{tab:measurements}.
The correlation matrix in {\scshape{\tt HEPfit}} is calculated following Ref.~\cite{Caldwell:2008fw}. Detailed information about the correlation between the parameters and the complete covariance matrix of the fit are given in Appendix~\ref{sec:cov_matrix}. Global (marginalized) limits obtained in the fit are shown as blue bars, the individual limits from  single-parameter fit in red. The (local) minimum of the $\chi^2$ are shown as triangles.}
\label{fig:results}
\end{figure}

Generally, the fit yields good results even when all operator coefficients are varied simultaneously.
The individual limit on $C_{tW}/\Lambda^2$ is very tight and the constraints remains very strong in the ten-parameter fit.
Several observables also have similar sensitivity to $C_{tB}/\Lambda^2$ and $C_{\varphi t}/\Lambda^2$.
The global limits are therefore not degraded too much compared to
the individual limits. 

For the operators that affect bottom-quark production
in $e^+e^-$ collisions, the individual limits from the $Z$-pole measurements are very tight. To disentangle the contributions of different operators, the fit must use several observables.
Given the large hierarchy in sensitivities observed in \autoref{fig:individual_limits}, the global limits are typically much weaker than the individual ones. Even so, tight constraints
of order $1~\tev^{-2}$ are obtained for $C_{\varphi Q}^1/\Lambda^2$ and $C_{\varphi Q}^3/\Lambda^2$ and $C_{\varphi b}/\Lambda^2$.

Comparing these limits to those obtained by other groups, we find that 
our fit yields better results.
In particular, the inclusion of the $Z$-pole data leads to considerably tighter limits on $C_{\varphi Q}^3$, compared to  Ref.~\cite{Hartland:2019bjb}.

\subsection{Impact of \texorpdfstring{$\Lambda^{-4}$}{1/Lambda4} terms}
\label{subsec:lambda4}

The results of the nominal fit are based on a parameterization according to \autoref{eq:param_eq}, that includes $\Lambda^{-2}$ and $\Lambda^{-4}$ terms.
The fit finds multiple allowed regions for several operator coefficients. These local minima are a result of the $\Lambda^{-4}$ terms in the parameterization. Two regions, roughly equidistant from the SM prediction, are found for $C_{bW}/\Lambda^2$, $C_{bB}/\Lambda^2$ and $C_{\varphi tb}/\Lambda^2$. 

In \autoref{tab:quadratic} the nominal results are compared to a fit based on
a parameterization that includes only the $\Lambda^{-2}$ terms.

\begin{table}[t]
    \centering
    \begin{tabular}{l|c@{\hspace{2mm}}c}
         & \hspace{-9mm} $\Lambda^{-2}$ and $\Lambda^{-4}$ terms &  $\Lambda^{-2}$ term only \\  \hline
    $C_{\varphi t}/\Lambda^{2}$    &   $(-17.99, -9.61),(-6.39, 0.79) $ &    $(-1.44,+5.34) $   \\
    $C_{\varphi Q}^3/\Lambda^{2}$   &    $(-1.61, -0.15) $&    $(-0.70, +0.52) $    \\
       $C_{\varphi Q}^1/\Lambda^{2}$  &     $(-0.99, 1.39) $&    $(-0.59, +0.65) $   \\
       $C_{tW}/\Lambda^{2} $     &     $(-0.34, 0.64) $ &    $(-0.42, +0.24) $      \\
       $C_{tB}/\Lambda^{2}$    &     $(-8.37, -2.83) (-2.00, -1.60) (-0.36, 6.36) $&     $(-9.3, +38.14) $         \\
	$C_{t\varphi}/\Lambda^2$ &   $(-4.57, -0.23) $ &   $(-5.12, -0.28) $	\\
       $C_{\varphi b}/\Lambda^{2}$ &       $(-5.40, 0.20) $&   $(-0.63, +0.15) $   \\
       $C_{bW}/\Lambda^{2}$     &      $(-1.56, 1.40) $&  ---         \\
       $C_{bB}/\Lambda^{2}$   &     $(-33.48, -6.12), (14.44, 22.76)  $ &  ---  \\ 
       $C_{\varphi tb}/\Lambda^{2}$   &      $(-3.09, 3.49) $ &  ---  \\
    \end{tabular}
\caption{The 68\% probability intervals on the dimension-six operator coefficients in units of $\tev^{-2}$.
These results are obtained with a fit to LHC and LEP/SLC data for two parameterizations of the dependence of the observables on dimension-six operator coefficients.
The first column lists the results from the fit based on the nominal parameterization, which includes terms proportional to $\Lambda^{-2}$ and $\Lambda^{-4}$ terms.
The second column is obtained with a fit based on a parameterization that only includes $\Lambda^{-2}$ terms.
The coefficient $C_{t\varphi}$ is marginalized over in the fit, but discussed separately in \autoref{sec:yukawa}.}
\label{tab:quadratic}
\end{table}

For several operator coefficients the inclusion of $\Lambda^{-4}$ terms is expected to have profound impact on the result.
In the dependence of the $t\bar{t}X$ on $C_{tB}$, the $\Lambda^{-2}$ term is suppressed, and the  $\Lambda^{-4}$ terms dominate the sensitivity when limits are saturated.
The bound on $C_{tB}$ is therefore severely degraded when the $\Lambda^{-4}$ terms are dropped.
This is not the case for $C_{tW}$, for which the bound is dominated by the measurements of the helicity fractions in top-quark decay and of the single top-quark production cross section.

For the bottom-quark dipole operators $C_{bW}$, $C_{bB}$ as well as for $C_{\varphi tb}$, the interferences with SM amplitudes vanish in the $m_b = 0$ approximation.
The fit based only on $\Lambda^{-2}$ terms can therefore not bound these operators. 

The correlations between the different operator coefficients propagate the effect of the $\Lambda^{-4}$ terms to other operators. If the fit is repeated excluding $O_{bW}$, $O_{bB}$ and $O_{\varphi tb}$ the results obtained with the two parameterizations are very similar for all operators except $O_{tB}$ and $O_{\varphi t}$.

The importance of $\Lambda^{-4}$ terms indicates that the validity of the EFT expansion should be carefully verified.
When recasting these results in a concrete BSM scenario, one must verify that the dimension-eight operators that are ignored here are subdominant in comparison with dimension-six ones.

\section{Future collider prospects}
\label{sec:prospects}

This section presents the prospects to improve the precision of the determination of top and bottom-quark EW couplings during the high-luminosity phase of the LHC or at a future electron-positron collider.

\subsection{High-luminosity phase of the LHC}
\label{subsec:hllhc}

At the time of writing ATLAS and CMS have collected approximately 140~\ifb{} of $pp$ collisions 
at a center-of-mass energy of 13~\tev{} in Run 2. After a long shutdown (LS2), LHC Run 3 is 
expected to deliver a total of 300~\ifb{} per experiment at the nominal energy ($\sqrt{s}= 14~\tev$).
Between 2023 and 2026, an upgrade of the LHC accelerator complex~\cite{ApollinariG.:2017ojx} 
and detectors will allow operation at five to seven times the nominal LHC luminosity.
The HL-LHC phase will bring the total integrated luminosity to 3~\iab{} by 2037.

The expected precision for SM measurements after the full 3~\iab{} is presented
in a series of Yellow Reports. The chapter on top-quark physics~\cite{Azzi:2019yne} does not 
provide a quantitative basis for all measurements included in our study. We therefore adopt two simple scenarios to project existing measurements, that are
loosely inspired by the scenarios prepared for the Higgs chapter of the HL-LHC Yellow
Report~\cite{Cepeda:2019klc}. The ``S1'' scenario envisages that the statistical uncertainty scales with the inverse square root
of the integrated luminosity. The systematic uncertainties, in measurements and predictions, do not change. The ``S2'' scenario envisages an improvement of a factor two for the theory uncertainty, while the statistical uncertainty and the 
experimental systematic uncertainty scale with the inverse square root of the integrated luminosity. For the measurements included in the fit, this scenario thus implies a reduction of the experimental uncertainty by a factor 6-10. At that point, the comparison with the SM is generally limited by the theory uncertainty,
that has the more modest improvement.

It is instructive to compare the S2 scenario to more detailed
projections. ATLAS and CMS have provided detailed prospect studies for some analyses~\cite{CMS:2018rcv}. Other groups have published independent prospect studies, see  
in particular Ref.~\cite{Schulze:2016qas} 
for $t\bar{t}Z$ production and Ref.~\cite{Birman:2016jhg} for top-quark decay.

The production of a top-quark pair in association with a gauge boson plays an important role in the fit. 
In the ATLAS and CMS measurements we consider, the theory uncertainty (typically of the order of 10\%) is similar in size to the experimental uncertainty.
In the S2 scenario, the experimental uncertainties are improved very substantially.
The theory uncertainties are then expected to be limiting by the end of the HL-LHC.
This indeed seems the most likely scenario.
The factor two improvement in the theory uncertainty envisaged in the S2 scenario could well be achieved by improving the description from the current NLO to NNLO in QCD, which seems feasible on the time scale of the HL-LHC programme. 

A promising avenue for many of the associated production processes is a differential analysis.
In the current data set, the precision is still very limited for rare processes. 
However, with a hundred-fold increase in the data sample, differential analyses at the HL-LHC 
are expected to provide powerful constraints~\cite{Rontsch:2014cca, Schulze:2016qas}. This is 
particularly relevant for the dipole operators. In \autoref{fig:ctb_diff_sens}, the
sensitivity of the differential $ pp \rightarrow t \bar{t} \gamma$ cross section is seen
to increase strongly with the transverse momentum of the photon $p_{\mathrm{T}}$. A
shape analysis of the spectrum may yield a powerful constraint, possibly even exceeding the 
prospects of the S2 scenario.

The case of the $W$-boson helicity fraction measurement in top-quark decays is an example where the S2 scenario is probably overly optimistic.
The theory uncertainty is currently significantly below the experimental precision, so that it does not limit the precision for this projection.
The strong improvement in the precision envisaged by the S2 scenario is optimistic in comparison with the outlook in Ref.~\cite{Birman:2016jhg}. 
In practice, the impact on the overall prospects is limited. The measurements in top-quark decay are most relevant for the constraint on $C_{tW}/\Lambda^2$, that is sensitive to several other measurements.
In case the measurements in top-quark decay should fail to improve as expected in S2, other measurements (such as single top-quark production with a $Z$ boson) can take over its role in the global fit.
We expect, therefore, that the overall results presented in this section are not affected too much, even if the top-quark decay measurements improve less than envisaged.

\begin{figure}
    \centering
    \includegraphics[width=0.8\textwidth]{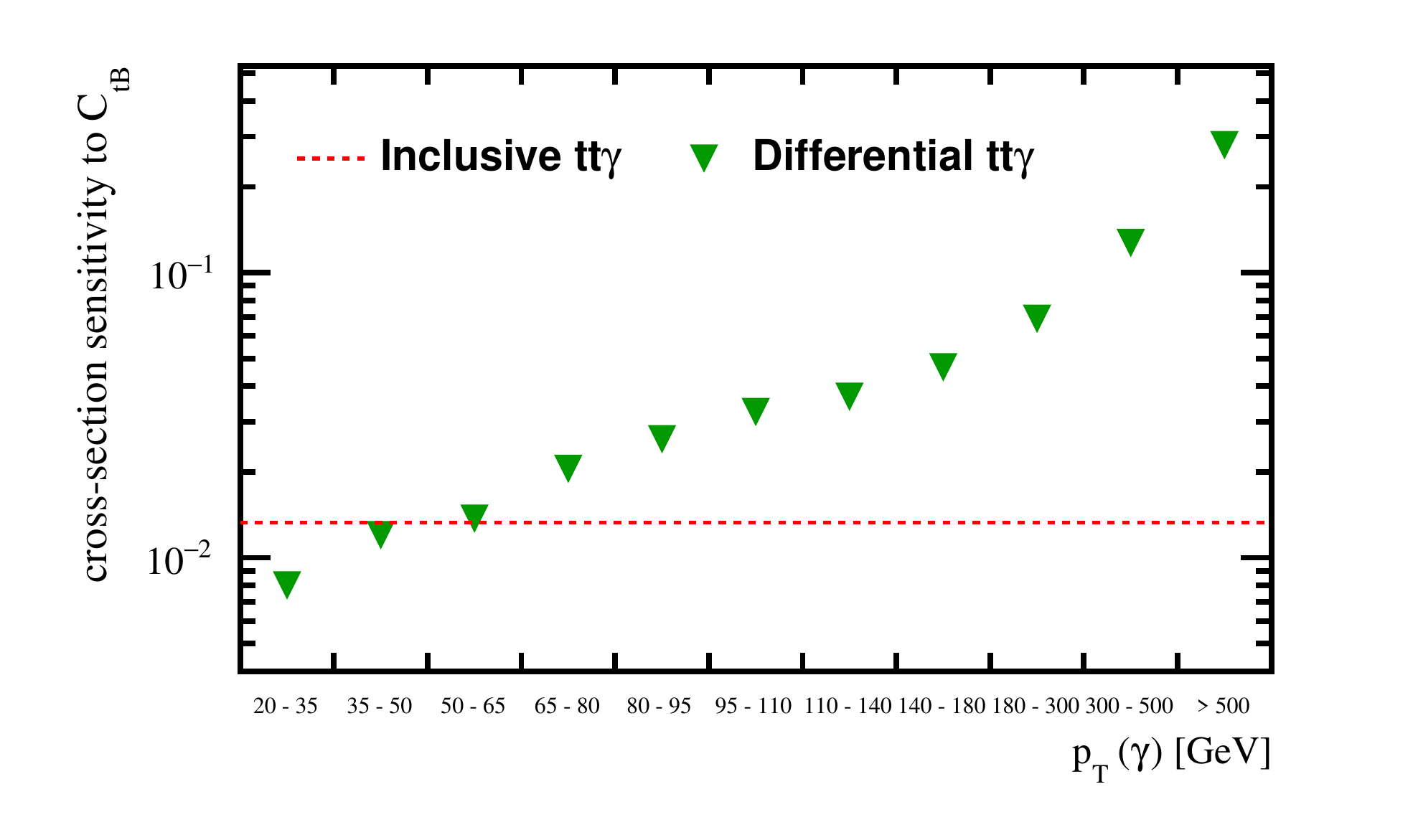}
    \caption{The sensitivity of the differential $pp\rightarrow t\bar{t}\gamma$ cross section to the operator coefficient $C_{tB}/\Lambda^2$. The sensitivity is defined as the relative change in the cross section due to a unit change in $C_{tB}/\Lambda^2$.}
    \label{fig:ctb_diff_sens}
\end{figure}

\subsection{Future \texorpdfstring{$e^+e^-$}{e+e-} collider: ILC}

At an electron-positron collider bottom and top-quark pair production through the exchange of a photon or $Z$ boson are among the dominant processes.
A future high-energy $e^+e^-$ collider thus provides an ideal laboratory to characterize the $Z/\gamma\:  b  \bar{b}$ and $Z/\gamma\: t  \bar{t}$ vertices.
Single top-quark production could also bring valuable constraining power~\cite{Fuster:2015jva} but no quantitative prospect is currently available.
So we do not consider this process.

The potential of the ILC for the measurement of the EW couplings of the bottom quark is studied in detail in Refs.~\cite{Bilokin:2017lco, adrian}.
These studies consider measurements of the cross-section and forward-backward asymmetry in the nominal ILC running scenario~\cite{Barklow:2015tja}, with an integrated luminosity of 2000~\ifb{} at $\sqrt{s}= 250~\gev$.
The electron and positron beams are polarized, with a polarization of $\pm80\%$ and $\pm30\%$, respectively.
The luminosity is divided equally among the left-right and right-left configurations.
The authors perform a full-simulation study, including the relevant SM backgrounds, and a realistic jet charge identification strategy based on the use of Kaon and vertex-charge tags.
We adopt the uncertainty estimates of Ref.~\cite{adrian}, that include statistical and systematic uncertainties. 

For the 500~\gev{} run a complete analysis does not yet exist. We adopt the acceptance times efficiency estimate of 25\% based on full simulation by the same authors. The statistical uncertainties for the cross-section and forward-backward asymmetry for the left-right and right-left beam polarizations at $\sqrt{s}=$ 500~\gev{} are estimated assuming a total integrated luminosity of 4~\iab.

To produce top-quark pairs, an $e^+e^-$ collider must be operated at a center-of-mass energy above twice the top-quark mass. 
Runs above the pair-production threshold are envisaged in the CLIC initial program and in later stages of the ILC and FCCee.
Beam polarization, foreseen in ILC and CLIC, allows to disentangle the photon and $Z$-boson vertices~\cite{AguilarSaavedra:2012vh, Amjad:2015mma}.
In a multi-parameter EFT fit, the initial-state polarization is helpful to simultaneously constrain the coefficients of $O_{tB}$ and $O_{tW}$~\cite{Durieux:2018tev}.
For the sake of brevity we focus on the ILC scenario.\footnote{The potential of the 500~\gev{} ILC and the initial stage at $\sqrt{s}=$ 380~\gev{} of the CLIC project~\cite{CLIC:2016zwp, Abramowicz:2018rjq, Charles:2018vfv} is found to be very similar for the relevant two-fermion operators, when rescaled by the appropriate integrated luminosity~\cite{Durieux:2018ekg}.}
We again consider the nominal operating scenario~\cite{Barklow:2015tja}, with an integrated luminosity of 4~\iab{} at $\sqrt{s}=$ 500~\gev{} with two different beam polarizations, $P\left(e^-,e^+\right) = (-0.8, +0.3)$ and $P\left(e^-,e^+\right) = (+0.8, -0.3)$. 
  
The projections for the $e^-e^+ \rightarrow t\, \bar{t}$ process are based on the statistically optimal observables  motivated in Ref.~\cite{Durieux:2018tev}.
These observables are optimized to fully exploit the $bW^+\bar{b}W^-$ differential information (in the narrow top-quark width approximation) and extract the tightest constraints on parameters with linear dependence.
In our case, these optimal observables place bounds on subset of operators that affect the top-quark EW couplings; $C_{\varphi t}$, $C_{\varphi Q}^-$, $C_{tW}$ and $C_{tB}$.
Ref.~\cite{Durieux:2018tev} demonstrates that at least two center-of-mass energies are needed if one wants to constrain all two-fermion and four-fermion operators coefficients simultaneously.
The experimental uncertainties are studied in full simulation in Ref.~\cite{Amjad:2015mma, Abramowicz:2018rjq}.
Statistical uncertainties are estimated including the relevant branching ratios for the lepton+jets final state, the effect of the luminosity spectrum and a $t\bar{t}$ reconstruction efficiency of $50\%$.
This yields an effective efficiency of $~10\%$ that multiplies the $e^{+}e^{-} \rightarrow t \bar{t}$ cross-section (see Ref.~\cite{Durieux:2018tev} for more details).

\subsection{Global fit on prospects}

\begin{figure*}[]
    \centering
    {\includegraphics[scale=0.8]{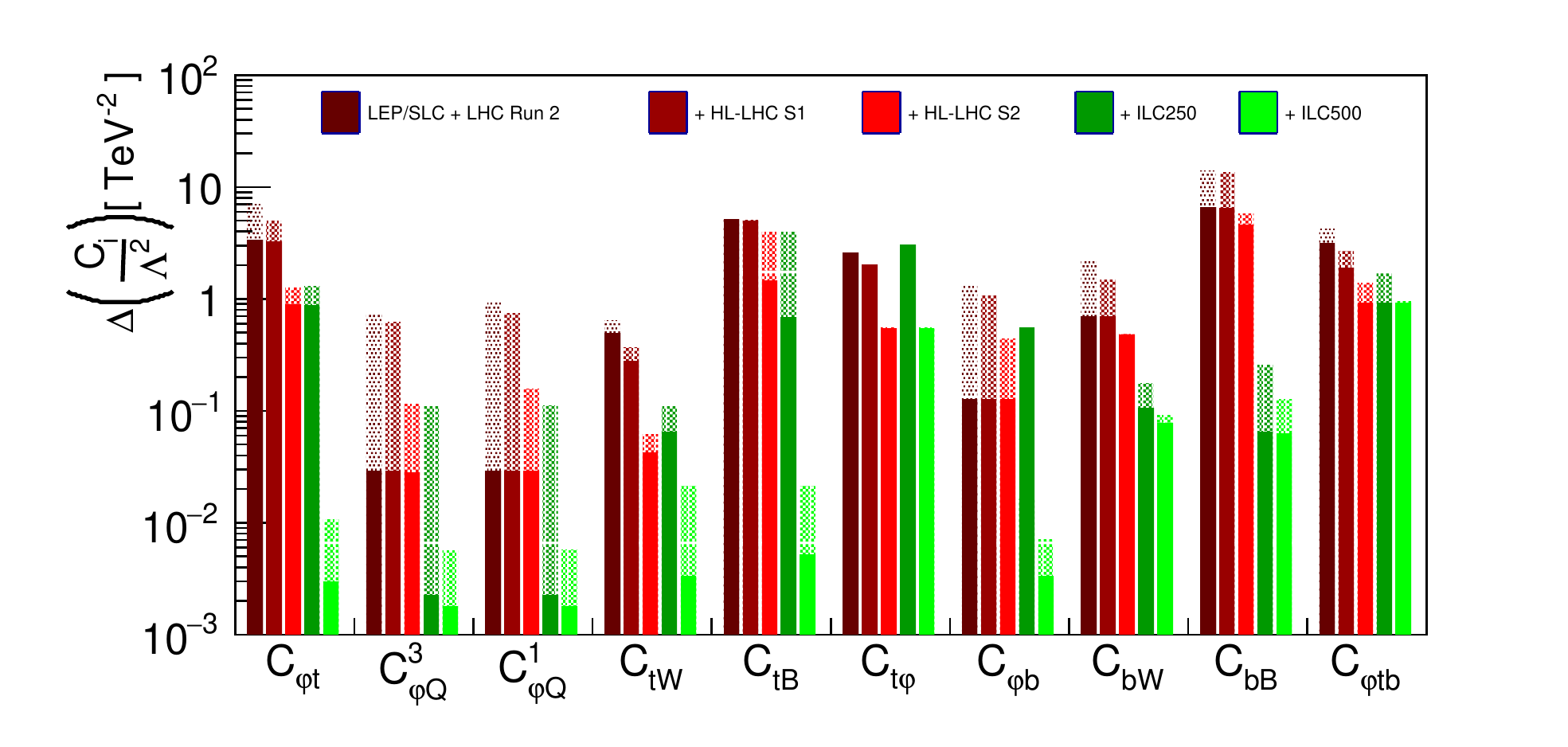}}
    \caption{Prospects for the precision of the Wilson coefficients in future high-luminosity operation of the LHC and at a high-energy $e^+e^-$ collider.
Assumptions on the operating scenarios and details of the uncertainty estimates are given in text.
The solid section of the bars represents the individual constraints, where each parameter is fitted in isolation, the full length indicates the marginalized constraint in a ten-parameter fit. The complete covariance matrices of the fits that are presented in this figure are available in Appendix~\ref{sec:cov_matrices}.}
    \label{fig:manhattan_plot}
\end{figure*}

In~\autoref{fig:manhattan_plot}, we present the global fit results for the future collider scenarios introduced in the previous sections. The complete covariance matrices for all the fits are provided in Appendix~\ref{sec:cov_matrices}. In~\autoref{fig:manhattan_plot} the uncertainty $\Delta C_i$ on the operator coefficients is shown. This uncertainty is estimated as half of the 68\% probability interval. In order to compare all projects on an equal footing, the central value of all measurements, including the existing LHC and LEP/SLC results, is set to the SM value. 
For each Wilson coefficient, the first vertical bar represents the current data. In the second and third bars, the measurements envisaged in the S1 or S2 scenario for the HL-LHC are added. The fourth bar includes the LEP/SLC data, the data of the HL-LHC S2 scenario and the ILC run at $\sqrt{s}=$ 250~\gev. The fifth bar adds also the 500~\gev{} run at the ILC. 
A discusstion of the extraction of the top-quark Yukawa coupling is postponed to  \autoref{sec:yukawa}.

For the first HL-LHC scenario, S1, we find that, due the conservative assumptions on systematic uncertainties, the bounds on the Wilson coefficient improve only marginally.
In the S2 scenario, almost all limits are considerably tighter.
For the dipole operator $O_{tB}$ the constraint remains very poor due the limited sensitivity of the LHC observables.
This could be improved by the addition of the differential $t\bar{t}\gamma$ measurement, as discussed in \autoref{subsec:hllhc}. 

The individual and marginalized limits for the operators that affect only the top-quark sector are very similar. Most operators are constrained from several angles, by different LHC observables (see~\autoref{fig:individual_limits}).
This limits the correlation in the global fit. In the bottom-quark sector, the sensitivity is dominated
by the $R_b$ measurement, giving rise to a strong correlation and considerably larger differences between individual and marginalized limits.

Adding the $e^+e^- \rightarrow b\,\bar{b} $ data at $\sqrt{s}=$ 250~\gev{} provides an improvement for the pure bottom-quark operators by an order of magnitude. The top-quark operators improve somewhat as well, due to a reduction of the correlation with the bottom-quark operators. 

Finally, we consider the ILC500 scenario.
At this energy, the sensitivity to the bottom-quark operators is very similar to that at $\sqrt{s}= 250~\gev$.
As the $b\bar{b}$ production cross section decreases with the center-of-mass energy, the addition of the 500~\gev{} data does not provide an important improvement on the bottom-quark coefficients limits. 

On the contrary, the addition of the $e^+e^- \rightarrow t\,\bar{t}$ data leads to a very pronounced improvement of the constraints on the top-quark operator coefficients, by one or two orders of magnitude. The direct access to the $Z/\gamma\:t\bar{t}$ vertex provides very tight constraints. Also the bounds on $C_{\varphi Q}^1/\Lambda^2$ and $C_{\varphi Q}^3/\Lambda^2$ are expected to improve by an order of magnitude. The combination of high-precision constraints on the two linear combinations ($C_{\varphi Q}^1 + C_{\varphi Q}^3$, that affects bottom-quark pair production, and the difference, $C_{\varphi Q}^1 - C_{\varphi Q}^3$, that affects top-quark pair production)
finally lift the degeneracy that affects the LHC/LEP/SLC fit of \autoref{sec:results}.

\subsection{Validity of the EFT framework}

In \autoref{subsec:lambda4}, the terms of order $\Lambda^{-4}$ were found to have a considerable impact on the fit to current LHC and LEP/SLC data.
This limits the generality of the interpretation to extensions of the SM where the contribution of the dimension-eight terms we have ignored is less important than that of the dimension-six operators we have included.
With the increasing precision of the measurements at the LHC and at future facilities, this tension in the EFT description is expected to decrease.

In the second HL-LHC scenario, S2, the difference between the nominal fit and a fit based on 
a parameterization that only considers the $\Lambda^{-2}$ terms is indeed reduced significantly.
In fact, for most of the observables the former gives better constraints (by a factor 3 at most) due to the fact that the observables depend on less parameters because of the vanishing $\Lambda^{-4}$ terms for $C_{bW}$, $C_{bB}$ and $C_{\varphi t b }$ in the $m_b\rightarrow 0$ limit. 
However, the $\Lambda^{-4}$ term still plays an important role for $C_{tB}$ due to the suppression of the linear term explained in \autoref{sec:eft}.

The high-precision measurements in $e^+e^-$ collisions improve the bounds by at least an order 
of magnitude and bring most operator coefficients safely into the range where the EFT expansion
is valid in full generality. The difference between the nominal fit and the fit based on
only $\Lambda^{-2}$ terms is reduced to less than 20\%.

\subsection{Four-fermion operators of the form \texorpdfstring{$e^+e^-Q\,\bar{Q}$}{e+e-QQbar}}
\label{subsec:fourfermion}

In this subsection, we discuss the perspective for an extension of the fit to the complete set of CP-conserving dimension-six operators that affect the bottom and top-quark EW couplings. 

The two-lepton-two-quark operators contributing to $e^+e^-t\,\bar{t}$ and $e^+ e^-b\,\bar{b}$ (as well as $\nu e^-t\bar{b}$) interactions are the following:
\begin{equation}
\begin{array}{rl@{\,}cc}
	O_{lq}^1
		&\equiv \frac12
		&\bar{\ges q}\gamma_\mu \ges q
		&\bar{\ges l}\gamma^\mu \ges l
	,\\
	O_{lq}^3
		&\equiv \frac12
		&\bar{\ges q}\tau^I\gamma_\mu \ges q
		&\bar{\ges l}\tau^I\gamma^\mu \ges l
	,\\
	O_{lu}
		&\equiv \frac12
		&\bar{\ges u}\gamma_\mu \ges u
		&\bar{\ges l}\gamma^\mu \ges l
	,\\
	O_{ld}
		&\equiv \frac12
		&\bar{\ges d}\gamma_\mu \ges d
		&\bar{\ges l}\gamma^\mu \ges l
	,\\
	O_{eq}
		&\equiv \frac12
		&\bar{\ges q}\gamma_\mu \ges q
		&\bar{\ges e}\gamma^\mu \ges e
	,\\
	O_{eu}
		&\equiv \frac12
		&\bar{\ges u}\gamma_\mu \ges u
		&\bar{\ges e}\gamma^\mu \ges e
	,\\
	O_{ed}
		&\equiv \frac12
		&\bar{\ges d}\gamma_\mu \ges d
		&\bar{\ges e}\gamma^\mu \ges e
	,
	\end{array}
\quad
\begin{array}{@{}rlcc@{}}
	O_{lequ}^T
		&\equiv
		&\bar{\ges q}\sigma^{\mu\nu} \ges u
		&\epsilon\bar{\ges l}\sigma_{\mu\nu} \ges e
	,
	\end{array}
\quad
\begin{array}{@{}rlcc@{}}
	O_{lequ}^S
		&\equiv
		&\bar{\ges q}\ges u
		&\epsilon\;
		\bar{\ges l}\ges e
	,\\
	O_{ledq}
		&\equiv
		&\bar{\ges d}\ges q
		&\bar{\ges l}\ges e
	,
\end{array}
\label{eq:op_2q2l}
\end{equation}
where $\ges l \equiv (\pmns \nu_L, e_L)^T$, $ \ges e \equiv e_R$, and $\pmns$\ is the Pontecorvo-Maki-Nakagawa-Sakata~\cite{Pontecorvo:1957cp, Maki:1962mu, Pontecorvo:1967fh} matrix. We define $O_{lq}^{+} = O_{lq}^{1} + O_{lq}^{3}$ which mediates $b\bar{b}$ production and $O_{lq}^{-} = O_{lq}^{1} - O_{lq}^{3}$ for $t\bar{t}$ production in $e^+ e^-$ collisions. 

The seven operators in the left column of \autoref{eq:op_2q2l} have vector Lorentz structures similar to SM gauge interactions.
Three further scalar and tensor operators, have non-standard Lorentz structures and can effectively be constrained with specialized observables~\cite{Durieux:2018tev} and runs with left-left or right-right beam polarization~\cite{Barklow:2015tja}.
In the following, we therefore focus on the seven vector operators.

The primary handle to constrain the two-fermion and four-fermion operators in a global fit is the energy dependence. The sensitivity to four-fermion operators grows very strongly with energy, while that to the two-fermion operators is essentially flat.

At hadron colliders, the four-fermion operators of $e^+e^-t\,\bar{t}\;$ form can, at least in principle, be constrained by a differential analysis of the cross section of the $pp \rightarrow t\,\bar{t}\,e^+e^-$ process versus the invariant mass and transverse momentum of the $e^+e^-$ system~\cite{Bylund:2016phk}.
The fit can then disentangle the photon, $Z$-boson, and the contact interaction contributions.
No such analysis has been made public, so far.

A future $e^+e^-$ collider with multiple energy stages is expected to provide a powerful bound on the four-fermion operator coefficients.
In Ref.~\cite{Durieux:2018tev}, a ten-parameter fit of the two-fermion and four-fermion operator coefficients that affect the EW couplings of the top quark is shown to provide stringent bounds when at least two well-separated energy stages are available.

To estimate the effect of the inclusion of the four-fermion operators, we extend the fit with seven additional degrees of freedom.
At the same time, the prospects for measurements at $\sqrt{s}=1~\tev$, with an integrated luminosity of 8~\iab{}, are added to the HL-LHC+ILC250+ILC500 scenario.
For the top-quark operators we again adopt the projections of Ref.~\cite{Durieux:2018tev}, for bottom-quark operators, statistical uncertainties on the cross-section and $A_{FB}$ are propagated, assuming a conservative acceptance times selection efficiency of 10\%.

The results of this extended fit are shown in \autoref{tab:global_fit_1000}.
The marginalized 68\% probability bounds are compared to those obtained in the ten-parameter fit (i.e.\ the results labeled ILC500 in \autoref{fig:manhattan_plot}). 
\begin{table}[t]
    \centering
    \begin{tabular}{l|cc}
         & 10-parameter fit & 17-parameter fit \\
         & ILC250 + ILC500 &  + ILC1000 \\   \hline
    $C_{\varphi t}/\Lambda^{2}$    &  0.01  & 0.09\\
    $C_{\varphi Q}^3/\Lambda^{2}$   &  0.006 & 0.04\\
    $C_{\varphi Q}^1/\Lambda^{2}$  & 0.006 &0.04\\
       $C_{tW}/\Lambda^{2} $     &  0.02 &  0.014\\
       $C_{tB}/\Lambda^{2}$    &    0.02 & 0.015 \\
       $C_{t \varphi}/\Lambda^{2}$    &  0.55  & 0.55\\
       $C_{\varphi b}/\Lambda^{2}$ & 0.007& 0.007 \\
       $C_{bW}/\Lambda^{2}$     &  0.09 &  0.17\\
       $C_{bB}/\Lambda^{2}$   & 0.13 &  0.17  \\
       $C_{\varphi tb}/\Lambda^{2}$ & 0.95 & 1.94 \\
       $C_{eu}/\Lambda^{2}$   & --- &  0.0005  \\
       $C_{ed}/\Lambda^{2}$   & --- &  0.0005  \\
       $C_{eq}/\Lambda^{2}$   & --- &  0.0004  \\
       $C_{lu}/\Lambda^{2}$   & --- &  0.0005  \\
       $C_{ld}/\Lambda^{2}$   & --- &  0.0009  \\
       $C_{lq}^{-}/\Lambda^{2}$   & --- &  0.0005  \\
       $C_{lq}^{+}/\Lambda^{2}$   & --- &  0.0005 
    \end{tabular}
    \caption{The marginalized 68\% probability bounds on the dimension-six operator coefficients in units of $\tev^{-2}$.
The results in the first column are based on a ten-parameter fit  on pseudo-data from two ILC runs, with an integrated luminosity of 2~\iab{} at 250~\gev{} and 4~\iab{} at $\sqrt{s}= 500~\gev$.
These results are identical to those of the ILC500 entry in \autoref{fig:manhattan_plot}.
The second column presents the results of the seventeen-parameter fit.
It includes an additional run, with an integrated luminosity of 8~\iab{} at $\sqrt{s}= 1~\tev$ and seven additional degrees of freedom corresponding to two-lepton-two-third-generation-quark operators.
}
    \label{tab:global_fit_1000}
\end{table}

This seventeen-parameter fit yields excellent limits on the four-fermion operators, below $10^{-3}$ $\tev^{-2}$.
The bounds agree with those of Ref.~\cite{Durieux:2018tev} when the larger integrated luminosity in the 1~\tev{} scenario is accounted for.

The bounds on the dipole operators are similar to those of the ten-parameter fit: the bounds on the coefficients $C_{tW}/\Lambda^2$ and $C_{tB}/\Lambda^2$ of the top-quark dipole operators improve somewhat, as the sensitivity of the optimal observables grows with increasing center-of-mass energy.
The bound on $C_{bW}/\Lambda^2$ derives from cross-section and $A_{FB}$ in $e^+e^-\to b\,\bar{b}$ measurements.
It does therefore not improve at higher center-of-mass energies and moreover suffers somewhat from the introduction of additional $e^+e^-b\,\bar{b}$ degrees of freedom.

The largest difference between the two fits is found for the two-fermion operators that modify the left-handed couplings of the top and bottom-quark to the $Z$ boson or the right-handed coupling of the top quark to the $Z$ boson.
The presence of the four-fermion operators degrades the excellent limits on $C_{\varphi t}/\Lambda^2$ and $C_{\varphi Q}^{1,3}/\Lambda^2$ by a factor eight.

We conclude, therefore, that a global EFT fit, including all dimension-six operators that affect the top and bottom-quark EW interactions, is feasible provided data is collected at two sufficiently distinct centre-of-mass energies above the top-quark pair production threshold.

\section{The top-quark Yukawa coupling}
\label{sec:yukawa}

In this section, we extract the top-quark Yukawa coupling from LHC data and the prospects for measurements at the HL-LHC and ILC.

\subsection{Indirect and direct bounds}

The top-quark Yukawa coupling is one of the most intriguing parameters in the SM. With a value close to 1 it is the largest of all Yukawa couplings.
New physics scenarios such as two-Higgs-doublet models, supersymmetric scenarios with small $\tan\beta$, and composite Higgs models~\cite{Dawson:2013bba} could lead to sizeable shifts from the SM prediction.
A precise and robust measurement is therefore one of the main targets of high-energy physics experiments in the next decades. 

The measurements of the Higgs boson decays and production rates other than $t\bar{t}H$ yield indirect constraints on the top-quark Yukawa coupling.
A model-dependent bound can be derived from the loop-induced $gg \rightarrow H$, $H\rightarrow Z\gamma$ and $H\rightarrow \gamma \gamma$ rates.
In the SM, the top-quark loop is the dominant contribution to these rates, but the effective couplings to the photon and the gluon could also receive contributions from new particles.
In the $\kappa$ fit framework employed in early Higgs coupling fits, these BSM contributions are assumed to be absent and the $gg \rightarrow H$ and $H\rightarrow \gamma \gamma$ rates yield a tight constraint on the factor $\kappa_{t} = \kappa_c = \kappa_u$ that multiplies the Yukawa couplings of the up-type quarks. The legacy result of LHC Run 1 is $\kappa_t = 1.40^{+0.24}_{-0.21}$~\cite{Khachatryan:2016vau}.
Significantly sharper results are available from Run 2 measurements~\cite{ATLAS-CONF-2019-005,Sirunyan:2018koj}. These indirect bounds tend to weaken considerably in a global fit. 

An $e^+ e^-$ collider also offers several handles on the top-quark Yukawa coupling. The same indirect methods are available at center-of-mass energies below the $t\bar{t}H$ production threshold.
In the $\kappa$ framework with $\kappa_u = \kappa_c = \kappa_t$, the precise determination of the $H \rightarrow c\bar{c}$ decay rate yields a tight bound on that parameter.
The Yukawa coupling can also be extracted indirectly from the measurement of the $Hgg$ and $H\gamma \gamma$ couplings, with 1\% precision after 2~\iab{} at $\sqrt{s}=$ 250~\gev~\cite{Boselli:2018zxr}.
A global effective field theory (EFT) analysis of the indirect sensitivity of Higgs and diboson measurements to EW top-quark couplings, including the top-quark Yukawa coupling, is performed in Ref.~\cite{Durieux:2018ggn}.
It is found that differential measurements are crucial to simultaneously disentangle all tree and loop-level contributions.

Several attempts have been made to disentangle the contributions of different operators that contribute to the $gg \rightarrow H$ (and $H \rightarrow \gamma\gamma$) rates (see Ref.~\cite{Azatov:2016xik} and references therein) with additional probes, such as boosted Higgs+jet production, di-Higgs boson production, off-shell Higgs production.
None of these seem sufficiently sensitive to lift the degeneracy between the operator that modifies the top-quark Yukawa coupling and operators representing $Hgg$ (or $H\gamma\gamma$) contact interactions. 

Therefore, we focus on the direct bound from $t\bar{t}H$ production in this paper.

\subsection{Associated \texorpdfstring{$t\bar{t}H$}{ttbarH} production at the LHC}

The observation of the associated production process of a top-quark pair with a Higgs boson~\cite{Aaboud:2018urx} provides a direct demonstration of the interaction of the Higgs boson with the top quark.
The ratio $\mu_{t\bar{t}H}$ of the measured cross section and the SM prediction is determined with a precision approaching 20\%.
With an uncertainty of 8\%, the NLO QCD prediction in the SM is also relatively precise.
The extraction of the top-quark Yukawa coupling from the $pp \rightarrow t\bar{t}H$ rate could thus yield a competitive and robust result, provided all other EFT contributions are sufficiently well constrained.

The fit presented in \autoref{sec:results} includes the LHC measurement of the $pp \rightarrow t\bar{t}H$ production cross section. A single-parameter fit yields an individual 68\% probability bound on the operator coefficient $C_{t \varphi}$ that shifts the value of the top-quark Yukawa coupling: 
 \begin{displaymath}
  C_{t \varphi}/\Lambda^{2} \in [-4.4, 0]\, \tev^{-2} \hskip 1cm \text{(individual).}
 \end{displaymath}
 Due to a small quadratic term in the dependence of the $t\bar{t}H$ cross section on $C_{t \varphi}$, the fit finds a second minimum very far from the SM value. Here we only treat the minimum which is closer to the SM value.
The bound becomes only slightly weaker in the ten-parameter fit: 
\begin{displaymath}
    C_{t \varphi}/\Lambda^{2} \in  [-4.6,-0.2] \, \tev^{-2}  \hskip 1cm \text{(marginalized).}
\end{displaymath} 
The individual and marginalized results are very close to each other, an indication that the constraint from the $t\bar{t}H$ rate is very robust against the effect of the operators that modify top-quark EW couplings.
The dependence of $pp \rightarrow t\bar{t}H$ on other top-quark EW operators arises mainly from $q\bar{q}$-initiated production which is subdominant compared to the $gg$-initiated process.
The correlation of $C_{t\varphi}/\Lambda^2$ with $C_{tW}/\Lambda^2$, $C_{\varphi Q}^3/\Lambda^2$, $C_{tB}/\Lambda^2$ and $C_{bW}/\Lambda^2$ is small, below 0.1\%.
We note, however, that including four-fermion $q\bar{q}t\bar{t}$ operators can have a significant impact on the extraction of the top-quark Yukawa coupling from $pp \rightarrow t\bar{t}H$ measurement.

\subsection{HL-LHC prospects}

The fit is repeated on projections to assess the expected precision after the complete data set collected during the high-luminosity phase of the LHC.
As before, we focus on the S2 scenario, based on an integrated luminosity of 3~\iab{} at $\sqrt{s}=$ 14~\tev. In this scenario, the statistical uncertainty on the $t\bar{t}H$ cross section becomes negligible and the precision is primarily limited by the precision of the theory prediction (currently 8\% and assumed to improve to 4\%). The precision of the global fit improves considerably, reducing the 68\% probability interval to $[-0.55,+0.55]$.
This result agrees with the S2 prospects in Ref.~\cite{Cepeda:2019klc}.

\subsection{ILC prospects}

The direct measurement of the Yukawa coupling in $e^+e^- \rightarrow t\bar{t}H$ production requires operation at a center-of-mass energy above the $t\bar{t}H$ production threshold.
The cross section turns on sharply at around $\sqrt{s}=$ 500~\gev{}.
The unpolarized cross section reaches a maximum of 2~fb at a center-of-mass energy of approximately 800~\gev.
The $t\bar{t}H$ production rate is two orders of magnitude lower than that for top-quark pair production rate, which forms the most important background for the $H\rightarrow b\bar{b}$ analysis.
The cross section of the irreducible $t\bar{t}b\bar{b}$ background, either from associated $t\bar{t}Z$ production or a hard gluon splitting to a $b\bar{b}$ pair, is similar to that of the signal.  

Full-simulation studies of the potential of the linear collider~\cite{Abramowicz:2018rjq,Price:2014oca,Yonamine:2011jg,Gay:2006vs,Juste:1999af} have been performed at center-of-mass energies from 500~\gev{} to several \tev{}s.
They include realistic descriptions of the $t\bar{t}$ and $t\bar{t}Z$ backgrounds, of the detector response, flavour tagging and jet clustering.

Projections for the nominal ILC programme~\cite{Barklow:2015tja}, with 4~\iab{} of integrated luminosity collected at 500~\gev{} are presented in Ref.~\cite{Fujii:2015jha}.
An uncertainty of 13\% is expected on the $t\bar{t}H$ cross section, limited by statistics.
As the nominal ILC energy is very close to the $t\bar{t}H$ production threshold, operation at a slightly higher energy improves the precision considerably.
Increase of the center-of-mass energy by 10\% (i.e.\ to $\sqrt{s}=$ 550~\gev) enhances the cross section by a factor of four and the precision on the Yukawa coupling by a factor two, for the same integrated luminosity~\cite{Fujii:2015jha}.

We base our projection for 1~\tev{} operation on the analysis of Ref.~\cite{Price:2014oca}  of $t\bar{t}H$ production followed by $H\rightarrow b\bar{b}$ decay.
The expected uncertainty on the $t\bar{t}H$ cross section for an integrated luminosity of 8~\iab{} is of 3.2\%, obtained by scaling the signal and background yields with a flat luminosity factor.

To match the statistical precision, the systematic uncertainties must be controlled to a challenging level. At 1~\tev{} the signal efficiency and background yield must be known to approximately 1\%, which seems feasible with  data-driven estimation in control regions. The theory uncertainty in the cross section at $\sqrt{s}=$ 1~\tev{} must be reduced to the level of 1-2\%, a factor two with respect to currently available calculations~\cite{Nejad:2016bci}.
On the other hand, it is likely that the analysis can be further improved, by reoptimizing the selection, with the inclusion of other Higgs decay channels and of the $\tau$-lepton plus jets final state.
Significant additional improvements are possible with improved jet clustering algorithms and the use of kinematic fits.

\subsection{Summary of results}

In \autoref{tab:yukawa_summary}, we present the individual and marginalized 68\% probability bounds on $C_{t\varphi}/\Lambda^2$ from the fits to LEP/SLC+LHC data and to the future collider scenarios. For comparison to the literature, the same results are also provided in terms of the precision with which the Yukawa coupling can be extracted, using the simple relation:
\begin{equation}
\delta y_t  = - \frac{C_{t\varphi}v^2}{\Lambda^2}.
    \label{eq:yukawa_eft}
\end{equation}
The results of the first four columns correspond to the ten-parameter fit that we used to obtain the results of \autoref{fig:manhattan_plot}.
The results for the scenario with ILC runs at two different center-of-mass energies in the last column were obtained with the extended seventeen-parameter fit presented in \autoref{subsec:fourfermion}.

\begin{table*}[h!]
\centering
\resizebox{1\linewidth}{!}{\begin{tabular}{lccccc}
\hline
scenario      & LHC Run 2   &  HL-LHC S2    &  ILC500   &   ILC550 &    ILC500  \\ 
              & +LEP/SLC    &  +LEP/SLC     &           &          &   +ILC1000  \\ 
$\sqrt{s}$, $\int \cal{L}$   &   13\tev, 36~\ifb & 14\tev, 3~\iab  & 500~\gev, 4~\iab & 550~\gev, 4~\iab & +1~\tev, +8~\iab  \\ \hline
\multicolumn{6}{l}{\em 68\% probability interval for effective operator coefficient $C_{t\varphi}/\Lambda^2$ $[\tev^{-2}]$} \\ 
individual    & $[-4.4,+0.0]$    &  $[-0.55,+0.55]$   &   $[-1.06,+1.06]$    &    $[-0.50,0.50]$    & $[-0.27,+0.27]$ \\
 marginalized & $[-4.6,-0.2]$ &  $[-0.55,+0.55]$   &   $[-1.07,+1.07]$   &   $[-0.52,+0.52]$  & $[-0.32,+0.32]$ \\ \hline
 \multicolumn{6}{l}{\em corresponding relative uncertainty on top-quark Yukawa coupling $\Delta y_{t}/y_{t}$ [\%]} \\
individual    & 13.2    &  3.3   &   6.4    &    3.0    & 1.62 \\
 marginalized & 13.2    &  3.3   &   6.4   &     3.1    &  1.96   \\
 \hline
 \end{tabular}}

\caption{The 68\% probability intervals for $C_{t\varphi}/\Lambda^{2}$ and the corresponding precision on the top-quark Yukawa coupling. The results of the first four columns correspond to the ten-parameter fit that we used to obtain the results of \autoref{fig:manhattan_plot}. The results for the scenario with ILC runs at two different center-of-mass energies in the last column were obtained with the extended seventeen-parameter fit presented in \autoref{subsec:fourfermion}.}
\label{tab:yukawa_summary}
\end{table*}

Before turning to a discussion of the global fit results, we compare the individual limits to the literature. The HL-LHC result in \autoref{tab:yukawa_summary} agrees with the HL-LHC projection of Ref.~\cite{Cepeda:2019klc}. The ILC results at 500~\gev{} agree ---by construction--- with the summary of the Higgs/EW group for the 2020 update of the European strategy for particle physics in Ref.~\cite{deBlas:2019rxi}. The results for operation at 550~\gev{} and 1~\tev{} extend the study to 
higher energy.

We find that in nearly all cases the individual and marginalized results agree very closely. 
This implies that the operators that modify the top-quark EW couplings do not affect the extraction of the top-quark Yukawa coupling.

In the LHC and HL-LHC fits, despite the relatively poor constraints on the EW couplings of the top quark, the bounds on $C_{t\varphi}/\Lambda^2$ are not affected by the presence of the additional degrees of freedom.
In this case, it is important to note, however, that the operators that affect the QCD interactions of the top quark, such as $C_{tG}/\Lambda^2$ and four-fermion operators of the form $q\bar{q}t\bar{t}$, are not included in the fit.
These can in principle be constrained using precise measurements of the differential $t\bar{t}$ cross section.
A recent global fit of the top-quark sector on LHC data~\cite{Hartland:2019bjb} finds, however, that the marginalized limit on $C_{t \varphi}$ is approximately a factor 10 weaker than the individual limit, due to strong correlations between operator coefficients.
The addition of Tevatron results or future differential measurements could help reducing this degeneracy.
It is nevertheless likely that a combination of $pp\to t\,\bar{t}$ and $pp\to t\,\bar{t}X$ measurements could be needed to constrain simultaneously all $q\bar{q}t\bar{t}$ operators.
In this respect, the extraction of the top-quark Yukawa coupling at future lepton colliders so far seems more robust.

At a future $e^+e^-$ collider, we indeed find that the contamination of both four-fermion and two-fermion operators in $e^-e^+ \rightarrow t\bar{t}H$ is limited due to the very tight constraints on these coefficients deriving from $e^-e^+ \rightarrow t\,\bar{t}$ production.
Even in the most challenging case, the ILC scenario at 1~\tev{} with a precision on the top-quark Yukawa coupling of 1.6\% and sixteen competing operator coefficients, the marginalized bound is only about 20\% weaker than the individual bound.
The extraction of the top-quark Yukawa is then very clean in this case.
We also note that the measurement of $e^-e^+ \rightarrow t\bar{t}H$ in addition to $e^-e^+ \rightarrow t\,\bar{t}$ does not improve significantly the constraints on operators other than the top-quark Yukawa one.
Only a $14\%$ improvement is observed on $C_{tW}$.

The results in \autoref{tab:yukawa_summary} demonstrate that the bounds on the Wilson coefficient $C_{t\varphi}/\Lambda^2$ that shifts the top-quark Yukawa coupling from measurements of the $t\bar{t}H$ production are robust in the presence of the operators that affect the top and bottom-quark EW couplings. A precise measurement of this rate is therefore an ideal complement to more indirect bounds from $gg\rightarrow H$ production and $H\rightarrow \gamma \gamma$, $H\rightarrow gg$ and $H \rightarrow Z\gamma$ decay.


\section{Conclusions}
\label{sec:conclusions}

We have performed a fit to existing data of the dimension-six two-fermion operator coefficients affecting the electro-weak couplings of the bottom and top quarks.
We combine LEP/SLC data on bottom-quark production at the $Z$ pole with LHC data on top-quark pair production in association with bosons, on single top-quark production and on $W$-boson helicity fraction in top-quark decay. 

The results of the fit are given in \autoref{tab:quadratic}.
All 68\% probability intervals include the Standard Model prediction.
The bound is well below $1~\tev^{-2}$ for the coefficient of the top-quark electro-weak dipole operator $C_{tW}/\Lambda^2$ that is constrained by charged-current interactions.
Very tight bounds are also obtained for the coefficients $C_{\varphi Q}^1/\Lambda^2$ and $C_{\varphi Q}^3/\Lambda^2$ that modify the left-handed couplings of the bottom and top quark to the $Z$ boson.
The combination of LHC data with that of LEP and SLC is very powerful to disentangle these operator coefficients that affect both top and bottom-quark physics. We are therefore able to present the tightest constraints on these operators to date. 

The LHC has limited sensitivity to the operator coefficient $C_{\varphi t}$ that modifies the right-handed coupling of the top quark to the $Z$ boson and coefficients of order $10^1$ are still allowed.
The same is true for the electro-weak dipole operators $C_{tB}/\Lambda^2$ and $C_{bB}/\Lambda^2$.
Inclusion of measurements of the $B_s \rightarrow \mu^+\mu^-$ and $b \rightarrow s \gamma$ decay rates may help to improve those bounds.
For some operator coefficients, we moreover note that the fit results depend strongly on the presence of the terms proportional to $\Lambda^{-4}$ (due to the contribution of dimension-six operators squared).
Care is therefore required to re-interpret these bounds in terms of concrete extensions of the Standard Model. 

We assess the potential of future measurements to improve the current bounds. The main result is presented in \autoref{fig:manhattan_plot}.
The remaining LHC program, including the high-luminosity phase, can sharpen most bounds by a factor two to three, provided the uncertainties on the Standard Model predictions are improved by a factor two and experimental systematics evolve with luminosity in the same way as the statistical uncertainties.
An electron-positron collider with a center-of-mass energy that exceeds the top-quark pair production threshold, can greatly improve the bounds.
The nominal ILC operating scenario~\cite{Barklow:2015tja} with runs at $\sqrt{s}=250~\gev$ and 500~\gev{} is expected to improve on the HL-LHC bounds by one or two orders of magnitude. 

The precision measurements at a future lepton collider also reduce the importance of terms of order $\Lambda^{-4}$ and brings the EFT expansion into the regime where the bounds are valid in full generality. Finally, we show that with a further run at higher energy the ILC can constrain the coefficients of the four-fermion operators that are not included in our baseline fit.

Finally, we present prospects for the extraction of the top-quark Yukawa coupling from the associated production processes $pp\rightarrow t\bar{t}H$ and $e^+e^- \rightarrow t\bar{t}H$ in \autoref{tab:yukawa_summary}.
The current precision of order 10\% is expected to improve by more than a factor three in the HL-LHC S2 scenario.
The ILC can achieve a similar precision when operated at 550~\gev{} and can exceed this precision by a further factor two for a 1~\tev{} energy upgrade with 8~\iab.
These results are found to be robust in a multi-parameter fit that includes the degrees of freedom corresponding to operators that modify the EW couplings of the bottom and top quark.
They may however not be robust against the inclusion of $q\bar{q}t\bar{t}$ operators~\cite{Hartland:2019bjb} that are not considered in this work.
For BSM scenarios in which such operators are relevant, it remains to be demonstrated that their impact on $pp\rightarrow t\bar{t}H$ can be controlled to a sufficient level with other measurements.

The electro-weak couplings of the third-generation quarks form one of the uncharted corners of the Standard Model.
These couplings are a sensitive probe of broad classes of extensions of the Standard Model.
It is therefore very exciting to see meaningful bounds in a multi-parameter fit on LEP/SLC and LHC data.
Further progress at the LHC, and especially at a future electron-positron collider can probe subtle contributions from physics beyond the Standard Model at scales well beyond the direct reach of the collider.

\section*{Acknowledgements}

This work builds on the effort of the LHC collaborations and the ILC and CLIC simulation studies. We would like to acknowledge the work of our colleagues in this place. We owe special thanks to Otto Eberhardt, Sunghoon Jung, Michael Peskin, Toni Pich, Junping Tian and Cen Zhang for their feedback on the project and write-up. 
MP and MV are supported by the Spanish national program for particle physics, project FPA2015-65652-C4-3-R (MINECO/FEDER), and PROMETEO grant 2018/060 of the Generalitat Valenciana. MP is supported by the “Severo Ochoa” Grant SEV-2014-0398-05, reference BES-2015-072974. The work of GD is supported in part at the Technion by a fellowship from the Lady Davis Foundation. VM and AP are supported by the Spanish Government and ERDF funds from the EU Commission [Grant FPA2017-84445-P], the Generalitat Valenciana [Grant Prometeo/2017/053], the Spanish Centro de Excelencia Severo Ochoa Programme [Grant SEV-2014-0398] and the DFG cluster of excellence “Origin and Structure of the Universe”. The work of VM and AP is funded by Ministerio de Ciencia, Innovación y Universidades, Spain [Grant FPU16/01911] and [Grant FPU15/05103] respectively.

\appendix

\begin{landscape}

\section{Observables parameterization}
\label{sec:parameterization}

In this appendix we show the diferent parameterizations of the observables we have used along this work. The details of these calculations are explained in \autoref{sec:fit}.

\subsection{LHC @13TeV}

\begin{equation}
\sigma_{t\bar{t}Z} [pb] = 0.59 + \left(\frac{1 \TeV}{\Lambda} \right)^2 \left( {\begin{array}{c}
   C_{\varphi t}  \\
   C_{\varphi Q}^{3}  \\
   C_{\varphi Q}^{1} \\
   C_{tW} \\
   C_{tB} \\
   C_{bW} \\
   C_{\varphi tb} \\
  \end{array} } \right)^T
  \left( {\begin{array}{c}
   0.041  \\
   -0.066  \\
   0.066 \\
   0.00068 \\
   0.00024 \\
   \cdot \\
   \cdot \\
  \end{array} } \right)
  + \left(\frac{1 \TeV}{\Lambda} \right)^4
  \left( {\begin{array}{c}
   C_{\varphi t}  \\
   C_{\varphi Q}^{3}  \\
   C_{\varphi Q}^{1} \\
   C_{tW} \\
   C_{tB} \\
   C_{bW} \\
   C_{\varphi tb} \\
  \end{array} } \right)^T
  \left( {\begin{array}{ccccccc}
   0.0024  & 0.0018 & 0.0018 & \cdot & \cdot& \cdot & \cdot \\
   \cdot  & 0.005 & \cdot & 0.0001 & -0.0001& \cdot & \cdot \\
   \cdot  & \cdot & 0.005 & -0.0001 & 0.0001& \cdot & \cdot \\
   \cdot  & \cdot & \cdot & 0.018 & -0.01& \cdot & \cdot \\
   \cdot  & \cdot & \cdot & \cdot & 0.0016& \cdot & \cdot \\
   \cdot  & \cdot & \cdot & \cdot & \cdot & 0.002 & \cdot \\
   \cdot  & \cdot & \cdot & \cdot & \cdot & \cdot & 0.0003 \\
  \end{array} } \right)
  \left( {\begin{array}{c}
   C_{\varphi t}  \\
   C_{\varphi Q}^{3}  \\
   C_{\varphi Q}^{1} \\
   C_{tW} \\
   C_{tB} \\
   C_{bW} \\
   C_{\varphi tb} 
  \end{array} } \right)
\end{equation}

\begin{equation}
\sigma_{t\bar{t}\gamma} [pb] = 2.18 + \left(\frac{1 \TeV}{\Lambda} \right)^2 \left( {\begin{array}{c}
   C_{\varphi Q}^{3} \\
   C_{tW} \\
   C_{tB} \\
   C_{bW} \\
  \end{array} } \right)^T
  \left( {\begin{array}{c}
   0.0034\\
   0.015 \\
   0.015 \\
  \end{array} } \right)
  + \left(\frac{1 \TeV}{\Lambda} \right)^4
  \left( {\begin{array}{c}
   C_{\varphi Q}^{3} \\
   C_{tW} \\
   C_{tB} \\
   C_{bW} \\
  \end{array} } \right)^T
  \left( {\begin{array}{cccc}
    0.0004 & \cdot & \cdot & \cdot\\
    \cdot & 0.007 & 0.014 & \cdot\\
    \cdot & \cdot & 0.007 & \cdot\\
    \cdot & \cdot & \cdot & 0.001\\
  \end{array} } \right)
  \left( {\begin{array}{c}
   C_{\varphi Q}^{3} \\
   C_{tW} \\
   C_{tB} \\
   C_{bW} \\
  \end{array} } \right)
\end{equation}

\begin{equation}
\sigma_{t\bar{t}H} [pb] = 0.4 + \left(\frac{1 \TeV}{\Lambda} \right)^2 \left( {\begin{array}{c}
   C_{\varphi Q}^{3} \\
   C_{tW} \\
   C_{tB} \\
   C_{t\varphi} \\
   C_{bW} \\
  \end{array} } \right)^T
  \left( {\begin{array}{c}
   0.0002\\
   0.0007 \\
   0.00014 \\
   -0.049 \\
   \cdot \\
  \end{array} } \right)
  + \left(\frac{1 \TeV}{\Lambda} \right)^4
  \left( {\begin{array}{c}
   C_{\varphi Q}^{3} \\
   C_{tW} \\
   C_{tB} \\
   C_{t\varphi} \\
   C_{bW} \\
  \end{array} } \right)^T
  \left( {\begin{array}{ccccc}
  \cdot & \cdot & \cdot & \cdot & \cdot \\
   \cdot & 0.00067 & \cdot & \cdot & \cdot\\
   \cdot & \cdot & \cdot & \cdot & \cdot \\
   \cdot & \cdot & \cdot & 0.0015 & \cdot \\
   \cdot & \cdot & \cdot & \cdot & 0.0001 \\
  \end{array} } \right)
  \left( {\begin{array}{c}
   C_{\varphi Q}^{3} \\
   C_{tW} \\
   C_{tB} \\
   C_{t\varphi} \\
   C_{bW} \\
  \end{array} } \right)
\end{equation}

\begin{equation}
\sigma_{t\bar{t}W} [pb] = 0.35 + \left(\frac{1 \TeV}{\Lambda} \right)^2 \left( {\begin{array}{c}
   C_{\varphi t}  \\
   C_{\varphi Q}^{3}  \\
   C_{\varphi Q}^{1} \\
   C_{tW} \\
  \end{array} } \right)^T
  \left( {\begin{array}{c}
  -0.00013 \\
   0.00036  \\
   -0.0003 \\
   0.0027 \\
  \end{array} } \right)
  + \left(\frac{1 \TeV}{\Lambda} \right)^4
  \left( {\begin{array}{c}
  C_{\varphi t}  \\
   C_{\varphi Q}^{3}  \\
   C_{\varphi Q}^{1} \\
   C_{tW} \\
  \end{array} } \right)^T
  \left( {\begin{array}{cccc}
  \cdot & \cdot & \cdot &  \cdot\\
   \cdot & \cdot & 0.00012 &  \cdot\\
   \cdot & \cdot &  \cdot & -0.00011 \\
   \cdot& \cdot & \cdot & 0.032 \\
  \end{array} } \right)
  \left( {\begin{array}{c}
  C_{\varphi t}  \\
   C_{\varphi Q}^{3}  \\
   C_{\varphi Q}^{1} \\
   C_{tW} \\
  \end{array} } \right)
\end{equation}

\begin{equation}
\sigma_{tq} [pb] = 44 + \left(\frac{1 \TeV}{\Lambda} \right)^2 \left( {\begin{array}{c}
   C_{\varphi Q}^{3}  \\
   C_{tW} \\
   C_{bW} \\
   C_{\varphi tb}  \\
  \end{array} } \right)^T
  \left( {\begin{array}{c}
   5.26  \\
   1.52 \\
   \cdot \\
   \cdot \\
  \end{array} } \right)
  + \left(\frac{1 \TeV}{\Lambda} \right)^4
  \left( {\begin{array}{c}
   C_{\varphi Q}^{3}  \\
   C_{tW} \\
   C_{bW} \\
   C_{\varphi tb}  \\
  \end{array} } \right)^T
  \left( {\begin{array}{cccc}
   0.16  & 0.1 &  \cdot & \cdot\\
   \cdot & 0.31 &  \cdot & \cdot\\
   \cdot & \cdot &  0.19 & -0.012\\
   \cdot & \cdot &  \cdot & 0.019\\
  \end{array} } \right)
  \left( {\begin{array}{c}
   C_{\varphi Q}^{3}  \\
   C_{tW} \\
   C_{bW} \\
   C_{\varphi tb}  \\
  \end{array} } \right)
\end{equation}

\begin{equation}
\sigma_{tW} [pb] = 13.5 + \left(\frac{1 \TeV}{\Lambda} \right)^2 \left( {\begin{array}{c}
   C_{\varphi Q}^{3}  \\
   C_{tW} \\
   C_{bW} \\
   C_{\varphi tb}  \\
  \end{array} } \right)^T
  \left( {\begin{array}{c}
   1.61  \\
   -0.74 \\
   \cdot \\
   \cdot \\
  \end{array} } \right)
  + \left(\frac{1 \TeV}{\Lambda} \right)^4
  \left( {\begin{array}{c}
   C_{\varphi Q}^{3}  \\
   C_{tW} \\
   C_{bW} \\
   C_{\varphi tb}  \\
  \end{array} } \right)^T
  \left( {\begin{array}{cccc}
   0.05  & -0.046 & \cdot & \cdot \\
   \cdot & 0.135 & \cdot & \cdot \\
   \cdot & \cdot & 0.14 & -0.022 \\
   \cdot & \cdot & \cdot & 0.017 \\
  \end{array} } \right)
  \left( {\begin{array}{c}
   C_{\varphi Q}^{3}  \\
   C_{tW} \\
   C_{bW} \\
   C_{\varphi tb}  
  \end{array} } \right)
\end{equation}

\begin{equation}
\sigma_{Ztq} [pb] = 0.48 + \left(\frac{1 \TeV}{\Lambda} \right)^2 \left( {\begin{array}{c}
   C_{\varphi t}  \\
   C_{\varphi Q}^{3}  \\
   C_{\varphi Q}^{1} \\
   C_{tW} \\
   C_{\varphi b} \\
   C_{bW}\\
   C_{bB}\\
   C_{\varphi tb}  \\
  \end{array} } \right)^T
  \left( {\begin{array}{c}
   0.0029  \\
   0.092  \\
   0.01 \\
   0.007 \\
   -0.0003 \\
   \cdot \\
   \cdot \\
   \cdot \\
  \end{array} } \right)
  + \left(\frac{1 \TeV}{\Lambda} \right)^4
  \left( {\begin{array}{c}
   C_{\varphi t}  \\
   C_{\varphi Q}^{3}  \\
   C_{\varphi Q}^{1} \\
   C_{tW} \\
   C_{\varphi b} \\
   C_{bW}\\
   C_{bB}\\
   C_{\varphi tb}  \\
  \end{array} } \right)^T
  \left( {\begin{array}{cccccccc}
   \cdot  & 0.0005 & -0.0005 & \cdot  & \cdot & \cdot & \cdot& \cdot\\
   \cdot  & 0.014 & 0.0017 & 0.001  & \cdot & -0.0002 & \cdot& \cdot\\
   \cdot  & \cdot & 0.001 & 0.0003  & \cdot & \cdot & \cdot& \cdot\\
   \cdot  & \cdot & \cdot & 0.016  & \cdot & \cdot & \cdot& \cdot\\
   \cdot  & \cdot & \cdot & \cdot  & \cdot & -0.0002 & \cdot& \cdot\\
   \cdot  & \cdot & \cdot & \cdot & \cdot & 0.012 & \cdot& \cdot\\
   \cdot  & \cdot & \cdot & \cdot & \cdot & \cdot & 0.003& \cdot\\
   \cdot  & \cdot & \cdot & \cdot & \cdot & \cdot & \cdot & 0.002\\
  \end{array} } \right)
  \left( {\begin{array}{c}
   C_{\varphi t}  \\
   C_{\varphi Q}^{3}  \\
   C_{\varphi Q}^{1} \\
   C_{tW} \\
   C_{\varphi b} \\
   C_{bW}\\
   C_{bB}\\
   C_{\varphi tb}  
  \end{array} } \right)
\end{equation}

\subsection{\texorpdfstring{$t \rightarrow W^{+}b$}{t>Wb}}

\begin{equation}
F_0 = 0.699 + \left(\frac{1 \TeV}{\Lambda} \right)^2 \left( {\begin{array}{c}
   C_{\varphi Q}^{3}  \\
   C_{tW} \\
   C_{bW} \\
   C_{\varphi tb}  \\
  \end{array} } \right)^T
  \left( {\begin{array}{c}
  00026  \\
   -0.04 \\
   0.00032 \\
   \cdot \\
  \end{array} } \right) + \left(\frac{1 \TeV}{\Lambda} \right)^4
  \left( {\begin{array}{c}
   C_{\varphi Q}^{3}  \\
   C_{tW} \\
   C_{bW} \\
   C_{\varphi tb}  \\
  \end{array} } \right)^T
  \left( {\begin{array}{cccc}
   0.00025  & \cdot & \cdot & \cdot \\
   \cdot & 0.00129 & \cdot & \cdot \\
   \cdot & \cdot & -0.0026 & \cdot  \\
   \cdot & \cdot & \cdot & 0.00104 \\
  \end{array} } \right)
  \left( {\begin{array}{c}
   C_{\varphi Q}^{3}  \\
   C_{tW} \\
   C_{bW} \\
   C_{\varphi tb}  
  \end{array} } \right)
\end{equation}

\begin{equation}
F_L = 0.301 + \left(\frac{1 \TeV}{\Lambda} \right)^2 \left( {\begin{array}{c}
   C_{\varphi Q}^{3}  \\
   C_{tW} \\
   C_{bW} \\
   C_{\varphi tb}  \\
  \end{array} } \right)^T
  \left( {\begin{array}{c}
  -0.00026  \\
  0.039 \\
   -0.00035 \\
   \cdot \\
  \end{array} } \right) + \left(\frac{1 \TeV}{\Lambda} \right)^4
  \left( {\begin{array}{c}
   C_{\varphi Q}^{3}  \\
   C_{tW} \\
   C_{bW} \\
   C_{\varphi tb}  \\
  \end{array} } \right)^T
  \left( {\begin{array}{cccc}
   -0.00025 & \cdot & \cdot & \cdot \\
   \cdot & -0.00129 & \cdot & \cdot \\
   \cdot & \cdot & -0.0016 & \cdot  \\
   \cdot & \cdot & \cdot & 0.00208 \\
  \end{array} } \right)
  \left( {\begin{array}{c}
   C_{\varphi Q}^{3}  \\
   C_{tW} \\
   C_{bW} \\
   C_{\varphi tb}  
  \end{array} } \right)
\end{equation}

\subsection{LEP/SLC @91GeV}

\begin{equation}
R_{b} = 0.21629 + \left(\frac{1 \TeV}{\Lambda} \right)^2 \left( {\begin{array}{c}
   C_{\varphi Q}^{3}  \\
   C_{\varphi Q}^{1} \\
   C_{\varphi b} \\
   C_{bW}\\
   C_{bB}\\
  \end{array} } \right)^T
  \left( {\begin{array}{c}
   0.023  \\
   0.023  \\
   -0.005 \\
   \cdot \\
   0.0007 \\
  \end{array} } \right)
  + \left(\frac{1 \TeV}{\Lambda} \right)^4
  \left( {\begin{array}{c}
   C_{\varphi Q}^{3}  \\
   C_{\varphi Q}^{1} \\
   C_{\varphi b} \\
   C_{bW}\\
   C_{bB}
  \end{array} } \right)^T
  \left( {\begin{array}{ccccc}
   \cdot & \cdot  & \cdot & \cdot & \cdot\\
   \cdot & \cdot  & \cdot & \cdot & \cdot\\
   \cdot & \cdot  & \cdot & \cdot & \cdot\\
   \cdot & \cdot & \cdot & 0.0008 & 0.0003\\
   \cdot & \cdot & \cdot & \cdot & \cdot\\
  \end{array} } \right)
  \left( {\begin{array}{c}
   C_{\varphi Q}^{3}  \\
   C_{\varphi Q}^{1} \\
   C_{\varphi b} \\
   C_{bW}\\
   C_{bB}
  \end{array} } \right)
\end{equation}

\begin{equation}
A_{FBLR}^{bb} = 0.66 + \left(\frac{1 \TeV}{\Lambda} \right)^2 \left( {\begin{array}{c}
   C_{\varphi Q}^{3}  \\
   C_{\varphi Q}^{1} \\
   C_{\varphi b} \\
   C_{bW}\\
   C_{bB}\\
  \end{array} } \right)^T
  \left( {\begin{array}{c}
   0.008  \\
   0.008  \\
   0.034 \\
   \cdot \\
   \cdot \\
  \end{array} } \right)
  + \left(\frac{1 \TeV}{\Lambda} \right)^4
  \left( {\begin{array}{c}
   C_{\varphi Q}^{3}  \\
   C_{\varphi Q}^{1} \\
   C_{\varphi b} \\
   C_{bW}\\
   C_{bB}\\
  \end{array} } \right)^T
  \left( {\begin{array}{ccccc}
   \cdot & \cdot  & \cdot & \cdot & \cdot\\
   \cdot & \cdot  & \cdot & \cdot & \cdot\\
   \cdot & \cdot  & \cdot & \cdot & 0.0007\\
   \cdot & \cdot & \cdot & 0.0023 & 0.0015\\
   \cdot & \cdot & \cdot & \cdot & 0.0002\\
  \end{array} } \right)
  \left( {\begin{array}{c}
   C_{\varphi Q}^{3}  \\
   C_{\varphi Q}^{1} \\
   C_{\varphi b} \\
   C_{bW}\\
   C_{bB}\\
  \end{array} } \right)
\end{equation}

\subsection{\texorpdfstring{$e^-e^+ \rightarrow b \bar{b}$}{e-e+>bbbar}}

\begin{equation}
\sigma_{-+, 250}[pb] = 3.29 + \left(\frac{1 \TeV}{\Lambda} \right)^2 \left( {\begin{array}{c}
   C_{\varphi Q}^{3}  \\
   C_{\varphi Q}^{1} \\
   C_{\varphi b} \\
   C_{bW}\\
   C_{bB}\\
   C_{ed}\\
   C_{eq}\\
   C_{ld}\\
   C_{lq}^{+}\\
  \end{array} } \right)^T
  \left( {\begin{array}{c}
   0.31  \\
   0.31  \\
   0.05 \\
   \cdot \\
   \cdot \\
   0.091 \\
   -0.064 \\
   0.71 \\
   3.77 \\
  \end{array} } \right)
  + \left(\frac{1 \TeV}{\Lambda} \right)^4
  \left( {\begin{array}{c}
   C_{\varphi Q}^{3}  \\
   C_{\varphi Q}^{1} \\
   C_{\varphi b} \\
   C_{bW}\\
   C_{bB}\\
   C_{ed}\\
   C_{eq}\\
   C_{ld}\\
   C_{lq}^{+}\\
  \end{array} } \right)^T
  \left( {\begin{array}{ccccccccc}
   \cdot & \cdot  & \cdot & \cdot & \cdot& \cdot  & \cdot & \cdot & \cdot\\
   \cdot & \cdot  & \cdot & \cdot & \cdot& \cdot  & \cdot & \cdot & \cdot\\
   \cdot & \cdot  & \cdot & \cdot & \cdot& \cdot  & \cdot & \cdot & \cdot\\
   \cdot & \cdot & \cdot & 0.12 & -0.057& \cdot  & \cdot & \cdot & \cdot\\
   \cdot & \cdot & \cdot & \cdot & 0.010& \cdot  & \cdot & \cdot & \cdot\\
   \cdot & \cdot  & \cdot & \cdot & \cdot& \cdot  & \cdot & \cdot & \cdot\\
   \cdot & \cdot  & \cdot & \cdot & \cdot& \cdot  & \cdot & \cdot & \cdot\\
   \cdot & \cdot  & \cdot & \cdot & \cdot& \cdot  & \cdot & \cdot & \cdot\\
   \cdot & \cdot  & \cdot & \cdot & \cdot& \cdot  & \cdot & \cdot & \cdot\\
  \end{array} } \right)
  \left( {\begin{array}{c}
   C_{\varphi Q}^{3}  \\
   C_{\varphi Q}^{1} \\
   C_{\varphi b} \\
   C_{bW}\\
   C_{bB}\\
   C_{ed}\\
   C_{eq}\\
   C_{ld}\\
   C_{lq}^{+}\\
  \end{array} } \right)
\end{equation}

\begin{equation}
\sigma_{+-, 250}[pb] = 1.02 + \left(\frac{1 \TeV}{\Lambda} \right)^2 \left( {\begin{array}{c}
   C_{\varphi Q}^{3}  \\
   C_{\varphi Q}^{1} \\
   C_{\varphi b} \\
   C_{bW}\\
   C_{bB}\\
   C_{ed}\\
   C_{eq}\\
   C_{ld}\\
   C_{lq}^{+}\\
  \end{array} } \right)^T
  \left( {\begin{array}{c}
   0.094  \\
   0.094  \\
   -0.11 \\
   \cdot \\
   \cdot \\
   1.61 \\
   -1.08 \\
   0.04 \\
   0.23 \\
  \end{array} } \right)
  + \left(\frac{1 \TeV}{\Lambda} \right)^4
  \left( {\begin{array}{c}
   C_{\varphi Q}^{3}  \\
   C_{\varphi Q}^{1} \\
   C_{\varphi b} \\
   C_{bW}\\
   C_{bB}\\
   C_{ed}\\
   C_{eq}\\
   C_{ld}\\
   C_{lq}^{+}\\
  \end{array} } \right)^T
  \left( {\begin{array}{ccccccccc}
   \cdot & \cdot  & \cdot & \cdot & \cdot& \cdot  & \cdot & \cdot & \cdot\\
   \cdot & \cdot  & \cdot & \cdot & \cdot& \cdot  & \cdot & \cdot & \cdot\\
   \cdot & \cdot  & \cdot & \cdot & \cdot& \cdot  & \cdot & \cdot & \cdot\\
   \cdot & \cdot & \cdot & 0.008 & 0.005& \cdot  & \cdot & \cdot & \cdot\\
   \cdot & \cdot & \cdot & \cdot & 0.038& \cdot  & \cdot & \cdot & \cdot\\
   \cdot & \cdot  & \cdot & \cdot & \cdot& \cdot  & \cdot & \cdot & \cdot\\
   \cdot & \cdot  & \cdot & \cdot & \cdot& \cdot  & \cdot & \cdot & \cdot\\
   \cdot & \cdot  & \cdot & \cdot & \cdot& \cdot  & \cdot & \cdot & \cdot\\
   \cdot & \cdot  & \cdot & \cdot & \cdot& \cdot  & \cdot & \cdot & \cdot\\
  \end{array} } \right)
  \left( {\begin{array}{c}
   C_{\varphi Q}^{3}  \\
   C_{\varphi Q}^{1} \\
   C_{\varphi b} \\
   C_{bW}\\
   C_{bB}\\
   C_{ed}\\
   C_{eq}\\
   C_{ld}\\
   C_{lq}^{+}\\
  \end{array} } \right)
\end{equation}

\begin{equation}
\sigma_{-+, 500}[pb] = 0.72 + \left(\frac{1 \TeV}{\Lambda} \right)^2 \left( {\begin{array}{c}
   C_{\varphi Q}^{3}  \\
   C_{\varphi Q}^{1} \\
   C_{\varphi b} \\
   C_{bW}\\
   C_{bB}\\
   C_{ed}\\
   C_{eq}\\
   C_{ld}\\
   C_{lq}^{+}\\
  \end{array} } \right)^T
  \left( {\begin{array}{c}
   0.064  \\
   0.064  \\
   0.012 \\
   \cdot \\
   \cdot \\
   0.095 \\
   -0.05 \\
   0.76 \\
   3.5 \\
  \end{array} } \right)
  + \left(\frac{1 \TeV}{\Lambda} \right)^4
  \left( {\begin{array}{c}
   C_{\varphi Q}^{3}  \\
   C_{\varphi Q}^{1} \\
   C_{\varphi b} \\
   C_{bW}\\
   C_{bB}\\
   C_{ed}\\
   C_{eq}\\
   C_{ld}\\
   C_{lq}^{+}\\
  \end{array} } \right)^T
  \left( {\begin{array}{ccccccccc}
   \cdot & \cdot  & \cdot & \cdot & \cdot& \cdot  & \cdot & \cdot & \cdot\\
   \cdot & \cdot  & \cdot & \cdot & \cdot& \cdot  & \cdot & \cdot & \cdot\\
   \cdot & \cdot  & \cdot & \cdot & \cdot& \cdot  & \cdot & \cdot & \cdot\\
   \cdot & \cdot & \cdot & 0.10 & -0.058& \cdot  & \cdot & \cdot & \cdot\\
   \cdot & \cdot & \cdot & \cdot & 0.010& \cdot  & \cdot & \cdot & \cdot\\
   \cdot & \cdot  & \cdot & \cdot & \cdot& \cdot  & \cdot & \cdot & \cdot\\
   \cdot & \cdot  & \cdot & \cdot & \cdot& \cdot  & \cdot & \cdot & \cdot\\
   \cdot & \cdot  & \cdot & \cdot & \cdot& \cdot  & \cdot & \cdot & \cdot\\
   \cdot & \cdot  & \cdot & \cdot & \cdot& \cdot  & \cdot & \cdot & \cdot\\
  \end{array} } \right)
  \left( {\begin{array}{c}
   C_{\varphi Q}^{3}  \\
   C_{\varphi Q}^{1} \\
   C_{\varphi b} \\
   C_{bW}\\
   C_{bB}\\
   C_{ed}\\
   C_{eq}\\
   C_{ld}\\
   C_{lq}^{+}\\
  \end{array} } \right)
\end{equation}

\begin{equation}
\sigma_{+-, 500}[pb] = 0.22 + \left(\frac{1 \TeV}{\Lambda} \right)^2 \left( {\begin{array}{c}
   C_{\varphi Q}^{3}  \\
   C_{\varphi Q}^{1} \\
   C_{\varphi b} \\
   C_{bW}\\
   C_{bB}\\
   C_{ed}\\
   C_{eq}\\
   C_{ld}\\
   C_{lq}^{+}\\
  \end{array} } \right)^T
  \left( {\begin{array}{c}
   0.02  \\
   0.02  \\
   -0.024 \\
   \cdot \\
   \cdot \\
   1.56 \\
   -0.84 \\
   0.046 \\
   0.2 \\
  \end{array} } \right)
  + \left(\frac{1 \TeV}{\Lambda} \right)^4
  \left( {\begin{array}{c}
   C_{\varphi Q}^{3}  \\
   C_{\varphi Q}^{1} \\
   C_{\varphi b} \\
   C_{bW}\\
   C_{bB}\\
   C_{ed}\\
   C_{eq}\\
   C_{ld}\\
   C_{lq}^{+}\\
  \end{array} } \right)^T
  \left( {\begin{array}{ccccccccc}
   \cdot & \cdot  & \cdot & \cdot & \cdot& \cdot  & \cdot & \cdot & \cdot\\
   \cdot & \cdot  & \cdot & \cdot & \cdot& \cdot  & \cdot & \cdot & \cdot\\
   \cdot & \cdot  & \cdot & \cdot & \cdot& \cdot  & \cdot & \cdot & \cdot\\
   \cdot & \cdot & \cdot & 0.006 & -0.002& \cdot  & \cdot & \cdot & \cdot\\
   \cdot & \cdot & \cdot & \cdot & 0.036& \cdot  & \cdot & \cdot & \cdot\\
   \cdot & \cdot  & \cdot & \cdot & \cdot& \cdot  & \cdot & \cdot & \cdot\\
   \cdot & \cdot  & \cdot & \cdot & \cdot& \cdot  & \cdot & \cdot & \cdot\\
   \cdot & \cdot  & \cdot & \cdot & \cdot& \cdot  & \cdot & \cdot & \cdot\\
   \cdot & \cdot  & \cdot & \cdot & \cdot& \cdot  & \cdot & \cdot & \cdot\\
  \end{array} } \right)
  \left( {\begin{array}{c}
   C_{\varphi Q}^{3}  \\
   C_{\varphi Q}^{1} \\
   C_{\varphi b} \\
   C_{bW}\\
   C_{bB}\\
   C_{ed}\\
   C_{eq}\\
   C_{ld}\\
   C_{lq}^{+}\\
  \end{array} } \right)
\end{equation}

\begin{equation}
\sigma_{-+, 1000}[pb] = 0.174 + \left(\frac{1 \TeV}{\Lambda} \right)^2 \left( {\begin{array}{c}
   C_{\varphi Q}^{3}  \\
   C_{\varphi Q}^{1} \\
   C_{\varphi b} \\
   C_{bW}\\
   C_{bB}\\
   C_{ed}\\
   C_{eq}\\
   C_{ld}\\
   C_{lq}^{+}\\
  \end{array} } \right)^T
  \left( {\begin{array}{c}
   0.015  \\
   0.015  \\
   0.003 \\
   \cdot \\
   \cdot \\
   0.093 \\
   -0.048 \\
   0.77 \\
   3.44 \\
  \end{array} } \right)
  + \left(\frac{1 \TeV}{\Lambda} \right)^4
  \left( {\begin{array}{c}
   C_{\varphi Q}^{3}  \\
   C_{\varphi Q}^{1} \\
   C_{\varphi b} \\
   C_{bW}\\
   C_{bB}\\
   C_{ed}\\
   C_{eq}\\
   C_{ld}\\
   C_{lq}^{+}\\
  \end{array} } \right)^T
  \left( {\begin{array}{ccccccccc}
   \cdot & \cdot  & \cdot & \cdot & \cdot& \cdot  & \cdot & \cdot & \cdot\\
   \cdot & \cdot  & \cdot & \cdot & \cdot& \cdot  & \cdot & \cdot & \cdot\\
   \cdot & \cdot  & \cdot & \cdot & \cdot& \cdot  & \cdot & \cdot & \cdot\\
   \cdot & \cdot & \cdot & 0.098 & -0.058& \cdot  & \cdot & \cdot & \cdot\\
   \cdot & \cdot & \cdot & \cdot & 0.011& \cdot  & \cdot & \cdot & \cdot\\
   \cdot & \cdot  & \cdot & \cdot & \cdot& \cdot  & \cdot & \cdot & \cdot\\
   \cdot & \cdot  & \cdot & \cdot & \cdot& \cdot  & \cdot & \cdot & \cdot\\
   \cdot & \cdot  & \cdot & \cdot & \cdot& \cdot  & \cdot & \cdot & \cdot\\
   \cdot & \cdot  & \cdot & \cdot & \cdot& \cdot  & \cdot & \cdot & \cdot\\
  \end{array} } \right)
  \left( {\begin{array}{c}
   C_{\varphi Q}^{3}  \\
   C_{\varphi Q}^{1} \\
   C_{\varphi b} \\
   C_{bW}\\
   C_{bB}\\
   C_{ed}\\
   C_{eq}\\
   C_{ld}\\
   C_{lq}^{+}\\
  \end{array} } \right)
\end{equation}

\begin{equation}
\sigma_{+-, 1000}[pb] = 0.052 + \left(\frac{1 \TeV}{\Lambda} \right)^2 \left( {\begin{array}{c}
   C_{\varphi Q}^{3}  \\
   C_{\varphi Q}^{1} \\
   C_{\varphi b} \\
   C_{bW}\\
   C_{bB}\\
   C_{ed}\\
   C_{eq}\\
   C_{ld}\\
   C_{lq}^{+}\\
  \end{array} } \right)^T
  \left( {\begin{array}{c}
   0.004  \\
   0.004  \\
   -0.006 \\
   \cdot \\
   \cdot \\
   1.55 \\
   -0.79 \\
   0.046 \\
   0.21 \\
  \end{array} } \right)
  + \left(\frac{1 \TeV}{\Lambda} \right)^4
  \left( {\begin{array}{c}
   C_{\varphi Q}^{3}  \\
   C_{\varphi Q}^{1} \\
   C_{\varphi b} \\
   C_{bW}\\
   C_{bB}\\
   C_{ed}\\
   C_{eq}\\
   C_{ld}\\
   C_{lq}^{+}\\
  \end{array} } \right)^T
  \left( {\begin{array}{ccccccccc}
   \cdot & \cdot  & \cdot & \cdot & \cdot& \cdot  & \cdot & \cdot & \cdot\\
   \cdot & \cdot  & \cdot & \cdot & \cdot& \cdot  & \cdot & \cdot & \cdot\\
   \cdot & \cdot  & \cdot & \cdot & \cdot& \cdot  & \cdot & \cdot & \cdot\\
   \cdot & \cdot & \cdot & 0.006 & -0.003& \cdot  & \cdot & \cdot & \cdot\\
   \cdot & \cdot & \cdot & \cdot & 0.035& \cdot  & \cdot & \cdot & \cdot\\
   \cdot & \cdot  & \cdot & \cdot & \cdot& \cdot  & \cdot & \cdot & \cdot\\
   \cdot & \cdot  & \cdot & \cdot & \cdot& \cdot  & \cdot & \cdot & \cdot\\
   \cdot & \cdot  & \cdot & \cdot & \cdot& \cdot  & \cdot & \cdot & \cdot\\
   \cdot & \cdot  & \cdot & \cdot & \cdot& \cdot  & \cdot & \cdot & \cdot\\
  \end{array} } \right)
  \left( {\begin{array}{c}
   C_{\varphi Q}^{3}  \\
   C_{\varphi Q}^{1} \\
   C_{\varphi b} \\
   C_{bW}\\
   C_{bB}\\
   C_{ed}\\
   C_{eq}\\
   C_{ld}\\
   C_{lq}^{+}\\
  \end{array} } \right)
\end{equation}

\begin{equation}
A^{FB}_{-+, 250}[\%] = 69.6 + \left(\frac{1 \TeV}{\Lambda} \right)^2 \left( {\begin{array}{c}
   C_{\varphi Q}^{3}  \\
   C_{\varphi Q}^{1} \\
   C_{\varphi b} \\
   C_{bW}\\
   C_{bB}\\
   C_{ed}\\
   C_{eq}\\
   C_{ld}\\
   C_{lq}^{+}\\
  \end{array} } \right)^T
  \left( {\begin{array}{c}
   0.3  \\
   0.3  \\
   -2.2 \\
   \cdot \\
   \cdot \\
   1.8 \\
   3.8 \\
   -29.5 \\
   8.57 \\
  \end{array} } \right)
  + \left(\frac{1 \TeV}{\Lambda} \right)^4
  \left( {\begin{array}{c}
   C_{\varphi Q}^{3}  \\
   C_{\varphi Q}^{1} \\
   C_{\varphi b} \\
   C_{bW}\\
   C_{bB}\\
   C_{ed}\\
   C_{eq}\\
   C_{ld}\\
   C_{lq}^{+}\\
  \end{array} } \right)^T
  \left( {\begin{array}{ccccccccc}
   \cdot & \cdot  & \cdot & \cdot & \cdot& \cdot  & \cdot & \cdot & \cdot\\
   \cdot & \cdot  & \cdot & \cdot & \cdot& \cdot  & \cdot & \cdot & \cdot\\
   \cdot & \cdot  & \cdot & \cdot & \cdot& \cdot  & \cdot & \cdot & \cdot\\
   \cdot & \cdot & \cdot & -2.2 & 1.02& \cdot  & \cdot & \cdot & \cdot\\
   \cdot & \cdot & \cdot & \cdot & -0.16& \cdot  & \cdot & \cdot & \cdot\\
   \cdot & \cdot  & \cdot & \cdot & \cdot& \cdot  & \cdot & \cdot & \cdot\\
   \cdot & \cdot  & \cdot & \cdot & \cdot& \cdot  & \cdot & \cdot & \cdot\\
   \cdot & \cdot  & \cdot & \cdot & \cdot& \cdot  & \cdot & \cdot & \cdot\\
   \cdot & \cdot  & \cdot & \cdot & \cdot& \cdot  & \cdot & \cdot & \cdot\\
  \end{array} } \right)
  \left( {\begin{array}{c}
   C_{\varphi Q}^{3}  \\
   C_{\varphi Q}^{1} \\
   C_{\varphi b} \\
   C_{bW}\\
   C_{bB}\\
   C_{ed}\\
   C_{eq}\\
   C_{ld}\\
   C_{lq}^{+}\\
  \end{array} } \right)
\end{equation}

\begin{equation}
A^{FB}_{+-, 250}[\%] = 35.9 + \left(\frac{1 \TeV}{\Lambda} \right)^2 \left( {\begin{array}{c}
   C_{\varphi Q}^{3}  \\
   C_{\varphi Q}^{1} \\
   C_{\varphi b} \\
   C_{bW}\\
   C_{bB}\\
   C_{ed}\\
   C_{eq}\\
   C_{ld}\\
   C_{lq}^{+}\\
  \end{array} } \right)^T
  \left( {\begin{array}{c}
   -7.7  \\
   -7.7  \\
   -4.5 \\
   \cdot \\
   \cdot \\
   62 \\
   119 \\
   -4.6 \\
   7.9 \\
  \end{array} } \right)
  + \left(\frac{1 \TeV}{\Lambda} \right)^4
  \left( {\begin{array}{c}
   C_{\varphi Q}^{3}  \\
   C_{\varphi Q}^{1} \\
   C_{\varphi b} \\
   C_{bW}\\
   C_{bB}\\
   C_{ed}\\
   C_{eq}\\
   C_{ld}\\
   C_{lq}^{+}\\
  \end{array} } \right)^T
  \left( {\begin{array}{ccccccccc}
   \cdot & \cdot  & \cdot & \cdot & \cdot& \cdot  & \cdot & \cdot & \cdot\\
   \cdot & \cdot  & \cdot & \cdot & \cdot& \cdot  & \cdot & \cdot & \cdot\\
   \cdot & \cdot  & \cdot & \cdot & \cdot& \cdot  & \cdot & \cdot & \cdot\\
   \cdot & \cdot & \cdot & -0.10 & -0.14& \cdot  & \cdot & \cdot & \cdot\\
   \cdot & \cdot & \cdot & \cdot & -0.95& \cdot  & \cdot & \cdot & \cdot\\
   \cdot & \cdot  & \cdot & \cdot & \cdot& \cdot  & \cdot & \cdot & \cdot\\
   \cdot & \cdot  & \cdot & \cdot & \cdot& \cdot  & \cdot & \cdot & \cdot\\
   \cdot & \cdot  & \cdot & \cdot & \cdot& \cdot  & \cdot & \cdot & \cdot\\
   \cdot & \cdot  & \cdot & \cdot & \cdot& \cdot  & \cdot & \cdot & \cdot\\
  \end{array} } \right)
  \left( {\begin{array}{c}
   C_{\varphi Q}^{3}  \\
   C_{\varphi Q}^{1} \\
   C_{\varphi b} \\
   C_{bW}\\
   C_{bB}\\
   C_{ed}\\
   C_{eq}\\
   C_{ld}\\
   C_{lq}^{+}\\
  \end{array} } \right)
\end{equation}

\begin{equation}
A^{FB}_{-+, 500}[\%] = 67.7 + \left(\frac{1 \TeV}{\Lambda} \right)^2 \left( {\begin{array}{c}
   C_{\varphi Q}^{3}  \\
   C_{\varphi Q}^{1} \\
   C_{\varphi b} \\
   C_{bW}\\
   C_{bB}\\
   C_{ed}\\
   C_{eq}\\
   C_{ld}\\
   C_{lq}^{+}\\
  \end{array} } \right)^T
  \left( {\begin{array}{c}
   0.2  \\
   0.2  \\
   -2.3 \\
   \cdot \\
   \cdot \\
   1.2 \\
   8 \\
   -139 \\
   40.3\\
  \end{array} } \right)
  + \left(\frac{1 \TeV}{\Lambda} \right)^4
  \left( {\begin{array}{c}
   C_{\varphi Q}^{3}  \\
   C_{\varphi Q}^{1} \\
   C_{\varphi b} \\
   C_{bW}\\
   C_{bB}\\
   C_{ed}\\
   C_{eq}\\
   C_{ld}\\
   C_{lq}^{+}\\
  \end{array} } \right)^T
  \left( {\begin{array}{ccccccccc}
   \cdot & \cdot  & \cdot & \cdot & \cdot& \cdot  & \cdot & \cdot & \cdot\\
   \cdot & \cdot  & \cdot & \cdot & \cdot& \cdot  & \cdot & \cdot & \cdot\\
   \cdot & \cdot  & \cdot & \cdot & \cdot& \cdot  & \cdot & \cdot & \cdot\\
   \cdot & \cdot & \cdot & -5.8 & 3.09& \cdot  & \cdot & \cdot & \cdot\\
   \cdot & \cdot & \cdot & \cdot & -0.47& \cdot  & \cdot & \cdot & \cdot\\
   \cdot & \cdot  & \cdot & \cdot & \cdot& \cdot  & \cdot & \cdot & \cdot\\
   \cdot & \cdot  & \cdot & \cdot & \cdot& \cdot  & \cdot & \cdot & \cdot\\
   \cdot & \cdot  & \cdot & \cdot & \cdot& \cdot  & \cdot & \cdot & \cdot\\
   \cdot & \cdot  & \cdot & \cdot & \cdot& \cdot  & \cdot & \cdot & \cdot\\
  \end{array} } \right)
  \left( {\begin{array}{c}
   C_{\varphi Q}^{3}  \\
   C_{\varphi Q}^{1} \\
   C_{\varphi b} \\
   C_{bW}\\
   C_{bB}\\
   C_{ed}\\
   C_{eq}\\
   C_{ld}\\
   C_{lq}^{+}\\
  \end{array} } \right)
\end{equation}

\begin{equation}
A^{FB}_{+-, 500}[\%] = 46.7 + \left(\frac{1 \TeV}{\Lambda} \right)^2 \left( {\begin{array}{c}
   C_{\varphi Q}^{3}  \\
   C_{\varphi Q}^{1} \\
   C_{\varphi b} \\
   C_{bW}\\
   C_{bB}\\
   C_{ed}\\
   C_{eq}\\
   C_{ld}\\
   C_{lq}^{+}\\
  \end{array} } \right)^T
  \left( {\begin{array}{c}
   -7  \\
   -7  \\
   -3.5 \\
   \cdot \\
   \cdot \\
   219 \\
   380 \\
   -26.6 \\
   28.4 \\
  \end{array} } \right)
  + \left(\frac{1 \TeV}{\Lambda} \right)^4
  \left( {\begin{array}{c}
   C_{\varphi Q}^{3}  \\
   C_{\varphi Q}^{1} \\
   C_{\varphi b} \\
   C_{bW}\\
   C_{bB}\\
   C_{ed}\\
   C_{eq}\\
   C_{ld}\\
   C_{lq}^{+}\\
  \end{array} } \right)^T
  \left( {\begin{array}{ccccccccc}
   \cdot & \cdot  & \cdot & \cdot & \cdot& \cdot  & \cdot & \cdot & \cdot\\
   \cdot & \cdot  & \cdot & \cdot & \cdot& \cdot  & \cdot & \cdot & \cdot\\
   \cdot & \cdot  & \cdot & \cdot & \cdot& \cdot  & \cdot & \cdot & \cdot\\
   \cdot & \cdot & \cdot & -0.67 & 0.09& \cdot  & \cdot & \cdot & \cdot\\
   \cdot & \cdot & \cdot & \cdot & -2.8& \cdot  & \cdot & \cdot & \cdot\\
   \cdot & \cdot  & \cdot & \cdot & \cdot& \cdot  & \cdot & \cdot & \cdot\\
   \cdot & \cdot  & \cdot & \cdot & \cdot& \cdot  & \cdot & \cdot & \cdot\\
   \cdot & \cdot  & \cdot & \cdot & \cdot& \cdot  & \cdot & \cdot & \cdot\\
   \cdot & \cdot  & \cdot & \cdot & \cdot& \cdot  & \cdot & \cdot & \cdot\\
  \end{array} } \right)
  \left( {\begin{array}{c}
   C_{\varphi Q}^{3}  \\
   C_{\varphi Q}^{1} \\
   C_{\varphi b} \\
   C_{bW}\\
   C_{bB}\\
   C_{ed}\\
   C_{eq}\\
   C_{ld}\\
   C_{lq}^{+}\\
  \end{array} } \right)
\end{equation}

\begin{equation}
A^{FB}_{-+, 1000}[\%] = 57 + \left(\frac{1 \TeV}{\Lambda} \right)^2 \left( {\begin{array}{c}
   C_{\varphi Q}^{3}  \\
   C_{\varphi Q}^{1} \\
   C_{\varphi b} \\
   C_{bW}\\
   C_{bB}\\
   C_{ed}\\
   C_{eq}\\
   C_{ld}\\
   C_{lq}^{+}\\
  \end{array} } \right)^T
  \left( {\begin{array}{c}
   0.48  \\
   0.48  \\
   -2.7 \\
   \cdot \\
   \cdot \\
   3.4 \\
   37.4 \\
   -593 \\
   145\\
  \end{array} } \right)
  + \left(\frac{1 \TeV}{\Lambda} \right)^4
  \left( {\begin{array}{c}
   C_{\varphi Q}^{3}  \\
   C_{\varphi Q}^{1} \\
   C_{\varphi b} \\
   C_{bW}\\
   C_{bB}\\
   C_{ed}\\
   C_{eq}\\
   C_{ld}\\
   C_{lq}^{+}\\
  \end{array} } \right)^T
  \left( {\begin{array}{ccccccccc}
   \cdot & \cdot  & \cdot & \cdot & \cdot& \cdot  & \cdot & \cdot & \cdot\\
   \cdot & \cdot  & \cdot & \cdot & \cdot& \cdot  & \cdot & \cdot & \cdot\\
   \cdot & \cdot  & \cdot & \cdot & \cdot& \cdot  & \cdot & \cdot & \cdot\\
   \cdot & \cdot & \cdot & -10 & 4.83& \cdot  & \cdot & \cdot & \cdot\\
   \cdot & \cdot & \cdot & \cdot & -0.7& \cdot  & \cdot & \cdot & \cdot\\
   \cdot & \cdot  & \cdot & \cdot & \cdot& \cdot  & \cdot & \cdot & \cdot\\
   \cdot & \cdot  & \cdot & \cdot & \cdot& \cdot  & \cdot & \cdot & \cdot\\
   \cdot & \cdot  & \cdot & \cdot & \cdot& \cdot  & \cdot & \cdot & \cdot\\
   \cdot & \cdot  & \cdot & \cdot & \cdot& \cdot  & \cdot & \cdot & \cdot\\
  \end{array} } \right)
  \left( {\begin{array}{c}
   C_{\varphi Q}^{3}  \\
   C_{\varphi Q}^{1} \\
   C_{\varphi b} \\
   C_{bW}\\
   C_{bB}\\
   C_{ed}\\
   C_{eq}\\
   C_{ld}\\
   C_{lq}^{+}\\
  \end{array} } \right)
\end{equation}

\begin{equation}
A^{FB}_{+-, 1000}[\%] = 34.6 + \left(\frac{1 \TeV}{\Lambda} \right)^2 \left( {\begin{array}{c}
   C_{\varphi Q}^{3}  \\
   C_{\varphi Q}^{1} \\
   C_{\varphi b} \\
   C_{bW}\\
   C_{bB}\\
   C_{ed}\\
   C_{eq}\\
   C_{ld}\\
   C_{lq}^{+}\\
  \end{array} } \right)^T
  \left( {\begin{array}{c}
   -6.8  \\
   -6.8  \\
   -3.6 \\
   \cdot \\
   \cdot \\
   826 \\
   1611 \\
   -120 \\
   6.5 \\
  \end{array} } \right)
  + \left(\frac{1 \TeV}{\Lambda} \right)^4
  \left( {\begin{array}{c}
   C_{\varphi Q}^{3}  \\
   C_{\varphi Q}^{1} \\
   C_{\varphi b} \\
   C_{bW}\\
   C_{bB}\\
   C_{ed}\\
   C_{eq}\\
   C_{ld}\\
   C_{lq}^{+}\\
  \end{array} } \right)^T
  \left( {\begin{array}{ccccccccc}
   \cdot & \cdot  & \cdot & \cdot & \cdot& \cdot  & \cdot & \cdot & \cdot\\
   \cdot & \cdot  & \cdot & \cdot & \cdot& \cdot  & \cdot & \cdot & \cdot\\
   \cdot & \cdot  & \cdot & \cdot & \cdot& \cdot  & \cdot & \cdot & \cdot\\
   \cdot & \cdot & \cdot & -1.34 & 0.1& \cdot  & \cdot & \cdot & \cdot\\
   \cdot & \cdot & \cdot & \cdot & -3.35& \cdot  & \cdot & \cdot & \cdot\\
   \cdot & \cdot  & \cdot & \cdot & \cdot& \cdot  & \cdot & \cdot & \cdot\\
   \cdot & \cdot  & \cdot & \cdot & \cdot& \cdot  & \cdot & \cdot & \cdot\\
   \cdot & \cdot  & \cdot & \cdot & \cdot& \cdot  & \cdot & \cdot & \cdot\\
   \cdot & \cdot  & \cdot & \cdot & \cdot& \cdot  & \cdot & \cdot & \cdot\\
  \end{array} } \right)
  \left( {\begin{array}{c}
   C_{\varphi Q}^{3}  \\
   C_{\varphi Q}^{1} \\
   C_{\varphi b} \\
   C_{bW}\\
   C_{bB}\\
   C_{ed}\\
   C_{eq}\\
   C_{ld}\\
   C_{lq}^{+}\\
  \end{array} } \right)
\end{equation}


\subsection{\texorpdfstring{$e^- e^+ \rightarrow t \bar{t} H $}{e-e+>ttbarH}}

\begin{table}[h!]
\begin{tabular}{lc|c|ccccccccc}
  Pol.  &  $\sqrt{s}$ [GeV]  &  SM   & $C_{\varphi t} $ & $C_{\varphi Q}^{-} $  & $ C_{tW}$ & $ C_{tB}$ & $ C_{t \varphi}$ & $ C_{lq}^{-}$ & $ C_{lu}$ & $ C_{eq}$ & $ C_{eu}$\\
-+  & 500  & 0.49  & -0.026 & -0.028 & 0.44 & 0.13 & -0.063 &-1.5&-1.4&-0.06&-0.04\\
 +-  & 500 & 0.23 & -0.015 & 0.014 & 0.016 & 0.18 & -0.03&-0.09&-0.09&-0.94&-0.98\\
 -+  & 550 & 1.9 & -0.1 & -0.1  & 1.8 & 0.56& -0.23&-7.12&-6.56&-0.24&-0.3\\
 +-  & 550 & 0.91 & 0.059 & 0.049  & 0.073 & 0.73& -0.11&-0.39&-0.42&-4.2&-4.7\\
  -+  & 1000 & 3.37 & -0.11 & -0.21  & 4.37 & 1.39& -0.39&-43&-29.5&-1.05&-1.7\\
 +-  & 1000 & 1.75 & 0.12 & 0.049  & 0.22 & 1.8& -0.2&-2.3&-1.77&-18.1&-29.9\\
\end{tabular}
\caption{Linear dependence of the cross-section [pb] for the process $e^-e^+ \rightarrow t\bar{t}H$. $P\left(e^-,e^+\right) =(- 0.8, + 0.3)$ is noted as -+, and $P\left(e^-,e^+\right) =(+ 0.8, - 0.3)$ is noted as +-.}
\end{table}

\end{landscape}

\section{Conversion to LHC TOP WG EFT conventions}
\label{sec:lhc_top_wg}%
The conversion between our conventions for top-quark operator coefficients and the LHC TOP WG standards of Ref.~\cite{AguilarSaavedra:2018nen} is the following:
\begin{equation}
\begin{pmatrix}
c_{\varphi t}  \\
c_{\varphi Q}^3\\
c_{\varphi Q}^-\\
c_{tW}  \\
c_{tZ}   \\
c_{t \varphi}  \\
c_{\varphi tb}  \\
c_{Qe}^{(1)}   \\
c_{tl}^{(1)}   \\
c_{te}^{(1)}   \\
c_{Ql}^{-(1)}  
\end{pmatrix}
=
\left(\begin{array}{*{11}{c}}
y_t^2	& 0	& 0	& 0	& 0	& 0	& 0	& 0 &0 &0&0	\\
0 & y_t^2	& 0	& 0 & 0	& 0	& 0	& 0 &0 &0&0	\\
0	& -y_t^2	& y_t^2	& 0	& 0	& 0	& 0	& 0 &0 &0&0	\\
0	& 0	& 0	& y_tg_W	& 0	& 0	& 0	& 0 &0 &0&0	\\
0	& 0	& 0	& y_tg_Wc_W	& -y_tg_Ys_W	& 0	& 0	& 0 &0 &0&0	\\
0	& 0	& 0	& 0	& 0	& 1	& 0	& 0 &0 &0&0	\\
0	& 0	& 0	& 0	& 0	& 0	& -y_t^2& 0 &0 &0&0	\\
0	& 0	& 0	& 0	& 0	& 0	& 0	& 1 &0 &0&0 \\
0	& 0	& 0	& 0	& 0	& 0	& 0	& 0 &1 &0&0	\\
0	& 0	& 0	& 0	& 0	& 0	& 0	& 0 &0 &1&0	\\
0	& 0	& 0	& 0	& 0	& 0	& 0	& 0 &0 &0&1 
\end{array}\right)
\begin{pmatrix}
C_{\varphi t} \\
C_{\varphi Q}^{3}  \\
C_{\varphi Q}^{1} \\
C_{tW}\\
C_{tB}\\
C_{t \varphi} \\
C_{\varphi tb} \\
C_{eu}\\
C_{eq}\\
C_{lu}\\
C_{lq}^{-}
\end{pmatrix}
.
\label{eq:WGconvTOP}
\end{equation}
No standard have been established for the bottom-quark operator coefficients $C_{\varphi b}$, $C_{bW}$, $C_{bB}$, $C_{ed}$, $C_{eq}$, $C_{ld}$, $C_{lq}^{+}$.
A conversion of our results to these standards is provided, for convenience, in the form of a Mathematica notebook at \url{https://arxiv.org/src/1907.10619/anc/cov_matrices.nb}.


\section{Covariance matrices}

\subsection{Present constraints}
\label{sec:cov_matrix}

\autoref{fig:corr_matrix_app} shows the projection of the fit posterior on planes formed by each pair of operator coefficient (top-quark Yukawa excluded) for \autoref{sec:results} analysis of LEP/SLC and LHC Run 2 data.
The associated mean values,uncertainties and correlation matrices, for $(C_{\varphi t}, C_{\varphi Q}^3, C_{\varphi Q}^1, C_{tW}, C_{tB}, C_{t\varphi}, C_{\varphi b}, C_{bW}, C_{bB}, C_{\varphi tb})/\Lambda^2$ in units of $\tev^{-2}$, are the following:
\begin{scriptsize}\begin{equation}
\begin{gathered}
\text{mean values} =
\left(\begin{array}{*{10}{c}}
1.8	& 0.018	& 0.022	& -0.1	& 5	& -2.5 & -0.11	& 0.75	& -0.6	& -1.89	
\end{array}\right)
\\\text{uncertainties} =
\left(\begin{array}{*{10}{c}}
3.2	& 0.76 & 1.35	& 0.48	& 3.6	& 3.6	& 3.15	& 1.52	& 14.4	& 3.4
\end{array}\right)
\\\text{corr}=
\left(\begin{array}{*{10}{@{}c@{\,}}}
1	&  -0.052	& 0.13	& -0.026	& -0.01	& 0.011	& 0.12	& -0.0089	& 0.064	& 0.0043	\\
-0.052	& 1	& -0.16	& 0.025	& 0.073	& 0.0029	& 0.28	& 0.051	& 0.14	& -0.00056	\\
0.13	& -0.16	& 1	& 0.11	& 0.046	& 0.0022	& 0.86	& 0.33	& 0.44	& 0.0074	\\
-0.026	& 0.025	& 0.11	& 1	& 0.027	& 0.015	& 0.13	& 0.016	& 0.064 & 0.019	\\
-0.01	& 0.073	& 0.046	& 0.027	& 1	& 0.0075	& 0.009	& 0.0042	& 0.0047	& 0.00099	\\
0.011	&0.0029	& 0.0022	& 0.015	& 0.0075	& 1	& 0.0054	& 0.0048	& -0.0019	&  -0.019	\\
0.12	& 0.28	& 0.86	& 0.13	& 0.009	& 0.0054	& 1	& 0.27	& 0.59	& -0.0097	\\
-0.0089	& 0.051	& 0.33	& 0.016	& 0.0042	& 0.0048	& 0.27	& 1	& 0.13	&  0.012	\\
0.064	& 0.14	&0.44	& 0.064	& 0.0047	&-0.0019	& 0.59	& 0.13	& 1	& -0.014	\\
0.0043	& -0.00056	& 0.0074	& 0.019	& 0.00099	& -0.019	& -0.0097	& 0.012	& -0.014	& 1\end{array}\right)
\end{gathered}
\label{eq:cov_matrix_0}
\end{equation}\end{scriptsize}%
Note that non-Gaussianities imply that the above information only permits an approximate reconstruction of the posterior probability.
A \texttt{Mathematica}  notebook with these numerical values and a conversion to LHC TOP WG standards is provided at \url{https://arxiv.org/src/1907.10619/anc/cov_matrices.nb}.

\begin{figure}[!ht]
    \includegraphics[width=1\textwidth]{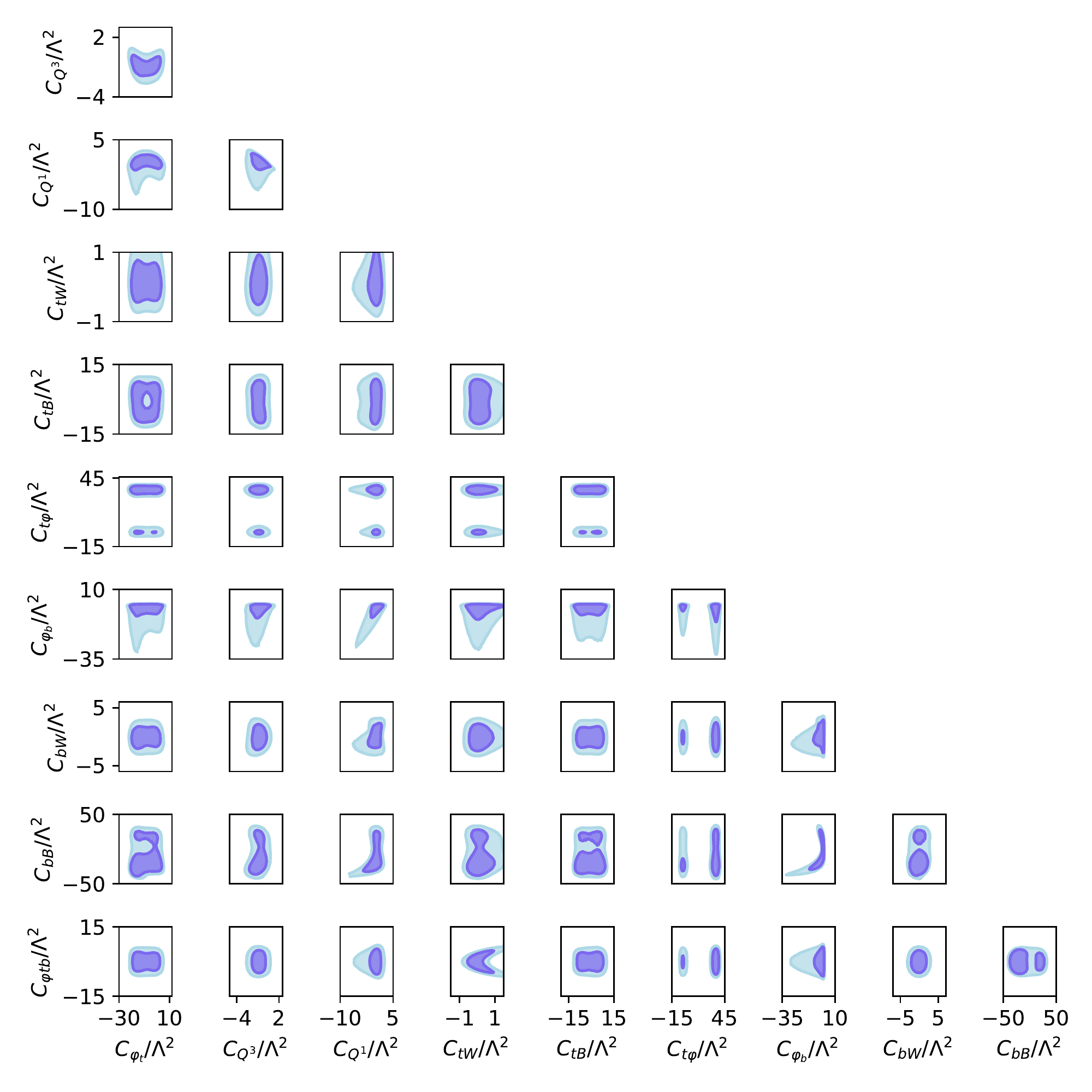}
    \caption{ Allowed regions after the fit to the LHC and LEP/SLC measurements in~\autoref{tab:measurements}.
The 68\% and 95\% probability regions are shown for each pair of the ten effective operator coefficients that affect the electro-weak interactions of the top and bottom quarks.
The top-quark Yukawa operator is marginalized over.
All Wilson coefficients are in units of $\tev^{-2}$.}
    \label{fig:corr_matrix_app}
\end{figure}
\twocolumn

\begin{landscape}

\subsection{Future prospects}
\label{sec:cov_matrices}

We provide here the uncertainties and correlation matrices of \autoref{sec:prospects} study of future prospects (summarized in~\autoref{fig:manhattan_plot}) on  $(C_{\varphi t}, C_{\varphi Q}^3, C_{\varphi Q}^1, C_{tW}, C_{tB}, C_{t\varphi}, C_{\varphi b}, C_{bW}, C_{bB}, C_{\varphi tb})/\Lambda^2$ operator coefficients in units of $\tev^{-2}$.
This information, together with a conversion to LHC TOP WG standards, is also provided for convenience in a Mathemtica notebook at \url{https://arxiv.org/src/1907.10619/anc/cov_matrices.nb}.

\subsubsection{LEP/SLC + LHC Run2}
\begin{scriptsize}\begin{equation}
\begin{gathered}
\text{uncertainties} =
\left(\begin{array}{*{10}{c}}
7.2	& 0.72	& 0.96	& 0.66	& 5.2	& 2.4	& 1.3	& 2.17	& 14	& 4.2
\end{array}\right)
\\\text{corr}=
\left(\begin{array}{*{10}{@{}c@{\,}}}
1 &-0.15 &0.15 &-0.024 &-0.0037 &-0.0055 &0.077 &0.0047 &0.034 \\
-0.15 & 1 &-0.35 &0.28 &-0.01 &0.0015 &0.25 & 0.062 &0.069  \\
0.15 &-0.35 &1 &-0.028 &-0.012 &0.012 & 0.74 &0.28 & 0.3  \\
-0.024 &0.28 &-0.028 &1 &-0.036 & 0.0078 &0.08 &0.056 &-0.0088 \\
-0.0037 &-0.01 &-0.012 &0.036 &1 &0.015 &-0.024 & -0.011 & -0.013 \\
-0.0055 &0.0015 &0.012 & 0.0078 &0.015 &  1 &0.015 &  7.9\times10^{-5} & 0.015 \\
0.077 & 0.25 & 0.74 & 0.08 & -0.024 &0.015 & 1 & 0.31 & 0.53 \\
0.0047 & 0.062 & 0.28 &0.056 & -0.011 & 7.9\times10^{-5} &0.31  & 1 & 0.2 \\
0.034 &0.069 &0.3 &-0.008 &-0.013 &0.015 &0.53 &0.2  &1
\end{array}\right)
\end{gathered}
\label{eq:cov_matrix_1}
\end{equation}\end{scriptsize}%
These values differ slightly from those of \autoref{eq:cov_matrix_0} since all measurements have here been assumed to reproduce the SM prediction.

\subsubsection{LEP/SLC + LHC S1}
\begin{scriptsize}\begin{equation}
\begin{gathered}
\text{uncertainties} =
\left(\begin{array}{*{10}{c}}
5.04	& 0.63	&  0.77	& 0.375	& 5.25	& 2.1	& 1.08	& 1.5	& 14	& 2.7
\end{array}\right)
\\\text{corr}=
\left(\begin{array}{*{10}{@{}c@{\,}}}
1 &  -0.2 2&  0.2 2&  -0.027&  -0.014&  -0.0014&  0.076&  0.0094&  0.029&  -0.0035 \\ 
 -0.2 2&  1&  -0.6&  0.0047&  -0.069&  0.004&  0.26&  0.04 2&  0.1&  0.0029 \\ 
 0.2 2&  -0.6&  1&  0.066&  0.06&  -0.0038 &  0.55&  0.2 2&  0.21&  -0.0069 \\ 
 -0.027&  0.0047&  0.066&  1&  -0.021&  0.0036&  0.093&  0.033&  0.033&  0.00086 \\ 
 -0.014&  -0.069&  0.06&  -0.021&  1&  0.0079&  -0.0011&  0.005 2&  -0.0053&  -0.0013 \\ 
 -0.0014&  0.004&  -0.0038 &  0.0036&  0.0079&  1&  -0.001 2&  0.0011&  -0.0061&  -0.0032 \\ 
 0.076&  0.26&  0.55&  0.093&  -0.0011&  -0.001 2&  1&  0.27&  0.5&  -0.0046 \\ 
 0.0094&  0.04 2&  0.2 2&  0.033&  0.005 2&  0.0011&  0.27&  1&  0.1 2&  -0.0086 \\ 
 0.029&  0.1&  0.21&  0.033&  -0.0053&  -0.0061&  0.5&  0.1 2&  1&  -0.0044 \\ 
 -0.0035&  0.0029&  -0.0069&  0.00086&  -0.0013&  -0.003 2&  -0.0046&  -0.0086&  -0.0044&  1 \\ 
\end{array}\right)
\end{gathered}
\label{eq:cov_matrix_2}
\end{equation}\end{scriptsize}

\end{landscape}
\begin{landscape}

\subsubsection{LEP/SLC + LHC S2}
\begin{scriptsize}\begin{equation}
\begin{gathered}
\text{uncertainties} =
\left(\begin{array}{*{10}{c}}
 1.28&  0.12	& 0.16	&  0.064	& 3.96	&  0.57	&  0.45	&  0.5	&  5.8	&  1.4
\end{array}\right)
\\\text{corr}=
\left(\begin{array}{*{10}{@{}c@{\,}}}
  1 &  0.18 &  -0.2 &  -0.022 &  0.28 &  0.0059 &  -0.047 &  -0.02 &  0.018 \\ 
 0.18 &   1 &  -0.6 1 &  0.012 &  -0.028 &  0.00066 &  0.16 &  -0.00026 &  0.048 \\ 
 -0.2 &  -0.6 1 &   1 &  0.0029 &  0.029 &  0.00095 &  0.6 &  0.068 &  0.05 \\ 
 -0.022 &  0.012 &  0.0029 &   1 &  -0.027 &  0.0027 &  0.0097 &  0.097 &  0.0002 \\ 
 0.28 &  -0.028 &  0.029 &  -0.027 &   1 &  0.019 &  0.017 &  -0.003 1 &  0.0096 \\ 
 0.0059 &  0.00066 &  0.00095 &  0.0027 &  0.019 &   1 &  0.0017 &  0.0028 &  0.0004 1 \\ 
 -0.047 &  0.16 &  0.6 &  0.0097 &  0.017 &  0.0017 &  1  &  0.054 &  0.13 \\ 
 -0.02 &  -0.00026 &  0.068 &  0.097 &  -0.003 1 &  0.0028 &  0.054 &   1 &  0.0047 \\ 
 0.018 &  0.048 &  0.05 &  0.0002 &  0.0096 &  0.0004 1 &  0.13 &  0.0047 &   1 \\  \end{array}\right)
\end{gathered}
\label{eq:cov_matrix_3}
\end{equation}\end{scriptsize}

\subsubsection{LEP/SLC + LHC S2 + ILC250}
\begin{scriptsize}\begin{equation}
\begin{gathered}
\text{uncertainties} =
\left(\begin{array}{*{10}{c}}
1.26	& 0.108	& 0.108	& 0.106	&3.75	& 0.54	& 0.0156	& 0.17	& 0.25	& 1.62
\end{array}\right)
\\\text{corr}=
\left(\begin{array}{*{10}{@{}c@{\,}}}
 1 & -0.17 &  0.17 & -0.34 &  0.23 &  0.0069 & -0.0032 &  0.0045 &  0.012 &  0.028 \\ 
-0.17 &  1 & -0.99 & -0.23 & -0.052 &  0.0037 & -0.015 & -0.00067 & -0.0089 & -0.0015 \\ 
 0.17 & -0.99 &  1 &  0.23 &  0.052 & -0.0038 & -0.034 &  0.0019 &  0.012 &  0.0021 \\ 
-0.34 & -0.23 &  0.23 &  1 & -0.11 &  0.00055 &  0.013 &  0.0095 & -0.0012 & -0.018 \\ 
 0.23 & -0.052 &  0.052 & -0.11 &  1 &  0.022 &  0.00018 & -0.0038 & -0.0015 &  0.0096 \\ 
 0.0069 &  0.0037 & -0.0038 &  0.00055 &  0.022 &  1 &  0.0024 & -0.00074 &  5.1 \times 10^{-5} & -0.00054 \\ 
-0.0032 & -0.015 & -0.034 &  0.013 &  0.00018 &  0.0024 &  1 & -0.017 & -0.041 & -0.01 \\ 
 0.0045 & -0.00067 &  0.0019 &  0.0095 & -0.0038 & -0.00074 & -0.017 &  1 &  0.29 & -0.0035 \\ 
 0.012 & -0.0089 &  0.012 & -0.0012 & -0.0015 &  5.1 \times 10^{-5} & -0.041 &  0.29 &  1 &  0.00034 \\ 
 0.028 & -0.0015 &  0.0021 & -0.018 &  0.0096 & -0.00054 & -0.01 & -0.0035 &  0.00034 &  1

\end{array}\right)
\end{gathered}
\label{eq:cov_matrix_4}
\end{equation}\end{scriptsize}

\end{landscape}
\begin{landscape}

\subsubsection{LEP/SLC + LHC S2 + ILC250 + ILC500}
\begin{scriptsize}\begin{equation}
\begin{gathered}
\text{uncertainties} =
\left(\begin{array}{*{10}{c}}
0.01	& 0.0057	& 0.0057	& 0.021	& 0.021	& 0.55	& 0.0070	& 0.092	& 0.13	& 0.95
\end{array}\right)
\\\text{corr}=
\left(\begin{array}{*{10}{@{}c@{\,}}}
  1  &  -0.75 &   0.75 &   0.94 &  -0.9 &   0.00015 &  -0.00067 &  -0.0019 &  -0.00062 &  -0.0013 \\ 
 -0.75 &   1  &  -0.73 &  -0.87 &   0.86 &  -0.0018 &   0.061  &   0.0028 &   0.0015 &   0.00096 \\ 
  0.75 &  -0.73 &   1  &   0.87 &  -0.86 &   0.0014 &   0.06 &   0.00063 &  -0.00025 &  -0.0009 \\ 
  0.94 &  -0.87 &   0.87 &   1  &  -0.97 &   0.00094 &  -0.00059 &  -0.0016 &  -0.00098 &  -0.0009 \\ 
 -0.9 &   0.86 &  -0.86 &  -0.97 &   1  &  -0.00083 &   0.00044 &   0.0018 &   0.0008 &   0.00052 \\ 
  0.00015 &  -0.0018 &   0.0014 &   0.00094 &  -0.00083 &   1  &  -0.0012 &   2.8*10^{-5} &  -0.0024 &   0.0025 \\ 
 -0.00067 &   0.061  &   0.06 &  -0.00059 &   0.00044 &  -0.0012 &   1  &  -0.0011  &  -0.0053 &   0.0022 \\ 
 -0.0019 &   0.0028 &   0.00063 &  -0.0016 &   0.0018 &   2.8*10^{-5} &  -0.0011  &   1  &   0.38 &  -0.00098 \\ 
 -0.00062 &   0.0015 &  -0.00025 &  -0.00098 &   0.0008 &  -0.0024 &  -0.0053 &   0.38 &   1  &   0.00087 \\ 
 -0.0013 &   0.00096 &  -0.0009 &  -0.0009 &   0.00052 &   0.0025 &   0.0022 &  -0.00098 &   0.00087 &   1

\end{array}\right)
\end{gathered}
\label{eq:cov_matrix_5}
\end{equation}\end{scriptsize}

\subsubsection{LEP/SLC + LHC S2 + ILC250 + ILC500 + ILC1000}
The study of \autoref{subsec:fourfermion} leads to the following uncertainties and correlation matrix on $(C_{\varphi t}, C_{\varphi Q}^3, C_{\varphi Q}^1, C_{tW}, C_{tB}, C_{t\varphi}, C_{\varphi b}, C_{bW}, C_{bB}$, $C_{\varphi tb}, C_{eu}, C_{ed}, C_{eq}, C_{lu},  C_{ld},     C_{lq}^-,     C_{lq}^+)/\Lambda^2$ operator coefficients in units of $\tev^{-2}$:
\begin{scriptsize}\begin{equation}
\begin{gathered}
\text{uncertainties} =
\left(\begin{array}{*{17}{c}}
0.09	& 0.04	& 0.04	& 0.014	& 0.015	& 0.55	& 0.007	& 0.18	& 0.18	& 0.9	& 0.0005	& 0.0005	& 0.0004	& 0.0005	& 0.0009	& 0.0005	& 0.0005
\end{array}\right)
\\\text{corr}=
\left(\begin{array}{*{17}{@{}c@{\,}}}
   1 &  -0.28&   0.28&   0.0044&  -0.0038&  -0.0075&  -0.0023&  -0.005 1&  -0.0059&   0.0014&   0.0014&   0.0026&   0.0013&  -0.00097 &   0.0024&  -0.0038&   0.00072\\ 
 -0.28&    1&  -0.99&  -0.0062&   0.005 1&   0.00065&   0.027 &   0.0067 &   0.016&   0.0038&  -0.0057 &   0.027 &   0.0063&  -0.0024&   0.0096&   0.0019&  -0.0006\\ 
  0.28&  -0.99&    1&   0.0068&  -0.0055&  -0.00058&   0.01 1&  -0.0052&  -0.014&  -0.0038&   0.0045&   0.027 &   0.024&   0.0029&   0.002&  -0.0018&  -0.0068\\ 
  0.0044&  -0.0062&   0.0068&    1&  -0.84&  -0.0023&   0.005 1&   0.014&   0.014&  -0.00054&   3.9*10^{-5} &   0.0099&  -0.0082&  -0.0014&   0.0062&  -0.0024&   0.0046\\ 
 -0.0038&   0.005 1&  -0.0055&  -0.84&    1&   0.0035&  -0.005&  -0.013&  -0.012&  -0.0014&  -0.00013&  -0.009&   0.0073&   0.0029&  -0.0059&   0.004&  -0.0038\\ 
 -0.0075&   0.00065&  -0.00058&  -0.0023&   0.0035&    1&   0.0035&  -0.0024&   0.00014&   0.0002&  -0.003 1&  -0.0026 &   0.0015&  -0.0025&   0.0038&  -0.00094&   0.0045 \\ 
 -0.0023&   0.027 &   0.01 1&   0.005 1&  -0.005&   0.0035&    1&  -0.0048 &   0.0037 &  -0.0033&   0.003&   0.15&  -0.1 1&  -0.0018&   0.029&   0.0013&   0.02 1\\ 
 -0.005 1&   0.0067 &  -0.0052&   0.014&  -0.013&  -0.0024&  -0.0048&    1&   0.23&   0.006 1&   0.0029&   0.0093&  -0.0018&   0.0048&  -0.02&   0.0033&  -0.028\\ 
 -0.0059&   0.016&  -0.014&   0.014&  -0.012&   0.00014&   0.0037 &   0.23&    1&   0.001 1&  -0.00066&   0.04 1&  -0.024&   0.00039&   0.0 1&  -0.0022&   0.008\\ 
  0.0014&   0.0038&  -0.0038&  -0.00054&  -0.0014&   0.0002&  -0.0033&   0.006 1&   0.001 1&    1&   0.0014&  -0.0036&   0.0055&  -0.0016&  -0.0026&  -0.00083&   0.0017 \\ 
  0.0014&  -0.0057 &   0.0045&   3.9*10^{-5} &  -0.00013&  -0.003 1&   0.003&   0.0029&  -0.00066&   0.0014&    1&   0.017 &  -0.035&  -0.0008&  -0.0096&   0.0022&  -0.0046\\ 
  0.0026&   0.027 &   0.027 &   0.0099&  -0.009&  -0.0026&   0.15&   0.0093&   0.04 1&  -0.0036&   0.017 &    1&  -0.38&  -0.00012&   0.047 &  -0.002&  -0.019\\ 
  0.0013&   0.0063&   0.024&  -0.0082&   0.0073&   0.0015&  -0.1 1&  -0.0018&  -0.024&   0.0055&  -0.035&  -0.38&    1&   0.0063&   0.076&   0.0029&  -0.0039\\ 
 -0.00097 &  -0.0024&   0.0029&  -0.0014&   0.0029&  -0.0025&  -0.0018&   0.0048&   0.00039&  -0.0016&  -0.0008&  -0.00012&   0.0063&    1&   0.0045&  -0.18&   0.0064\\ 
  0.0024&   0.0096&   0.002&   0.0062&  -0.0059&   0.0038&   0.029&  -0.02&   0.0 1&  -0.0026&  -0.0096&   0.047 &   0.076&   0.0045&    1&   0.0052&   0.6 1\\ 
 -0.0038&   0.0019&  -0.0018&  -0.0024&   0.004&  -0.00094&   0.0013&   0.0033&  -0.0022&  -0.00083&   0.0022&  -0.002&   0.0029&  -0.18&   0.0052&    1&   0.0053\\ 
  0.00072&  -0.0006&  -0.0068&   0.0046&  -0.0038&   0.0045&   0.02 1&  -0.028&   0.008&   0.0017 &  -0.0046&  -0.019&  -0.0039&   0.0064&   0.6 1&   0.0053&   1
  \end{array}\right)
\end{gathered}
\label{eq:cov_matrix_6}
\end{equation}\end{scriptsize}

\end{landscape}

\onecolumn{
\bibliographystyle{JHEP_2}
\bibliography{EW_EFT,top_yukawa,biblio,HEPfit_bib}

\providecommand{\href}[2]{#2}\begingroup\raggedright\begin{thebibliography}{100}

\bibitem{Aad:2012tfa}
{\scshape ATLAS} collaboration, \emph{{Observation of a new particle in the
  search for the Standard Model Higgs boson with the ATLAS detector at the
  LHC}}, \href{https://doi.org/10.1016/j.physletb.2012.08.020}{\emph{Phys.
  Lett.} {\bfseries B716} (2012) 1}
  [\href{https://arxiv.org/abs/1207.7214}{{\ttfamily 1207.7214}}].

\bibitem{Chatrchyan:2012xdj}
{\scshape CMS} collaboration, \emph{{Observation of a New Boson at a Mass of
  125 GeV with the CMS Experiment at the LHC}},
  \href{https://doi.org/10.1016/j.physletb.2012.08.021}{\emph{Phys. Lett.}
  {\bfseries B716} (2012) 30}
  [\href{https://arxiv.org/abs/1207.7235}{{\ttfamily 1207.7235}}].

\bibitem{Richard:2014upa}
F.~Richard, \emph{{Present and future constraints on top EW couplings}},
  \href{https://arxiv.org/abs/1403.2893}{{\ttfamily 1403.2893}}.

\bibitem{Durieux:2018ekg}
G.~Durieux and O.~Matsedonskyi, \emph{{The top-quark window on compositeness at
  future lepton colliders}},
  \href{https://doi.org/10.1007/JHEP01(2019)072}{\emph{JHEP} {\bfseries 01}
  (2019) 072} [\href{https://arxiv.org/abs/1807.10273}{{\ttfamily
  1807.10273}}].

\bibitem{Englert:2017dev}
C.~Englert and M.~Russell, \emph{{Top quark electroweak couplings at future
  lepton colliders}},
  \href{https://doi.org/10.1140/epjc/s10052-017-5095-z}{\emph{Eur. Phys. J.}
  {\bfseries C77} (2017) 535}
  [\href{https://arxiv.org/abs/1704.01782}{{\ttfamily 1704.01782}}].

\bibitem{Durieux:2018tev}
G.~Durieux, M.~Perelló, M.~Vos and C.~Zhang, \emph{{Global and optimal probes
  for the top-quark effective field theory at future lepton colliders}},
  \href{https://doi.org/10.1007/JHEP10(2018)168}{\emph{JHEP} {\bfseries 10}
  (2018) 168} [\href{https://arxiv.org/abs/1807.02121}{{\ttfamily
  1807.02121}}].

\bibitem{hepfitgit}
\url{https://github.com/silvest/HEPfit/blob/master/NewPhysics/src/NPSMEFT6dtopquark.cpp}.

\bibitem{deBlas:2016ojx}
J.~de~Blas, M.~Ciuchini, E.~Franco, S.~Mishima, M.~Pierini, L.~Reina et~al.,
  \emph{{Electroweak precision observables and Higgs-boson signal strengths in
  the Standard Model and beyond: present and future}},
  \href{https://doi.org/10.1007/JHEP12(2016)135}{\emph{JHEP} {\bfseries 12}
  (2016) 135} [\href{https://arxiv.org/abs/1608.01509}{{\ttfamily
  1608.01509}}].

\bibitem{Caldwell:2008fw}
A.~Caldwell, D.~Kollar and K.~Kroninger, \emph{{BAT: The Bayesian Analysis
  Toolkit}}, \href{https://doi.org/10.1016/j.cpc.2009.06.026}{\emph{Comput.
  Phys. Commun.} {\bfseries 180} (2009) 2197}
  [\href{https://arxiv.org/abs/0808.2552}{{\ttfamily 0808.2552}}].

\bibitem{ApollinariG.:2017ojx}
G.~Apollinari, I.~Béjar~Alonso, O.~Brüning, P.~Fessia, M.~Lamont, L.~Rossi
  et~al., \emph{{High-Luminosity Large Hadron Collider (HL-LHC)}},
  \href{https://doi.org/10.23731/CYRM-2017-004}{\emph{CERN Yellow Rep. Monogr.}
  {\bfseries 4} (2017) 1}.

\bibitem{Bambade:2019fyw}
P.~Bambade et~al., \emph{{The International Linear Collider: A Global
  Project}},  \href{https://arxiv.org/abs/1903.01629}{{\ttfamily 1903.01629}}.

\bibitem{Charles:2018vfv}
{\scshape CLICdp, CLIC} collaboration, \emph{{The Compact Linear Collider
  (CLIC) - 2018 Summary Report}},
  \href{https://doi.org/10.23731/CYRM-2018-002}{\emph{CERN Yellow Rep. Monogr.}
  {\bfseries 1802} (2018) 1}
  [\href{https://arxiv.org/abs/1812.06018}{{\ttfamily 1812.06018}}].

\bibitem{Abada:2019zxq}
{\scshape FCC} collaboration, \emph{{FCC-ee: The Lepton Collider}},
  \href{https://doi.org/10.1140/epjst/e2019-900045-4}{\emph{Eur. Phys. J. ST}
  {\bfseries 228} (2019) 261}.

\bibitem{CEPCStudyGroup:2018rmc}
{\scshape CEPC Study Group} collaboration, \emph{{CEPC Conceptual Design
  Report: Volume 1 - Accelerator}},
  \href{https://arxiv.org/abs/1809.00285}{{\ttfamily 1809.00285}}.

\bibitem{Bilokin:2017lco}
S.~Bilokin, R.~Pöschl and F.~Richard, \emph{{Measurement of b quark EW
  couplings at ILC}},  \href{https://arxiv.org/abs/1709.04289}{{\ttfamily
  1709.04289}}.

\bibitem{Abramowicz:2018rjq}
{\scshape CLICdp} collaboration, \emph{{Top-Quark Physics at the CLIC
  Electron-Positron Linear Collider}},
  \href{https://arxiv.org/abs/1807.02441}{{\ttfamily 1807.02441}}.

\bibitem{Amjad:2015mma}
M.~S. Amjad et~al., \emph{{A precise characterisation of the top quark
  electro-weak vertices at the ILC}},
  \href{https://doi.org/10.1140/epjc/s10052-015-3746-5}{\emph{Eur. Phys. J.}
  {\bfseries C75} (2015) 512}
  [\href{https://arxiv.org/abs/1505.06020}{{\ttfamily 1505.06020}}].

\bibitem{Buckley:2015nca}
A.~Buckley, C.~Englert, J.~Ferrando, D.~J. Miller, L.~Moore, M.~Russell et~al.,
  \emph{{Global fit of top quark effective theory to data}},
  \href{https://doi.org/10.1103/PhysRevD.92.091501}{\emph{Phys. Rev.}
  {\bfseries D92} (2015) 091501}
  [\href{https://arxiv.org/abs/1506.08845}{{\ttfamily 1506.08845}}].

\bibitem{deBeurs:2018pvs}
M.~de~Beurs, E.~Laenen, M.~Vreeswijk and E.~Vryonidou, \emph{{Effective
  operators in $t$-channel single top production and decay}},
  \href{https://doi.org/10.1140/epjc/s10052-018-6399-3}{\emph{Eur. Phys. J.}
  {\bfseries C78} (2018) 919}
  [\href{https://arxiv.org/abs/1807.03576}{{\ttfamily 1807.03576}}].

\bibitem{Hartland:2019bjb}
N.~P. Hartland, F.~Maltoni, E.~R. Nocera, J.~Rojo, E.~Slade, E.~Vryonidou
  et~al., \emph{{A Monte Carlo global analysis of the Standard Model Effective
  Field Theory: the top quark sector}},
  \href{https://doi.org/10.1007/JHEP04(2019)100}{\emph{JHEP} {\bfseries 04}
  (2019) 100} [\href{https://arxiv.org/abs/1901.05965}{{\ttfamily
  1901.05965}}].

\bibitem{AguilarSaavedra:2018nen}
J.~A. Aguilar-Saavedra, C.~Degrande, G.~Durieux, F.~Maltoni, E.~Vryonidou and
  C.~Zhang~(eds.), \emph{{Interpreting top-quark LHC measurements in the
  standard-model effective field theory}},
  \href{https://arxiv.org/abs/1802.07237}{{\ttfamily 1802.07237}}.

\bibitem{Contino:2016jqw}
R.~Contino, A.~Falkowski, F.~Goertz, C.~Grojean and F.~Riva, \emph{{On the
  Validity of the Effective Field Theory Approach to SM Precision Tests}},
  \href{https://doi.org/10.1007/JHEP07(2016)144}{\emph{JHEP} {\bfseries 07}
  (2016) 144} [\href{https://arxiv.org/abs/1604.06444}{{\ttfamily
  1604.06444}}].

\bibitem{Birman:2016jhg}
J.~L. Birman, F.~Déliot, M.~C.~N. Fiolhais, A.~Onofre and C.~M. Pease,
  \emph{{New limits on anomalous contributions to the $Wtb$ vertex}},
  \href{https://doi.org/10.1103/PhysRevD.93.113021}{\emph{Phys. Rev.}
  {\bfseries D93} (2016) 113021}
  [\href{https://arxiv.org/abs/1605.02679}{{\ttfamily 1605.02679}}].

\bibitem{Bernreuther:2017cyi}
W.~Bernreuther, L.~Chen, I.~García, M.~Perelló, R.~Poeschl, F.~Richard
  et~al., \emph{{CP-violating top quark couplings at future linear $e^+e^-$
  colliders}}, \href{https://doi.org/10.1140/epjc/s10052-018-5625-3}{\emph{Eur.
  Phys. J.} {\bfseries C78} (2018) 155}
  [\href{https://arxiv.org/abs/1710.06737}{{\ttfamily 1710.06737}}].

\bibitem{Cirigliano:2016njn}
V.~Cirigliano, W.~Dekens, J.~de~Vries and E.~Mereghetti, \emph{{Is there room
  for CP violation in the top-Higgs sector?}},
  \href{https://doi.org/10.1103/PhysRevD.94.016002}{\emph{Phys. Rev.}
  {\bfseries D94} (2016) 016002}
  [\href{https://arxiv.org/abs/1603.03049}{{\ttfamily 1603.03049}}].

\bibitem{Grzadkowski:2010es}
B.~Grzadkowski, M.~Iskrzynski, M.~Misiak and J.~Rosiek, \emph{{Dimension-Six
  Terms in the Standard Model Lagrangian}},
  \href{https://doi.org/10.1007/JHEP10(2010)085}{\emph{JHEP} {\bfseries 10}
  (2010) 085} [\href{https://arxiv.org/abs/1008.4884}{{\ttfamily 1008.4884}}].

\bibitem{AguilarSaavedra:2008zc}
J.~A. Aguilar-Saavedra, \emph{{A Minimal set of top anomalous couplings}},
  \href{https://doi.org/10.1016/j.nuclphysb.2008.12.012}{\emph{Nucl. Phys.}
  {\bfseries B812} (2009) 181}
  [\href{https://arxiv.org/abs/0811.3842}{{\ttfamily 0811.3842}}].

\bibitem{Zhang:2010dr}
C.~Zhang and S.~Willenbrock, \emph{{Effective-Field-Theory Approach to
  Top-Quark Production and Decay}},
  \href{https://doi.org/10.1103/PhysRevD.83.034006}{\emph{Phys. Rev.}
  {\bfseries D83} (2011) 034006}
  [\href{https://arxiv.org/abs/1008.3869}{{\ttfamily 1008.3869}}].

\bibitem{Cabibbo:1963yz}
N.~Cabibbo, \emph{{Unitary Symmetry and Leptonic Decays}},
  \href{https://doi.org/10.1103/PhysRevLett.10.531}{\emph{Phys. Rev. Lett.}
  {\bfseries 10} (1963) 531}.

\bibitem{Kobayashi:1973fv}
M.~Kobayashi and T.~Maskawa, \emph{{CP Violation in the Renormalizable Theory
  of Weak Interaction}}, \href{https://doi.org/10.1143/PTP.49.652}{\emph{Prog.
  Theor. Phys.} {\bfseries 49} (1973) 652}.

\bibitem{Alwall:2014hca}
J.~Alwall, R.~Frederix, S.~Frixione, V.~Hirschi, F.~Maltoni, O.~Mattelaer
  et~al., \emph{{The automated computation of tree-level and next-to-leading
  order differential cross sections, and their matching to parton shower
  simulations}}, \href{https://doi.org/10.1007/JHEP07(2014)079}{\emph{JHEP}
  {\bfseries 07} (2014) 079} [\href{https://arxiv.org/abs/1405.0301}{{\ttfamily
  1405.0301}}].

\bibitem{Bylund:2016phk}
O.~Bessidskaia~Bylund, F.~Maltoni, I.~Tsinikos, E.~Vryonidou and C.~Zhang,
  \emph{{Probing top quark neutral couplings in the Standard Model Effective
  Field Theory at NLO in QCD}},
  \href{https://doi.org/10.1007/JHEP05(2016)052}{\emph{JHEP} {\bfseries 05}
  (2016) 052} [\href{https://arxiv.org/abs/1601.08193}{{\ttfamily
  1601.08193}}].

\bibitem{Brivio:2017btx}
I.~Brivio, Y.~Jiang and M.~Trott, \emph{{The SMEFTsim package, theory and
  tools}}, {\emph{JHEP} {\bfseries 12} (2017) 070}
  [\href{https://arxiv.org/abs/1709.06492}{{\ttfamily 1709.06492}}].

\bibitem{deBlas:2019okz}
J.~de~Blas et~al., \emph{{$\texttt{HEPfit}$: a Code for the Combination of
  Indirect and Direct Constraints on High Energy Physics Models}},
  \href{https://arxiv.org/abs/1910.14012}{{\ttfamily 1910.14012}}.

\bibitem{hepfitsite}
{\scshape {\tt HEPfit}} collaboration, \emph{{\tt HEPfit}, a tool to combine
  indirect and direct constraints on high-energy physics},
  {\emph{\url{http://hepfit.roma1.infn.it}} }.

\bibitem{Ana:2019}
O.~Eberhardt, A.~Pe\~nuelas and A.~Pich, \emph{{Global fits in the Aligned
  Two-Higgs-Doublet model}}, {\emph{{In preparation}} }.

\bibitem{Victor:2019}
O.~Eberhardt, V.~Miralles and A.~Pich, \emph{{Global fits in the coloured
  scalar model}}, {\emph{{In preparation}} }.

\bibitem{deBlas:2018tjm}
J.~de~Blas, O.~Eberhardt and C.~Krause, \emph{{Current and Future Constraints
  on Higgs Couplings in the Nonlinear Effective Theory}},
  \href{https://doi.org/10.1007/JHEP07(2018)048}{\emph{JHEP} {\bfseries 07}
  (2018) 048} [\href{https://arxiv.org/abs/1803.00939}{{\ttfamily
  1803.00939}}].

\bibitem{Ciuchini:2019usw}
M.~Ciuchini, A.~M. Coutinho, M.~Fedele, E.~Franco, A.~Paul, L.~Silvestrini
  et~al., \emph{{New Physics in $b \to s \ell^+ \ell^-$ confronts new data on
  Lepton Universality}},  \href{https://arxiv.org/abs/1903.09632}{{\ttfamily
  1903.09632}}.

\bibitem{James:2004xla}
F.~James and M.~Winkler, \emph{{MINUIT User's Guide}},
  {\emph{\url{https://inspirehep.net/record/1258345}} (2004) }.

\bibitem{Aaboud:2018urx}
{\scshape ATLAS} collaboration, \emph{{Observation of Higgs boson production in
  association with a top quark pair at the LHC with the ATLAS detector}},
  \href{https://doi.org/10.1016/j.physletb.2018.07.035}{\emph{Phys. Lett.}
  {\bfseries B784} (2018) 173}
  [\href{https://arxiv.org/abs/1806.00425}{{\ttfamily 1806.00425}}].

\bibitem{Sirunyan:2018hoz}
{\scshape CMS} collaboration, \emph{{Observation of $\mathrm{t\overline{t}}$H
  production}},
  \href{https://doi.org/10.1103/PhysRevLett.120.231801}{\emph{Phys. Rev. Lett.}
  {\bfseries 120} (2018) 231801}
  [\href{https://arxiv.org/abs/1804.02610}{{\ttfamily 1804.02610}}].

\bibitem{Aaboud:2019njj}
{\scshape ATLAS} collaboration, \emph{{Measurement of the $t\bar{t}Z$ and
  $t\bar{t}W$ cross sections in proton-proton collisions at $\sqrt{s}=13$ TeV
  with the ATLAS detector}},
  \href{https://doi.org/10.1103/PhysRevD.99.072009}{\emph{Phys. Rev.}
  {\bfseries D99} (2019) 072009}
  [\href{https://arxiv.org/abs/1901.03584}{{\ttfamily 1901.03584}}].

\bibitem{Sirunyan:2017uzs}
{\scshape CMS} collaboration, \emph{{Measurement of the cross section for top
  quark pair production in association with a W or Z boson in proton-proton
  collisions at $\sqrt{s} =$ 13 TeV}},
  \href{https://doi.org/10.1007/JHEP08(2018)011}{\emph{JHEP} {\bfseries 08}
  (2018) 011} [\href{https://arxiv.org/abs/1711.02547}{{\ttfamily
  1711.02547}}].

\bibitem{CMS:2019nos}
{\scshape CMS} collaboration, \emph{{Measurement of top quark pair production
  in association with a Z boson in proton-proton collisions at $\sqrt{s}$ = 13
  TeV}}, {\emph{CMS-PAS-TOP-18-009} (2019) }.

\bibitem{Aaboud:2018hip}
{\scshape ATLAS} collaboration, \emph{{Measurements of inclusive and
  differential fiducial cross-sections of $t\bar{t}\gamma$ production in
  leptonic final states at $\sqrt{s}$ = 13 TeV in ATLAS}}, {\emph{Submitted to:
  Eur. Phys. J.} (2018) } [\href{https://arxiv.org/abs/1812.01697}{{\ttfamily
  1812.01697}}].

\bibitem{Sirunyan:2018zgs}
{\scshape CMS} collaboration, \emph{{Observation of single top quark production
  in association with a Z boson in proton-proton collisions at $\sqrt{s} =$ 13
  TeV}}, {\emph{Submitted to: Phys. Rev. Lett.} (2018) }
  [\href{https://arxiv.org/abs/1812.05900}{{\ttfamily 1812.05900}}].

\bibitem{Aaboud:2017ylb}
{\scshape ATLAS} collaboration, \emph{{Measurement of the production
  cross-section of a single top quark in association with a Z boson in
  proton–proton collisions at 13 TeV with the ATLAS detector}},
  \href{https://doi.org/10.1016/j.physletb.2018.03.023}{\emph{Phys. Lett.}
  {\bfseries B780} (2018) 557}
  [\href{https://arxiv.org/abs/1710.03659}{{\ttfamily 1710.03659}}].

\bibitem{Czarnecki:2010gb}
A.~Czarnecki, J.~G. Korner and J.~H. Piclum, \emph{{Helicity fractions of W
  bosons from top quark decays at NNLO in QCD}},
  \href{https://doi.org/10.1103/PhysRevD.81.111503}{\emph{Phys. Rev.}
  {\bfseries D81} (2010) 111503}
  [\href{https://arxiv.org/abs/1005.2625}{{\ttfamily 1005.2625}}].

\bibitem{Aaboud:2016hsq}
{\scshape ATLAS} collaboration, \emph{{Measurement of the W boson polarisation
  in $t\bar{t}$ events from pp collisions at $\sqrt{s}$ = 8 TeV in the
  lepton + jets channel with ATLAS}},
  \href{https://doi.org/10.1140/epjc/s10052-018-6520-7,
  10.1140/epjc/s10052-017-4819-4}{\emph{Eur. Phys. J.} {\bfseries C77} (2017)
  264} [\href{https://arxiv.org/abs/1612.02577}{{\ttfamily 1612.02577}}].

\bibitem{Khachatryan:2016fky}
{\scshape CMS} collaboration, \emph{{Measurement of the W boson helicity
  fractions in the decays of top quark pairs to lepton $+$ jets final states
  produced in pp collisions at $\sqrt s=$ 8TeV}},
  \href{https://doi.org/10.1016/j.physletb.2016.10.007}{\emph{Phys. Lett.}
  {\bfseries B762} (2016) 512}
  [\href{https://arxiv.org/abs/1605.09047}{{\ttfamily 1605.09047}}].

\bibitem{Chatrchyan:2013jna}
{\scshape CMS} collaboration, \emph{{Measurement of the W-boson helicity in
  top-quark decays from $t\bar{t}$ production in lepton+jets events in pp
  collisions at $\sqrt{s} =$ 7 TeV}},
  \href{https://doi.org/10.1007/JHEP10(2013)167}{\emph{JHEP} {\bfseries 10}
  (2013) 167} [\href{https://arxiv.org/abs/1308.3879}{{\ttfamily 1308.3879}}].

\bibitem{Aad:2012ky}
{\scshape ATLAS} collaboration, \emph{{Measurement of the W boson polarization
  in top quark decays with the ATLAS detector}},
  \href{https://doi.org/10.1007/JHEP06(2012)088}{\emph{JHEP} {\bfseries 06}
  (2012) 088} [\href{https://arxiv.org/abs/1205.2484}{{\ttfamily 1205.2484}}].

\bibitem{Sirunyan:2018rlu}
{\scshape CMS} collaboration, \emph{{Measurement of the single top quark and
  antiquark production cross sections in the $t$ channel and their ratio in
  proton-proton collisions at $\sqrt{s}=$ 13 TeV}},
  \href{https://arxiv.org/abs/1812.10514}{{\ttfamily 1812.10514}}.

\bibitem{Aaboud:2016ymp}
{\scshape ATLAS} collaboration, \emph{{Measurement of the inclusive
  cross-sections of single top-quark and top-antiquark $t$-channel production
  in $pp$ collisions at $\sqrt{s}$ = 13 TeV with the ATLAS detector}},
  \href{https://doi.org/10.1007/JHEP04(2017)086}{\emph{JHEP} {\bfseries 04}
  (2017) 086} [\href{https://arxiv.org/abs/1609.03920}{{\ttfamily
  1609.03920}}].

\bibitem{Sirunyan:2018lcp}
{\scshape CMS} collaboration, \emph{{Measurement of the production cross
  section for single top quarks in association with W bosons in proton-proton
  collisions at $ \sqrt{s}=13 $ TeV}},
  \href{https://doi.org/10.1007/JHEP10(2018)117}{\emph{JHEP} {\bfseries 10}
  (2018) 117} [\href{https://arxiv.org/abs/1805.07399}{{\ttfamily
  1805.07399}}].

\bibitem{Aaboud:2016lpj}
{\scshape ATLAS} collaboration, \emph{{Measurement of the cross-section for
  producing a W boson in association with a single top quark in pp collisions
  at $ \sqrt{s}=13 $ TeV with ATLAS}},
  \href{https://doi.org/10.1007/JHEP01(2018)063}{\emph{JHEP} {\bfseries 01}
  (2018) 063} [\href{https://arxiv.org/abs/1612.07231}{{\ttfamily
  1612.07231}}].

\bibitem{Aaboud:2017yqf}
{\scshape ATLAS} collaboration, \emph{{Analysis of the $Wtb$ vertex from the
  measurement of triple-differential angular decay rates of single top quarks
  produced in the $t$-channel at $\sqrt{s}$ = 8 TeV with the ATLAS detector}},
  \href{https://doi.org/10.1007/JHEP12(2017)017}{\emph{JHEP} {\bfseries 12}
  (2017) 017} [\href{https://arxiv.org/abs/1707.05393}{{\ttfamily
  1707.05393}}].

\bibitem{Aaboud:2017aqp}
{\scshape ATLAS} collaboration, \emph{{Probing the W tb vertex structure in
  t-channel single-top-quark production and decay in pp collisions at $
  \sqrt{s}=8 $ TeV with the ATLAS detector}},
  \href{https://doi.org/10.1007/JHEP04(2017)124}{\emph{JHEP} {\bfseries 04}
  (2017) 124} [\href{https://arxiv.org/abs/1702.08309}{{\ttfamily
  1702.08309}}].

\bibitem{Khachatryan:2014vma}
{\scshape CMS} collaboration, \emph{{Measurement of the W boson helicity in
  events with a single reconstructed top quark in pp collisions at $ \sqrt{s}=8
  $ TeV}}, \href{https://doi.org/10.1007/JHEP01(2015)053}{\emph{JHEP}
  {\bfseries 01} (2015) 053} [\href{https://arxiv.org/abs/1410.1154}{{\ttfamily
  1410.1154}}].

\bibitem{ALEPH:2005ab}
{\scshape ALEPH, DELPHI, L3, OPAL, SLD, LEP Electroweak Working Group, SLD
  Electroweak Group, SLD Heavy Flavour Group} collaboration, \emph{{Precision
  electroweak measurements on the $Z$ resonance}},
  \href{https://doi.org/10.1016/j.physrep.2005.12.006}{\emph{Phys. Rept.}
  {\bfseries 427} (2006) 257}
  [\href{https://arxiv.org/abs/hep-ex/0509008}{{\ttfamily hep-ex/0509008}}].

\bibitem{Aad:2014dvb}
{\scshape ATLAS} collaboration, \emph{{Measurement of differential production
  cross-sections for a $Z$ boson in association with $b$-jets in 7 TeV
  proton-proton collisions with the ATLAS detector}},
  \href{https://doi.org/10.1007/JHEP10(2014)141}{\emph{JHEP} {\bfseries 10}
  (2014) 141} [\href{https://arxiv.org/abs/1407.3643}{{\ttfamily 1407.3643}}].

\bibitem{Chatrchyan:2012vr}
{\scshape CMS} collaboration, \emph{{Measurement of the Z/$\gamma$*+b-jet cross
  section in pp collisions at $\sqrt{s}$ = 7 TeV}},
  \href{https://doi.org/10.1007/JHEP06(2012)126}{\emph{JHEP} {\bfseries 06}
  (2012) 126} [\href{https://arxiv.org/abs/1204.1643}{{\ttfamily 1204.1643}}].

\bibitem{Brod:2014hsa}
J.~Brod, A.~Greljo, E.~Stamou and P.~Uttayarat, \emph{{Probing anomalous $
  t\overline{t}Z $ interactions with rare meson decays}},
  \href{https://doi.org/10.1007/JHEP02(2015)141}{\emph{JHEP} {\bfseries 02}
  (2015) 141} [\href{https://arxiv.org/abs/1408.0792}{{\ttfamily 1408.0792}}].

\bibitem{Bissmann:2019gfc}
S.~Bißmann, J.~Erdmann, C.~Grunwald, G.~Hiller and K.~Kröninger,
  \emph{{Constraining top-quark couplings combining top-quark and
  $\boldsymbol{B}$ decay observables}},
  \href{https://arxiv.org/abs/1909.13632}{{\ttfamily 1909.13632}}.

\bibitem{Grzadkowski:2008mf}
B.~Grzadkowski and M.~Misiak, \emph{{Anomalous Wtb coupling effects in the weak
  radiative B-meson decay}}, \href{https://doi.org/10.1103/PhysRevD.84.059903,
  10.1103/PhysRevD.78.077501}{\emph{Phys. Rev.} {\bfseries D78} (2008) 077501}
  [\href{https://arxiv.org/abs/0802.1413}{{\ttfamily 0802.1413}}].

\bibitem{Drobnak:2011aa}
J.~Drobnak, S.~Fajfer and J.~F. Kamenik, \emph{{Probing anomalous tWb
  interactions with rare B decays}},
  \href{https://doi.org/10.1016/j.nuclphysb.2011.10.004}{\emph{Nucl. Phys.}
  {\bfseries B855} (2012) 82}
  [\href{https://arxiv.org/abs/1109.2357}{{\ttfamily 1109.2357}}].

\bibitem{Fox:2007in}
P.~J. Fox, Z.~Ligeti, M.~Papucci, G.~Perez and M.~D. Schwartz,
  \emph{{Deciphering top flavor violation at the LHC with $B$ factories}},
  \href{https://doi.org/10.1103/PhysRevD.78.054008}{\emph{Phys. Rev.}
  {\bfseries D78} (2008) 054008}
  [\href{https://arxiv.org/abs/0704.1482}{{\ttfamily 0704.1482}}].

\bibitem{Aebischer:2015fzz}
J.~Aebischer, A.~Crivellin, M.~Fael and C.~Greub, \emph{{Matching of gauge
  invariant dimension-six operators for $b\to s$ and $b\to c$ transitions}},
  \href{https://doi.org/10.1007/JHEP05(2016)037}{\emph{JHEP} {\bfseries 05}
  (2016) 037} [\href{https://arxiv.org/abs/1512.02830}{{\ttfamily
  1512.02830}}].

\bibitem{Feruglio:2018fxo}
F.~Feruglio, P.~Paradisi and O.~Sumensari, \emph{{Implications of scalar and
  tensor explanations of $R_{D^{(\ast)}}$}},
  \href{https://doi.org/10.1007/JHEP11(2018)191}{\emph{JHEP} {\bfseries 11}
  (2018) 191} [\href{https://arxiv.org/abs/1806.10155}{{\ttfamily
  1806.10155}}].

\bibitem{Endo:2018gdn}
M.~Endo, T.~Kitahara and D.~Ueda, \emph{{SMEFT top-quark effects on $\Delta
  F=2$ observables}},
  \href{https://doi.org/10.1007/JHEP07(2019)182}{\emph{JHEP} {\bfseries 07}
  (2019) 182} [\href{https://arxiv.org/abs/1811.04961}{{\ttfamily
  1811.04961}}].

\bibitem{Greiner:2011tt}
N.~Greiner, S.~Willenbrock and C.~Zhang, \emph{{Effective Field Theory for
  Nonstandard Top Quark Couplings}},
  \href{https://doi.org/10.1016/j.physletb.2011.09.026}{\emph{Phys. Lett.}
  {\bfseries B704} (2011) 218}
  [\href{https://arxiv.org/abs/1104.3122}{{\ttfamily 1104.3122}}].

\bibitem{Zhang:2012cd}
C.~Zhang, N.~Greiner and S.~Willenbrock, \emph{{Constraints on Non-standard Top
  Quark Couplings}},
  \href{https://doi.org/10.1103/PhysRevD.86.014024}{\emph{Phys. Rev.}
  {\bfseries D86} (2012) 014024}
  [\href{https://arxiv.org/abs/1201.6670}{{\ttfamily 1201.6670}}].

\bibitem{Vryonidou:2018eyv}
E.~Vryonidou and C.~Zhang, \emph{{Dimension-six electroweak top-loop effects in
  Higgs production and decay}},
  \href{https://doi.org/10.1007/JHEP08(2018)036}{\emph{JHEP} {\bfseries 08}
  (2018) 036} [\href{https://arxiv.org/abs/1804.09766}{{\ttfamily
  1804.09766}}].

\bibitem{Boselli:2018zxr}
S.~Boselli, R.~Hunter and A.~Mitov, \emph{{Prospects for the determination of
  the top-quark Yukawa coupling at future $e^+e^-$ colliders}},
  \href{https://doi.org/10.1088/1361-6471/ab2e5c}{\emph{J. Phys.} {\bfseries
  G46} (2019) 095005} [\href{https://arxiv.org/abs/1805.12027}{{\ttfamily
  1805.12027}}].

\bibitem{Durieux:2018ggn}
G.~Durieux, J.~Gu, E.~Vryonidou and C.~Zhang, \emph{{Probing top-quark
  couplings indirectly at Higgs factories}},
  \href{https://doi.org/10.1088/1674-1137/42/12/123107}{\emph{Chin. Phys.}
  {\bfseries C42} (2018) 123107}
  [\href{https://arxiv.org/abs/1809.03520}{{\ttfamily 1809.03520}}].

\bibitem{Sunghoon}
S.~Jung, J.~Lee, M.~Perello, J.~Tian and M.~Vos, \emph{{Higgs and top precision
  at future $e^+ e^-$ colliders with renormalization group mixing}}, {\emph{in
  preparation} (2019) }.

\bibitem{Cirigliano:2016nyn}
V.~Cirigliano, W.~Dekens, J.~de~Vries and E.~Mereghetti, \emph{{Constraining
  the top-Higgs sector of the Standard Model Effective Field Theory}},
  \href{https://doi.org/10.1103/PhysRevD.94.034031}{\emph{Phys. Rev.}
  {\bfseries D94} (2016) 034031}
  [\href{https://arxiv.org/abs/1605.04311}{{\ttfamily 1605.04311}}].

\bibitem{Sirunyan:2017nbr}
{\scshape CMS} collaboration, \emph{{Measurement of the associated production
  of a single top quark and a Z boson in pp collisions at $\sqrt{s} =$ 13
  TeV}}, \href{https://doi.org/10.1016/j.physletb.2018.02.025}{\emph{Phys.
  Lett.} {\bfseries B779} (2018) 358}
  [\href{https://arxiv.org/abs/1712.02825}{{\ttfamily 1712.02825}}].

\bibitem{CMS:2019bke}
{\scshape CMS} collaboration, \emph{{First constraints on invisible Higgs boson
  decays using $\mathrm{t}\bar{\mathrm{t}}\mathrm{H}$ production at
  $\sqrt{s}=13~\mathrm{TeV}$}}, .

\bibitem{Azzi:2019yne}
{\scshape HL-LHC, HE-LHC Working Group} collaboration, \emph{{Standard Model
  Physics at the HL-LHC and HE-LHC}},
  \href{https://arxiv.org/abs/1902.04070}{{\ttfamily 1902.04070}}.

\bibitem{Cepeda:2019klc}
{\scshape Physics of the HL-LHC Working Group} collaboration, \emph{{Higgs
  Physics at the HL-LHC and HE-LHC}},
  \href{https://arxiv.org/abs/1902.00134}{{\ttfamily 1902.00134}}.

\bibitem{CMS:2018rcv}
{\scshape CMS} collaboration, \emph{{Observation of single top quark production
  in association with a Z boson in proton-proton collisions at $\sqrt{s} =
  13~\mathrm{TeV}$}}, {\emph{CMS-PAS-TOP-18-008} }.

\bibitem{Schulze:2016qas}
M.~Schulze and Y.~Soreq, \emph{{Pinning down electroweak dipole operators of
  the top quark}},
  \href{https://doi.org/10.1140/epjc/s10052-016-4263-x}{\emph{Eur. Phys. J.}
  {\bfseries C76} (2016) 466}
  [\href{https://arxiv.org/abs/1603.08911}{{\ttfamily 1603.08911}}].

\bibitem{Rontsch:2014cca}
R.~R{\"o}ntsch and M.~Schulze, \emph{{Constraining couplings of top quarks to
  the Z boson in $ t\overline{t} $ + Z production at the LHC}},
  \href{https://doi.org/10.1007/JHEP09(2015)132,
  10.1007/JHEP07(2014)091}{\emph{JHEP} {\bfseries 07} (2014) 091}
  [\href{https://arxiv.org/abs/1404.1005}{{\ttfamily 1404.1005}}].

\bibitem{Fuster:2015jva}
J.~Fuster, I.~García, P.~Gomis, M.~Perelló, E.~Ros and M.~Vos, \emph{{Study
  of single top production at high energy electron positron colliders}},
  \href{https://doi.org/10.1140/epjc/s10052-015-3453-2}{\emph{Eur. Phys. J.}
  {\bfseries C75} (2015) 223}
  [\href{https://arxiv.org/abs/1411.2355}{{\ttfamily 1411.2355}}].

\bibitem{adrian}
S.~Bilokin, A.~Irles, R.~Pöschl and F.~Richard, \emph{{Measurement of $b$
  quark EW couplings at ILC250 with 2000~\ifb}}, {\emph{In preparation} }.

\bibitem{Barklow:2015tja}
T.~Barklow, J.~Brau, K.~Fujii, J.~Gao, J.~List, N.~Walker et~al., \emph{{ILC
  Operating Scenarios}},  \href{https://arxiv.org/abs/1506.07830}{{\ttfamily
  1506.07830}}.

\bibitem{AguilarSaavedra:2012vh}
J.~A. Aguilar-Saavedra, M.~C.~N. Fiolhais and A.~Onofre, \emph{{Top Effective
  Operators at the ILC}},
  \href{https://doi.org/10.1007/JHEP07(2012)180}{\emph{JHEP} {\bfseries 07}
  (2012) 180} [\href{https://arxiv.org/abs/1206.1033}{{\ttfamily 1206.1033}}].

\bibitem{CLIC:2016zwp}
{\scshape CLIC, CLICdp} collaboration, \emph{{Updated baseline for a staged
  Compact Linear Collider}},
  \href{https://arxiv.org/abs/1608.07537}{{\ttfamily 1608.07537}}.

\bibitem{Pontecorvo:1957cp}
B.~Pontecorvo, \emph{{Mesonium and anti-mesonium}}, {\emph{Sov. Phys. JETP}
  {\bfseries 6} (1957) 429}.

\bibitem{Maki:1962mu}
Z.~Maki, M.~Nakagawa and S.~Sakata, \emph{{Remarks on the unified model of
  elementary particles}}, \href{https://doi.org/10.1143/PTP.28.870}{\emph{Prog.
  Theor. Phys.} {\bfseries 28} (1962) 870}.

\bibitem{Pontecorvo:1967fh}
B.~Pontecorvo, \emph{{Neutrino Experiments and the Problem of Conservation of
  Leptonic Charge}}, {\emph{Sov. Phys. JETP} {\bfseries 26} (1968) 984}.

\bibitem{Dawson:2013bba}
S.~Dawson et~al., \emph{{Working Group Report: Higgs Boson}},  in
  \emph{{Community Summer Study 2013: Snowmass on the Mississippi (CSS2013)
  Minneapolis, MN, USA, July 29-August 6, 2013}}, 2013,
  \href{https://arxiv.org/abs/1310.8361}{{\ttfamily 1310.8361}},
  \href{http://inspirehep.net/record/1262795/files/arXiv:1310.8361.pdf}{http://inspirehep.net/record/1262795/files/arXiv:1310.8361.pdf}.

\bibitem{Khachatryan:2016vau}
{\scshape ATLAS, CMS} collaboration, \emph{{Measurements of the Higgs boson
  production and decay rates and constraints on its couplings from a combined
  ATLAS and CMS analysis of the LHC pp collision data at $ \sqrt{s}=7 $ and 8
  TeV}}, \href{https://doi.org/10.1007/JHEP08(2016)045}{\emph{JHEP} {\bfseries
  08} (2016) 045} [\href{https://arxiv.org/abs/1606.02266}{{\ttfamily
  1606.02266}}].

\bibitem{ATLAS-CONF-2019-005}
{\scshape ATLAS} collaboration, \emph{{Combined measurements of Higgs boson
  production and decay using up to $80$ fb$^{-1}$ of proton--proton collision
  data at $\sqrt{s}=$ 13 TeV collected with the ATLAS experiment}},  Tech. Rep.
  ATLAS-CONF-2019-005, CERN, Geneva, Mar, 2019.

\bibitem{Sirunyan:2018koj}
{\scshape CMS} collaboration, \emph{{Combined measurements of Higgs boson
  couplings in proton–proton collisions at $\sqrt{s}=13\,\text {Te}\text {V}
  $}}, \href{https://doi.org/10.1140/epjc/s10052-019-6909-y}{\emph{Eur. Phys.
  J.} {\bfseries C79} (2019) 421}
  [\href{https://arxiv.org/abs/1809.10733}{{\ttfamily 1809.10733}}].

\bibitem{Azatov:2016xik}
A.~Azatov, C.~Grojean, A.~Paul and E.~Salvioni, \emph{{Resolving gluon fusion
  loops at current and future hadron colliders}},
  \href{https://doi.org/10.1007/JHEP09(2016)123}{\emph{JHEP} {\bfseries 09}
  (2016) 123} [\href{https://arxiv.org/abs/1608.00977}{{\ttfamily
  1608.00977}}].

\bibitem{Price:2014oca}
T.~Price, P.~Roloff, J.~Strube and T.~Tanabe, \emph{{Full simulation study of
  the top Yukawa coupling at the ILC at $\sqrt{s}=$ 1 TeV}},
  \href{https://doi.org/10.1140/epjc/s10052-015-3532-4}{\emph{Eur. Phys. J.}
  {\bfseries C75} (2015) 309}
  [\href{https://arxiv.org/abs/1409.7157}{{\ttfamily 1409.7157}}].

\bibitem{Yonamine:2011jg}
R.~Yonamine, K.~Ikematsu, T.~Tanabe, K.~Fujii, Y.~Kiyo, Y.~Sumino et~al.,
  \emph{{Measuring the top Yukawa coupling at the ILC at $\sqrt{s}=500$ GeV}},
  \href{https://doi.org/10.1103/PhysRevD.84.014033}{\emph{Phys. Rev.}
  {\bfseries D84} (2011) 014033}
  [\href{https://arxiv.org/abs/1104.5132}{{\ttfamily 1104.5132}}].

\bibitem{Gay:2006vs}
A.~Gay, \emph{{Measurement of the top-Higgs Yukawa coupling at a Linear e+ e-
  Collider}}, \href{https://doi.org/10.1140/epjc/s10052-006-0161-y}{\emph{Eur.
  Phys. J.} {\bfseries C49} (2007) 489}
  [\href{https://arxiv.org/abs/hep-ph/0604034}{{\ttfamily hep-ph/0604034}}].

\bibitem{Juste:1999af}
A.~Juste and G.~Merino, \emph{{Top Higgs-Yukawa coupling measurement at a
  linear $e^+ e^-$ collider}},
  \href{https://arxiv.org/abs/hep-ph/9910301}{{\ttfamily hep-ph/9910301}}.

\bibitem{Fujii:2015jha}
K.~Fujii et~al., \emph{{Physics Case for the International Linear Collider}},
  \href{https://arxiv.org/abs/1506.05992}{{\ttfamily 1506.05992}}.

\bibitem{Nejad:2016bci}
B.~Chokoufé~Nejad, W.~Kilian, J.~M. Lindert, S.~Pozzorini, J.~Reuter and
  C.~Weiss, \emph{{NLO QCD predictions for off-shell $ t\overline{t} $ and $
  t\overline{t}H $ production and decay at a linear collider}},
  \href{https://doi.org/10.1007/JHEP12(2016)075}{\emph{JHEP} {\bfseries 12}
  (2016) 075} [\href{https://arxiv.org/abs/1609.03390}{{\ttfamily
  1609.03390}}].

\bibitem{deBlas:2019rxi}
J.~De~Blas et~al., \emph{{Higgs Boson Studies at Future Particle Colliders}},
  \href{https://arxiv.org/abs/1905.03764}{{\ttfamily 1905.03764}}.

\end{thebibliography}\endgroup
}

\end{document}